\documentclass[a4paper,11pt, openany]{book}

\pdfoutput=1
\usepackage{ifpdf}

\usepackage[USenglish]{babel}

\usepackage{float}
\usepackage{amssymb}
\usepackage{mathrsfs}
\usepackage{amsmath}
\usepackage{graphicx}
\usepackage{indentfirst}
\usepackage{textcomp}
\usepackage{gensymb}
\usepackage{adjustbox}
\usepackage{subcaption}
\usepackage{multirow}
\usepackage{mathtools}
\usepackage{cancel}

\usepackage[top=3cm, bottom=3cm, left=2cm, right=2cm]{geometry}
\usepackage{url}
\usepackage{epigraph}
\usepackage[compact]{titlesec}  
\usepackage[nodisplayskipstretch]{setspace}
\usepackage{hyperref}
\usepackage{listings}
\usepackage{xcolor}
\usepackage{setspace}
\definecolor{codegreen}{rgb}{0,0.6,0}
\definecolor{codegray}{rgb}{0.5,0.5,0.5}
\definecolor{codepurple}{rgb}{0.58,0,0.82}
\definecolor{backcolour}{rgb}{0.95,0.95,0.92}
\usepackage{lineno}
\usepackage[nottoc]{tocbibind}
\usepackage{pdfpages}
    
\lstdefinestyle{mystyle}{
	backgroundcolor=\color{backcolour},   
	commentstyle=\color{codegreen},
	keywordstyle=\color{magenta},
	numberstyle=\tiny\color{codegray},
	stringstyle=\color{codepurple},
	basicstyle=\ttfamily\footnotesize,
	breakatwhitespace=false,         
	breaklines=true,                 
	captionpos=b,                    
	keepspaces=true,                 
	numbers=left,                    
	numbersep=5pt,                  
	showspaces=false,                
	showstringspaces=false,
	showtabs=false,                  
	tabsize=2
}
\hypersetup{
	hidelinks
}
\titleformat{\chapter}[display]
{\bfseries\Huge}                                            
{\filright}
{1ex}{}[]    

\newcommand*{\BeginNoToc}{%
  \addtocontents{toc}{%
    \edef\protect\SavedTocDepth{\protect\the\protect\value{tocdepth}}%
  }%
  \addtocontents{toc}{%
    \protect\setcounter{tocdepth}{-10}%
  }%
}

\makeatletter
\newcommand{\emptypage}[1]{%
  \cleardoublepage
  \begingroup
  \let\ps@plain\ps@empty
  \pagestyle{empty}
  #1
  \cleardoublepage}
\makeatletter

\usepackage{fancyhdr}

\usepackage[backend=biber,style=nature]{biblatex}
\bibliography{bibliografia}

\long\def\mytitle{%
    \begin{titlepage}
        \begin{minipage}{0.125\textwidth}%
            \includegraphics[width=0.8\textwidth]{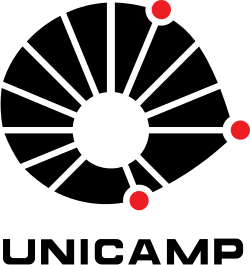}%
        \end{minipage}
        \begin{minipage}{0.7\textwidth}%
            \centering
            {\Large UNIVERSIDADE ESTADUAL DE CAMPINAS}\\[3pt]%
            {\Large Instituto de Física ``Gleb Wataghin''}\\[19pt]%
        \end{minipage}%
    \vspace{2cm}   

    \begin{center}
        {\Large \bfseries João Paulo Picchetti} \\[5pt]
        \vspace{3cm}

        \textsc{\LARGE{{Estudando o impacto do tamanho do nucleon em colisões de íons pesados relativísticos}}} \\[5pt]
        \vspace{2cm}
 
        \textsc{\LARGE{{\bfseries Studying the impact of the nucleon size in relativistic heavy-ion collisions}}} \\[5pt]
        \vfill

    {Campinas}\\[5pt]
    {2022}

    \end{center}
    \end{titlepage}
}

\begin{document}
	\renewcommand{\contentsname}{Contents}
	\renewcommand{\figurename}{Figure}
	\renewcommand{\lstlistingname}{Algorithm}
	\renewcommand{\thefootnote}{\fnsymbol{footnote}}
	\renewcommand{\chaptername}{Chapter}	
	\renewcommand{\bibsection}{\chapter{References}}
	\renewcommand{\tablename}{Table}
	\renewcommand{\listfigurename}{List of Figures}
	\setstretch{1.5}
	\doublespacing
	
    \mytitle

\thispagestyle{empty}
\setcounter{page}{1}

\let\cleardoublepage\clearpage

\newpage

    \setstretch{1.5}
    \doublespacing

\begin{titlepage}
\begin{center}

    {\Large \bfseries João Paulo Picchetti}\\[5pt]
    \vspace{1cm}
    
    \textsc{\LARGE{{\bfseries Studying the impact of the nucleon size in relativistic heavy-ion collisions}}} \\[5pt]
     \vspace{1cm}

    \textsc{\LARGE{{Estudando o impacto do tamanho do nucleon em colisões de íons pesados relativísticos}}} \\[5pt]
    \vspace{1cm}

    \begin{flushright}
    {Dissertação apresentada ao Instituto de Física \\``Gleb Wataghin'' da Universidade Estadual de Campinas \\ como parte dos requisitos exigidos para obtenção do \\ título de Mestre em física, na área de física.} \\[1cm]
    \end{flushright}
    
    \begin{flushright}
    {Dissertation presented to the ``Gleb Wataghin'' \\ Institute of Physics of the University of Campinas \\ in partial fulfillment of the requeriments for the \\ degree of Master in Physics, in the area of Physics.} \\[1cm]
    \end{flushright}

    \begin{flushleft}
    \Large{\textbf{Supervisor}: Jun Takahashi }
    \end{flushleft}

    \begin{flushleft}
    \small{
    ESTE TRABALHO CORRESPONDE À VERSÃO FINAL \\
    DA DISSERTAÇÃO DEFENDIDA PELO ALUNO\\
    JOÃO PAULO PICCHETTI, E ORIENTADO\\
    PELO PROF. DR. JUN TAKAHASHI.}
    \end{flushleft}
    \vspace{1cm}
    
    {Campinas}\\[5pt]
    {2022}
    \end{center}
    \pagenumbering{arabic} 
    \setcounter{page}{2}
    \end{titlepage}
    
\newpage






\thispagestyle{empty}
\newpage

\section*{\Huge{Abstract}}
\noindent\rule{\textwidth}{.1pt}\\[1ex]

Under the extreme conditions of temperature generated in relativistic heavy-ion collisions, a fascinating fluid-like state of matter where quarks and gluons are no longer confined is formed, the Quark Gluon Plasma (QGP). The most modern computational approaches are multi-stage (hybrid) simulations, in which different models are used in a chain structure, each one dedicated to the description of a specific stage of the collision.  

The hydrodynamic stage of the simulation requires an energy density profile of the system as an initial condition. In the process of converting the two colliding nuclei in such an energy distribution, some specification about the  \textit{nucleon size} inevitably has to be made. Nucleons are usually modeled as bidimensional Gaussians, and the Gaussian width (the nucleon-width) is a free parameter of the simulation. A best-fit value of the nucleon-width can be inferred by Bayesian Analyses, where the model is confronted with experimental data. 

Some of the most recent analyses have obtained surprisingly large values for the nucleon width parameter, exceeding in over 50 \% the current value for the proton charged radius. This motivates the development of a better understanding of the role played by this parameter inside the simulation. 

In this work, we perform simulations of relativistic heavy-ion collisions using a state-of-the-art hybrid simulation chain, using three different values of the nucleon width inside the initial condition generator T$_{\text{R}}$ENTo, and systematically investigate its effects on the initial condition characteristics and observables. The nucleon-width strongly affects the eccentricity harmonics and the gradients in the initial condition. The mean $p_{T}$ of particles in the simulation using $w$ = 0.5 fm is much larger than experimental data. We associate this to the combination of stronger gradients in the initial condition and the coupling of a conformal pre-equilibrium dynamics to the hydrodynamic simulation.

    \textbf{Keywords}: Heavy-ion phenomenology, Heavy-ion collisions, High-energy nuclear physics.

\thispagestyle{empty}
\newpage


\emptypage\tableofcontents
\thispagestyle{empty}

\newpage




\makeatletter
\let\savedchap\@makechapterhead
\def\@makechapterhead{\vspace*{-5cm}\savedchap}

\pagestyle{fancy} 
    \fancyhf{} 
    \fancyhead[R]{\thepage}
    \fancyhead[L]{\leftmark}

    \fancypagestyle{plain}{%
    \renewcommand{\headrulewidth}{0pt}%
    \fancyhf{}%
    \fancyhead[R]{\thepage}%
    }


\chapter{Introduction}
\noindent\rule{\textwidth}{.1pt}\\[1ex]

While at Earth-like conditions quarks are confined inside the protons and neutrons, which are in their turn bound together in the atomic nuclei, at sufficiently high energies a phase transition occurs to a fluid-like state of matter where quarks and gluons are no longer confined, the Quark Gluon Plasma (QGP).

This chapter is structured as an introduction to high-energy physics, focusing on the topics which are relevant for the physics of relativistic heavy-ion collisions and for the development of this work. Section 1.1 is an introduction to the Standard Model of particle physics, while Section 1.2 provides an overview of the most important aspects of Quantum Chromodynamics (QCD), the theory of the strong interaction. Section 1.3 is dedicated to a brief discussion about the QCD phase diagram and phase transitions in QCD. 

\section{The Standard Model}
\label{SEC-SM}

Particle physics studies the fundamental building blocks of the Universe and the interactions between them. While the gravitational interaction is the subject of Einstein's Theory of General Relativity, the other three fundamental forces can be studied in a single unified framework, which is the Standard Model of Particle Physics (SM) \cite{MarkThomson}. These are the electromagnetic force, which is the only long range SM interaction (and therefore, the only one that we are able to actually experience daily), and also the strong and weak forces, which act on a short range and are relevant only in the scale of quantum mechanics. The strong force is responsible for holding the protons and neutrons together in atomic nuclei, while the weak force is related to processes such as the $\beta$-decay.

\begin{figure}[h]
	\centering
	\includegraphics[width=8cm]{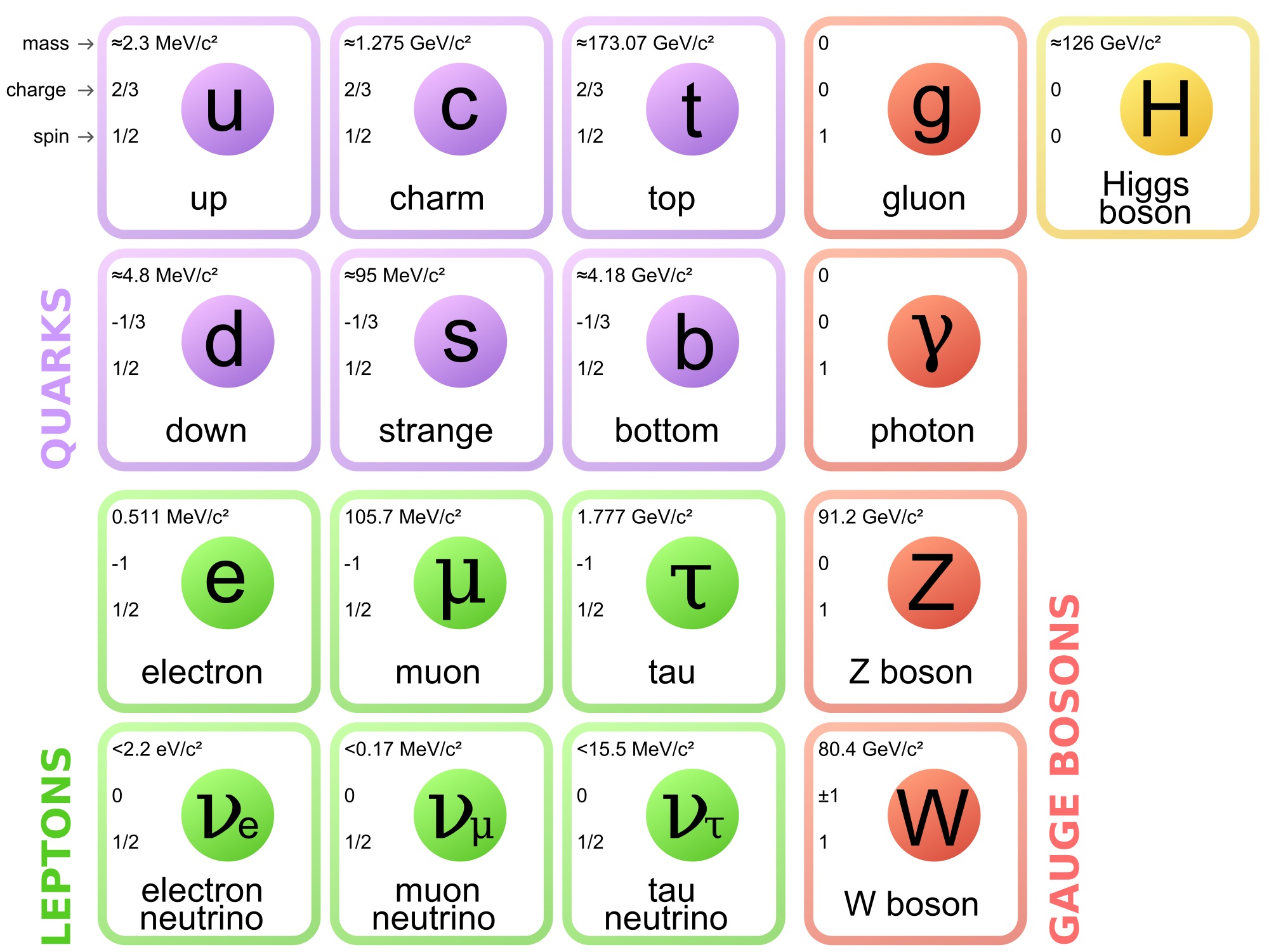}
	\caption{Standard Model particles. Figure taken from \cite{SM_Website}.}
	\label{FIG-SM}
\end{figure}

The formal description of particles and their interactions is the object of Quantum Field Theory (QFT) \cite{Schwartz}, which combines quantum mechanics, special relativity and classical field theory. In QFT, particles are viewed as excited states of quantum fields, and the interactions are mediated by the exchange of virtual bosons. Historically, QFT was first developed as a quantum theory of the electromagnetic interaction \cite{QED1}, and the term Quantum Electrodynamics (QED) was first used by Paul Dirac in 1927. 

In the SM, matter is made of twelve elementary spin-1/2 fermions which interact through the fundamental forces. To interact via one of the forces, a particle must posses the respective interaction charge. All the elementary fermions interact through the weak force, while particles with electric charge participate in the electromagnetic interaction. The charge of the strong interaction is called \textit{color charge}. The twelve fundamental fermions are separated in two groups of six based on whether they have color charge or not. The six fermions which are color charged are called quarks, while the ones which are not are known as leptons. 

The elementary fermions interact by exchanging spin-1 bosons (the gauge bosons). For this reason, the gauge bosons are usually referred to as the \textit{force carriers} (or mediators) of the SM. The photon ($\gamma$) is the force carrier of the electromagnetic interaction, the gluon ($g$) is the mediator of the strong interaction, and the weak interaction is mediated by the neutral $Z$ boson and also by the charged $W^{\pm}$ bosons. In addition, all massive SM particles are coupled to the Higgs field. The Higgs boson was theoretically proposed by Peter Higgs in 1964 \cite{Higgs} and experimentally discovered in 2012 both by the ATLAS \cite{20121} and CMS \cite{CMS:2012qbp, CMS:2013btf} Collaborations. Table 1.1 shows a schematic view of which of the elementary fermions participate in each interaction.

\begin{table}[h]
    \centering
    	\begin{tabular}{| c | c  c  c |} 
    		\hline
    		& Strong & Electromagnetic & Weak \\ 
    		\hline
    		& & & \\
    		Quarks (u, d, c, s, t, b) & \checkmark & \checkmark & \checkmark \\ 
    		& & & \\
    		Charged leptons ($e^{-}$, $\mu^{-}$, $\tau^{-}$) &  & \checkmark & \checkmark \\ 
    		& & & \\		
    		Neutrinos ($\nu_{e}$, $\nu_{\mu}$, $\nu_{\tau}$) &  &  & \checkmark \\ 
    		& & & \\
    		\hline
    	\end{tabular}
    \caption{Standard Model particles and interactions in which they participate}
\end{table}

There is a mass hierarchy between the elementary fermions in the SM, that separates them in three generations. The first generation consists on the u and d quarks, which form the protons (uud) and neutrons (udd), and also the electron ($e^{-}$) and the electron neutrino ($\nu_{e}$). These particles dominate the low-energy world in which we live in and, with the exception of the electron neutrino, are the building blocks of atoms. Being the constituents of atomic nuclei, protons and neutrons are also known as \textit{nucleons}. For each particle in the first generation, there is a heavier one, which differs from its first generation partner only by the mass. The second generation is formed by the c and s quarks, the muon ($\mu^{-}$) and the muon neutrino ($\nu_{\mu}$), while the third generation is composed by the t and b quarks, the tau ($\tau^{-}$) and the tau neutrino ($\nu_{\tau}$). Each elementary fermion has also a corresponding anti-particle, which is identical except for the electric charge, which has the opposite sign.

\section{Quantum Chromodynamics (QCD)}
\label{SEC-QCD}

Quantum Chromodynamics (QCD) is the QFT that studies the strong interaction \cite{QCDReview1, QCDReview2}. In opposition to the electric charge, which comes in two types (positive and negative), color charge comes in six types: red, green, blue, and the corresponding anti-colors. In QCD, quarks interact by exchanging gluons, which are themselves color charged and also interact with each other through the same mechanism. This is an expression of the fact that QCD is a non-Abelian field theory, which means that the generators of the theory's symmetry group, in the case SU(3), don't commute. These gluon-gluon interactions appear in Feynman diagrams as three-gluon and four-gluon vertices, for which there are no analogues in QED. The fact that the force carriers of QCD are themselves color charged is one the things that make it so distinct from the electroweak part of the SM, and also so complicated. There are $N_{c}^{2} -  1 = 8$ ``types'' of gluons (where $N_{c} = 3$ is the number of colors), meaning eight possible color states a gluon may have, known as the ``color octet'' \cite{MarkThomson}.

\begin{figure}[h]
	\centering
	\includegraphics[width=10cm]{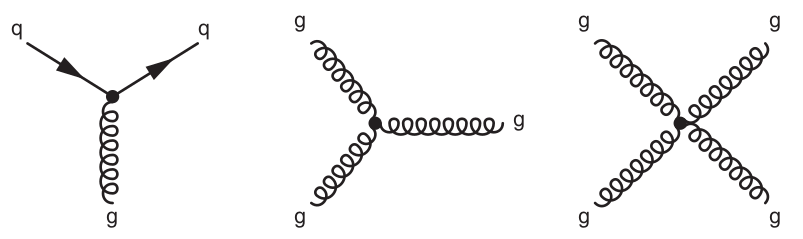}
	\caption{The basic QCD interaction vertices: the quark-gluon vertex, and the three-gluon and four-gluon vertices, for which there are no analogues in QED. Figure taken from from \cite{MarkThomson}.}
	\label{FIG-SM}
\end{figure}

A remarkable feature of QCD is that the coupling constant of the theory ($\alpha_{s}$) is strongly dependent on the energy scale of the interaction or, in other words, the four-momentum of the exchanged gluon ($Q$). The energy dependence of the coupling constant is dictated by the $\beta$-function \cite{Deur}:
\begin{equation}
Q^{2}\frac{\partial}{\partial Q^{2}}\frac{\alpha_{s}}{4 \pi} = \beta (\alpha_{s}),
\end{equation}
which can be expressed as a perturbative series. The first term of the series is:
\begin{equation}
\beta_{0} = 11 - \frac{2}{3}n_{f},
\end{equation}
where $n_{f}$ is the number of quark flavors considered at the energy scale in question. Then, at $\beta_{0}$ order, there is an analytic solution to Equation (1.1):
\begin{equation}
\alpha_{s}(Q^{2}) = \frac{4 \pi}{\beta_{0} \ln{(Q^{2}/\Lambda_{QCD}^{2})}},
\end{equation}
where $\Lambda_{\text{QCD}}$ = 220 MeV. Figure 1.3 shows different measurements of the strong coupling constant $\alpha_{s}(Q^{2})$ as a function of $Q$. The coupling constant is larger for low $Q^{2}$ and decreases significantly with increasing energy. 
\begin{figure}[h]
	\centering
	\includegraphics[width=8cm]{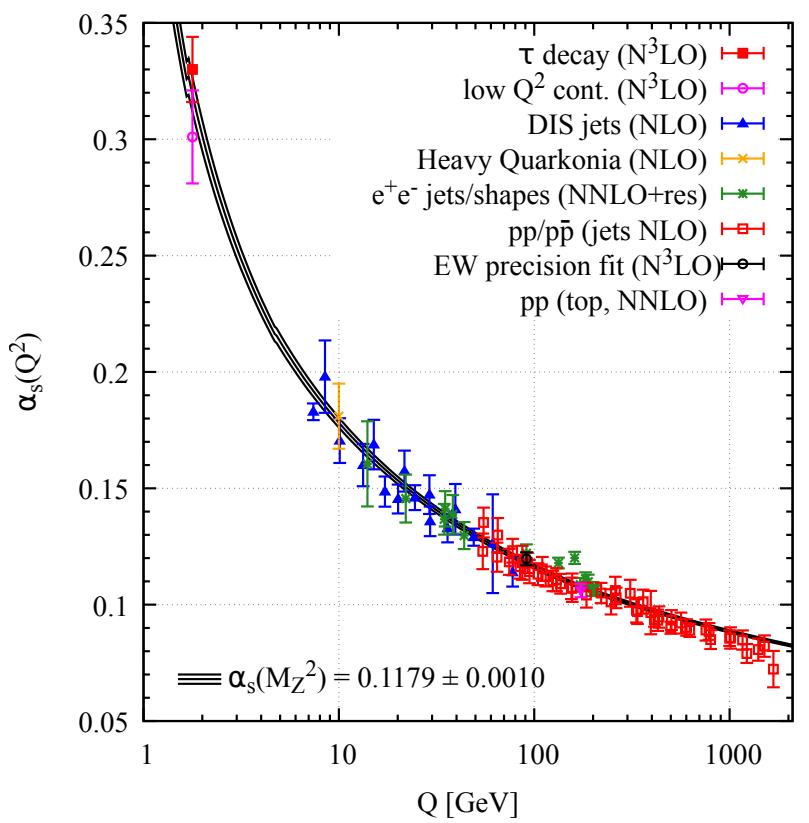}
	\caption{The QCD coupling constant $\alpha_{s}(Q^{2})$ as a function of $Q$ from different experiments. The order of perturbation theory used in the extraction of $\alpha_{s}$ is also indicated. From \cite{PDG}.}
	\label{FIG-ALPHA_S}
\end{figure}

Due to this behaviour of the coupling constant, the characteristic energy scale $\Lambda_{\text{QCD}}$ roughly separates QCD in two regimes. At higher energies, the coupling constant is small, and perturbation theory can be applied (perturbative QCD), while at lower energies perturbation theory is not applicable (non-perturbative QCD). In general, interactions with low momentum transfer are called \textit{soft}, while interactions whit high momentum transfer are referred to as \textit{hard} interactions. 

This strong energy dependence of the coupling constant (known as running of $\alpha_{s}$) leads to two main properties of QCD:

\begin{itemize}
    \item \textbf{Color confinement:} color charged particles are never observed as free particles. Quarks and gluons are always observed as confined bound states of three quarks (baryons) or a quark and an anti-quark (mesons), which are the hadrons. In nature, hadrons are always found in a color singlet state, and it is not possible to separate them into their colored constituents. Although color confinement is qualitatively well established, an analytical demonstration of it is still an open problem. 
    \item \textbf{Asymptotic freedom:} at very high energies, quarks and gluons interact weakly. This means that inside the hadrons, when they are very close to each other, quarks coexist as almost free particles. Asymptotic freedom was discovered by David Gross and Frank Wilczek \cite{PhysRevLett.30.1343}, and independently by David Politzer \cite{DAVIDPOLITZER1974129} in 1973, which awarded them the Nobel Prize in 2004. 
\end{itemize}

It seems reasonable to say that QCD began with the invention of the quark model in 1964 by Murray Gell-Mann \cite{GELLMANN1964214}, with the proposition that baryons and mesons were not elementary, but rather bound states of elementary particles, precisely the quarks. In 1969, Richard Feynman introduced the ideia that at very high energies, hadrons are not only formed by its valence quarks, but contain a collection of gluons and virtual pairs of quarks, which came to be known as \textit{partons} \cite{Feynman:1969wa}. Later on in 1973 Gell-Mann, Harald Fritzsch and Heinrich Leutwyler introduced color as the source of the ``strong fields'' \cite{Fritzsch_Gell-Mann}.

\section{The QCD Phase Diagram}
\label{SEC-Phase_Diagram}

While at low temperatures and densities quarks are confined inside hadrons, at very high energies or densities, lattice calculations QCD predict the existence of a phase transition to a state where quarks and gluons are deconfined \cite{Aoki, Alford:1997zt}. This state of matter, where quarks and gluons themselves are the degrees of freedom, is called Quark Gluon Plasma (QGP).

The idea of a QCD phase diagram was first proposed by N. Cabbibo and G. Parisi in 1975 \cite{Parisi}. It consists of a diagram that shows in one axis the baryonic density (or alternatively, the baryonic chemical potential $\mu_{B}$) and in the other, the temperature. Figure 1.4 shows a more modern view of the QCD phase diagram.

\begin{figure}[h]
	\centering
	\includegraphics[width=12cm]{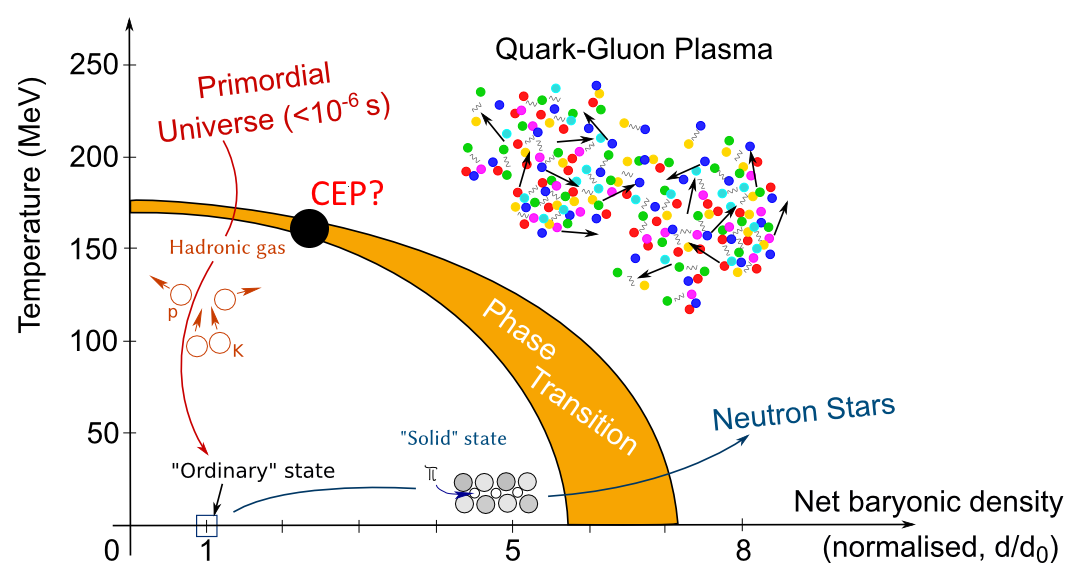}
	\caption{The QCD Phase Diagram: baryonic density is shown in the horizontal axis, and temperature in the vertical axis. The orange band represents the phase transition region from hadronic matter to QGP. Figure taken from \cite{Maire}.}
	\label{FIG-Phase_Diagram}
\end{figure}

It is known since 2006 from lattice calculations that the phase transition between hadronic matter and QGP in the region of high temperature and low baryonic density is a smooth crossover \cite{Aoki}. On the other hand, in the region of low temperature and non-zero baryonic potential it is believed to exist a first order phase transition from hadronic matter to QGP, where the first derivatives of the thermodynamic fields are discontinuous \cite{Alford:1997zt}. An active topic of research is the search for a critical end point (CEP) above the phase transition line, where the phase transition changes its order, which is the objective of the Beam Energy Scam (BES) program at RHIC, for example \cite{Tlusty}.

Figure 1.5 shows pressure, energy density and entropy density as a function of temperature (results from lattice QCD at zero baryonic chemical potential). There is an abrupt (yet smooth) increase in the number of degrees of freedom of the system around the temperature of 150 MeV, which corresponds to the crossover phase transition between nuclear matter and quark matter.

\begin{figure}[h!]
	\centering
	\includegraphics[width=8cm]{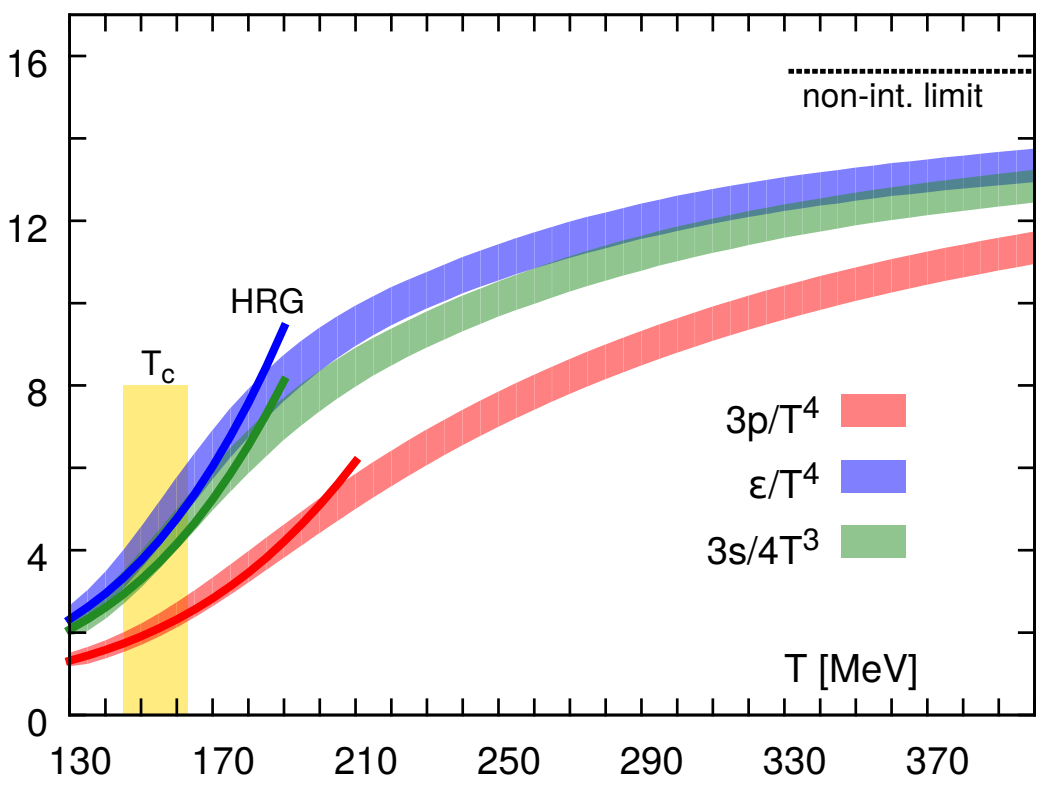}
	\caption{Pressure, energy density and entropy density as a function of temperature, obtained from lattice calculations. Results obtained from a hadron resonance gas (HRG) model calculation are shown as solid lines. Figure obtained from \cite{Ding}.}
	\label{FIG-Phase_Diagram}
\end{figure}

Lattice QCD calculations are more accurately performed at $\mu_{B}$ = 0. At non-vanishing baryonic density, lattice calculations are not so accurate, which accounts for the broadening of the orange band in Figure 1.4 near the horizontal axis of the phase diagram. The region of the diagram of (almost) vanishing baryonic chemical potential and high temperature (above the phase transition temperature) can be accessed in heavy ion collisions experiments. These were the conditions believed to exist in the whole Universe in the first moments after the Big Bang. On the other hand, the conditions of low temperature and high baryonic potential are believed to exist in the interior of very dense neutron stars.

\chapter{Relativistic heavy-ion collisions}
\label{CHAP-HIC}
\noindent\rule{\textwidth}{.1pt}\\[1ex]

The extreme temperatures necessary for the QCD phase transition to occur can be achieved in relativistic heavy-ion collisions. Chapter 2 aims to provide an introduction to the field of relativistic heavy-ion collisions. Section 2.1 briefly discusses the history of relativistic heavy-ion collisions experiments. Section 2.2 gives an overview of the space-time evolution of a heavy-ion collision, and its main stages. Section 2.3 presents the main kinematic variables and the coordinate system used in heavy-ion experiments, and introduces most of the basic vocabulary which will be used in the rest of this work. Section 2.4 is devoted the the main experimental observables used to characterize the QGP, and evidences of the QGP formation. Section 2.5 talks about the modelling of the initial state of the collision, and introduces concepts that will be necessary for what follows.

\section{Historical overview}

The field of relativistic heavy-ion collisions is strongly interdisciplinary. It makes connections between nuclear physics, which relies mostly on effective models, and particle physics, which can be studied from first principles under the formalism of QFT. As a system of thousands of particles is formed, thermodynamic quantities such as temperature, pressure and entropy are used in the theoretical description of heavy-ion collisions, providing also a connection to the world of statistical physics. Moreover, the conditions created in heavy-ion accelerators were believed to exist in the early stages of the Universe, making the field of heavy-ion collisions also relevant for cosmology \cite{CHEN201438}.

Nuclear physics went through a revolution in the 1970s and early 1980s, as the first particle accelerators were adapted to accelerate heavy nuclei. The most famous example is probably the coupling of the SuperHilac linear accelerator to the Bevatron in Berkeley, forming the Bevalac \cite{Stock:2004iim}. At the same time, the Dubna Syncrophasotron was also modified to accelerate heavy ions \cite{Baldin_1983}.

Collisions with energy exceeding 10 GeV per nucleon were first made at the Alternating Gradient Synchrotron (AGS) at the Brookhaven National Laboratory (BNL) in the year of 1986, and in 1987 at the Super Proton Synchrotron (SPS) at the European Organization for Nuclear Research (CERN). Later on, an experimental breakthrough took place with the construction of the Relativistic Heavy Ion (RHIC) \cite{Baym:2001in}, designed to accelerate gold nuclei at the center-of-mass energy of $\sqrt{s_{\text{NN}}}$ = 200 GeV. After a first run in the year of 2000, where the maximum energy of $\sqrt{s_{\text{NN}}}$ = 130 GeV was achieved, three runs were performed at the full capacity of $\sqrt{s_{\text{NN}}}$ = 200 GeV between 2001 and 2004. Contradicting the expectation of the community that the deconfined state formed in such collisions would be a weakly interacting system of quarks and gluons (similar to a gas), RHIC data showed that it was indeed much more similar to a strongly interacting fluid. The findings of the first runs at RHIC were summarized in the famous ``White Papers'' \cite{PHOBOS:2004zne, BRAHMS:2004adc, STAR:2005gfr, PHENIX:2004vcz}, published by each of RHIC's collaborations (PHOBOS, BRAHMS, STAR and PHENIX). Together, these four articles have over 11.500 citations. 

Between 1998 and 2008 the Large Hadron Collider (LHC) was built by CERN near the border of France and Switzerland. With a 27 kilometers circumference, it is the most energetic particle collider built to this day, achieving the center-of-mass energy of $\sqrt{s_{\text{NN}}}$ = 5.02 TeV in Pb-Pb collisions \cite{ALICE_Multiplicity}, and reaching the maximum energy of $\sqrt{s}$ = 13 TeV in p-p collisions \cite{ALICE:2019mmy}.  

\section{Space-time evolution of relativistic heavy-ion collisions}
\label{SEC-Space_time}

After being accelerated almost at the speed of light in circular trajectories and in opposite directions, two highly relativistic nuclei collide inside the accelerator beam pipe. Much less than a second later, an enormity of particles is detected, containing many hadron species, leptons and photons. What happened between the collision and the observation of this final-state particles? A heavy-ion collision is a complicated process, which can be divided in a few main stages:
\begin{itemize}
    \item \textbf{Pre-equilibrium:} in the first moments of the collision, just after the two Lorentz contracted nuclei overlap and quarks and gluons become deconfined, the system is far from equilibrium. At this point, the energy density across all of the system is much larger than 500 $\text{MeV}/\text{fm}^{3}$, the typical energy density inside hadrons \cite{Busza}. Most of the initial interactions between partons are soft, but rare scatterings with very large momentum transfer also occur, leading to the formation of jets \cite{Connors:2017ptx, Brewer:2020tbb}. In this initial moments after the collision, the system expands violently, close to the speed of light. 
    \item \textbf{Hydrodynamic evolution:} approximately 1 fm/c after the two nuclei collided, the system has been driven to local equilibrium. At this point, the QGP behaves as a strongly interacting, almost ideal fluid, with an extremely low shear viscosity to entropy ratio of $\eta/s \approx 1/4\pi$ \cite{Heinz:2013th}. This is the hydrodynamic stage of the collision. In general, hydrodynamics is an effective theory based on conservation laws which describes the dynamics of a system over long times and long distances, and is applicable when the microscopic distance between particles ($\ell$) is much smaller than the macroscopic length scale of the system ($L$).
    \item \textbf{Particlization:} about 10 fm/c after the beginning of the hydrodynamic evolution, the QGP has expanded and consequently, cooled down. At some point, the energy density of the system is small enough so that quarks and gluons recombine into hadrons. 
    \item \textbf{Hadronic phase:} in the final stage of the collision, the system is in a hadron resonance gas (HRG) phase. The final state hadrons, which are mainly pions (the lightest hadron), are boosted in the radial direction by the expansion of the system, and propagate until they reach the detectors. In this process, resonances might decay into more stable particles and the particles interact with each other, both elastically and inelastically. At some point, the decays and inelastic scatterings cease, and each particle species abundance stays approximately stable. This stage of the hadronic phase is known as \textit{chemical freeze-out}. Later, as the system continues to expand and becomes more dilute, the elastic scatterings cease, and the final momentum distribution of the particles is fixed, which defines the \textit{kinetic freeze-out}. After kinetic freeze-out, the final state hadrons propagate as free particles until they reach the detectors of the experiment.  
\end{itemize}

\begin{figure}[h]
	\centering
	\includegraphics[width=10cm]{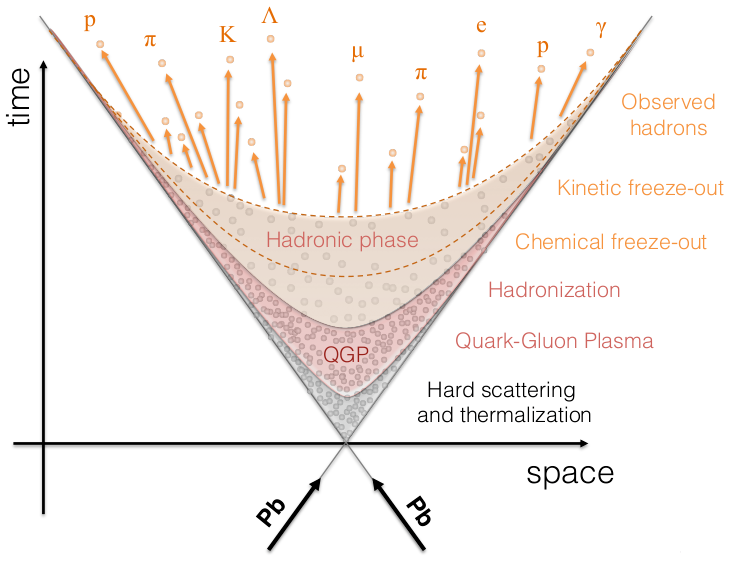}
	\caption{Space-time evolution of a heavy ion collision and its stages: initial scatterings and thermalization are followed by the hydrodynamic stage, where the QGP expands as a fluid. At some point, quarks and gluons recombine into hadrons (hadronization) which, after chemical and kinetic freeze-out, are observed.}
	\label{FIG-Space-time_Diagram}
\end{figure}

\section{Basic kinematic variables, coordinate system and centrality}
\label{SEC-Basic}

The coordinate system usually employed in studies of relativistic heavy-ion collisions is shown in Figure 2.2, on top of a sketch of a typical collider type experiment, in this case, the ALICE experiment at the LHC. The $z$-axis lies is the direction of the beam. In symmetric collisions, the two colliding nuclei travel at the same speed in the positive and negative $z$ directions, so that it is specially convenient to perform calculations in the center-of-mass frame (which, in this case, coincides with the laboratory frame). The $x$-axis is oriented parallel to the ground, an the $y$-axis is placed along the vertical direction. In analogy with spherical coordinates, a polar angle $\theta$ and an azimuthal angle $\phi$ are defined.

\begin{figure}[h]
	\centering
	\includegraphics[width=8cm]{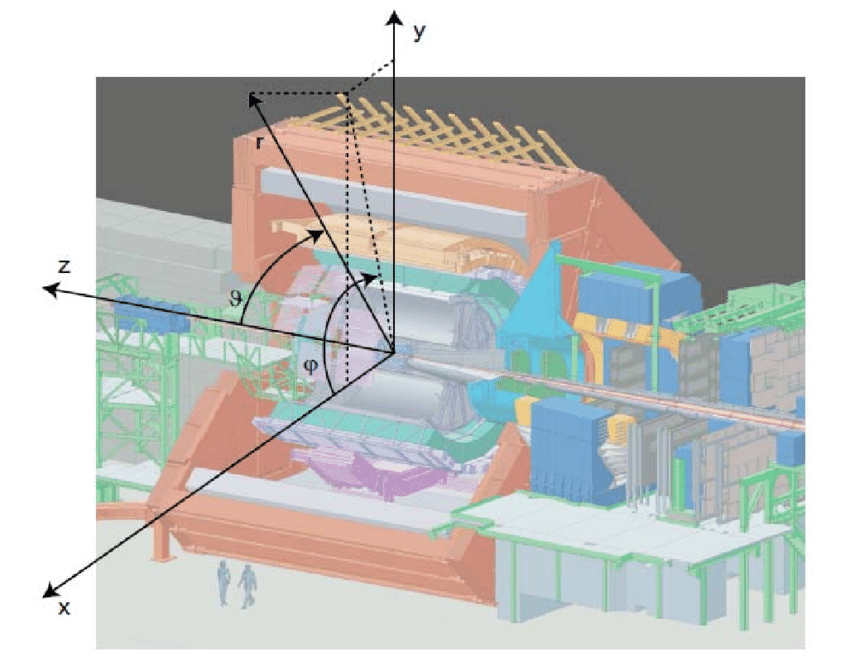}
	\caption{Coordinate system used in this work, shown on top of a sketch of the ALICE experiment. Figure obtained from \cite{Costanza}.}
	\label{FIG-System}
\end{figure}

The QGP itself is extremely short-lived and never observed, so it is necessary to extract information about its characteristics solely based on the particles that reach the experiment's detector. A few kinematic variables are usually employed to characterize the final state particles. Transverse momentum ($p_{T}$) is the projection of the particle's momentum in the transverse plane of the collision:
\begin{equation}
p_{T} = \sqrt{p_{x}^{2} + p_{y}^{2}},
\end{equation}
High-$p_{T}$ particles in the final state emerge from the QGP travelling fast and almost perpendicular to the beam direction. The rapidity variable ($y$) can be used to characterize the particle's momentum in the longitudinal direction:
\begin{equation}
y = \frac{1}{2} \ln{ \left (\frac{E + p_{z}}{E - p_{z}} \right)}
\end{equation}
Rapidity is close to zero when a particle has little longitudinal momentum, that is, when it is moving approximately perpendicular to the direction of the beam. Alternatively, when a particle has large $p_{z}$, then $|y| \rightarrow \infty$. Rapidity has the convenient property of being additive under longitudinal boosts. Alternatively, the pseudo-rapidity variable ($\eta$) can be used: 
\begin{equation}
\eta = \frac{1}{2} \ln{ \left (\frac{|\Vec{p}| + p_{z}}{|\Vec{p}| - p_{z}} \right)} = - \ln{\left(\tan{\frac{\theta}{2}} \right)}
\end{equation}
Pseudo-rapidity is zero perpendicular to the beam axis ($\theta = 90^{\circ}$) and $|\eta| \rightarrow \infty$ close to the beam axis ($\theta = 0^{\circ}$ or $\theta = 180^{\circ}$). Pseudo-rapidity is specially useful because it is only necessary to know the particle's angle of emission $\theta$, while more details about the particle, such as mass and momentum, are necessary to know its rapidity. For this reason, rapidity is used when dealing with identified particle species, and pseudo-rapidity is used when dealing with charged particles. In the relativistic limit, where $E \approx |\vec{p}|$, the rapidity and pseudo-rapidity coincide: $y \approx \eta$. Experimental measurements are usually performed in a particular region of pseudo-rapidity or rapidity, as $|y| < 0.5$ or $|\eta| < 0.8$, for example.

Being extended objects, the two nuclei might collide ``head-on'', with their centers aligned, or they might collide in a way such that their centers are dislocated and the overlap between them is only partial. The variable that quantifies the amount of overlap between the projectiles is the impact parameter ($b$), defined as the distance between the centers of the two colliding nuclei. In practice, the impact parameter is not known for each collision, and it is necessary to use another quantity, analogous to the impact parameter, to classify collisions. For this purpose, the concept of centrality is used.
\begin{figure}[h]
	\centering
	\includegraphics[width=7cm]{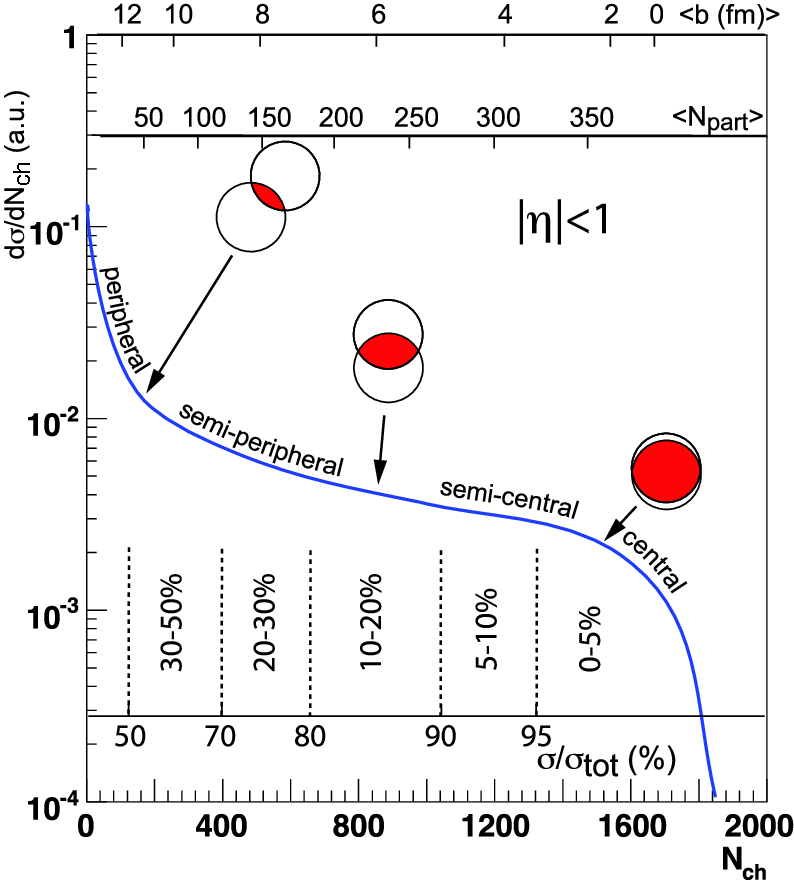}
	\caption{Illustration of the concept of centrality. The impact parameter, number of participants, and number of charged particles are shown in the horizontal axis. The interaction cross-section is shown in the vertical axis, in arbitrary units. From \cite{Betz}.}
\end{figure}

The concept of centrality is based on the intuitive idea that collisions with greater overlap area between the colliding nuclei (small impact parameter) should produce more particles, while in collisions with a smaller overlap area, less particles should be observed in the final state. In this way, it is possible to classify collisions (events) based on the number of detected particles, and arrange them in centrality classes. Out of all events, the 10 \% which produced more particles constitute the 0 - 10 \% centrality class. The 20 \% ones that produced more particles, removing those from the 0 - 10 \% class, constitute the 10 - 20 \% class, and so on.

Events where the nuclei collide ``head-on'' are called \textit{central} collisions, and as the impact parameter grows the collisions are said to be more \textit{peripheral}. Collisions where the impact parameter is greater than twice the nuclear radius, so that there is no overlap of hadronic matter, are called \textit{ultra-peripheral} \cite{LbL}. Figure 2.3 illustrates the concept of centrality: the blue line shows the nucleus-nucleus cross-section distribution as a function of the number of charged particles (the impact parameter is also indicated) and the dashed lines show the centrality percentiles. Nucleons that collide with at least another (the darker circles in Figure 2.11) are called \textit{participants}, while the remaining are known as \textit{spectators}. Central collisions are characterized by a large number of participants, while in more peripheral collisions, there are less participant nucleons. 

\section{Experimental observables and evidences of QGP formation}
\label{SEC-Observables_and_Evidences}

\subsection{Multiplicity}

Multiplicity is a very straightforward observable, defined simply as the number of charged particles detected in the final state of the collision, usually measured in a given pseudo-rapidity region. Nevertheless, multiplicity in the central pseudo-rapidity region is an important observable to study the bulk properties of the QGP formed in heavy-ion collisions. Table 2.1 shows, for each centrality class, the charged-particle multiplicity at mid-rapidity and the mean number of participants for Pb-Pb collisions at $\sqrt{s_{\text{NN}}}$ = 2.76 TeV, as measured by the ALICE Collaboration \cite{ALICE_Mult}. The number of participant nucleons in each centrality class was obtained from Glauber Model estimates (see Section 2.5.1).

\begin{table}[h!]
\centering
\begin{tabular}{c | c | c}
Centrality &  $dN_{\text{ch}}/d\eta$   &  $N_{\text{part}}$ \\ \hline
0 - 5 \%  &  1601 $\pm$ 60  & 382.8 $\pm$ 3.1  \\ 
5 - 10 \%    & 1294 $\pm$ 49  &  329.7 $\pm$ 4.6 \\ 
10 - 20 \%  & 966 $\pm$ 37  & 260.5 $\pm$ 4.4  \\ 
20 - 30 \%  & 649 $\pm$ 23  & 186.4 $\pm$ 3.9 \\ 
30 - 40 \%   & 426 $\pm$ 15  & 128.9 $\pm$ 3.3  \\ 
40 - 50 \%   & 261 $\pm$ 9  & 85.0 $\pm$ 2.6 \\ 
50 - 60 \%   & 149 $\pm$ 6 & 52.8 $\pm$ 2.0  \\ 
60 - 70 \%   & 76 $\pm$ 4 & 30.0 $\pm$ 1.3 \\ 
70 - 80 \%   & 35 $\pm$ 2 & 15.8 $\pm$ 0.6 \\ 
\end{tabular}
\caption{Charged-particle multiplicity density at mid-rapidity and mean number of participant nucleons (obtained from Glauber Model estimates) for each centrality class in Pb-Pb collisions at $\sqrt{s_{\text{NN}}}$ = 2.76 TeV, as measured by the ALICE experiment. Data from \cite{ALICE_Mult}.}
\end{table}

The number of charged particles observed in the final state is much larger than the original number of participant nucleons, and decreases as the collisions become more peripheral. This should be expected, since for more peripheral collisions there is a decrease in the overlap area between the two nuclei. To study bulk particle production, it is more convenient to calculate the multiplicity \textit{per participant pair}, as in Figure 2.4. This also allows comparisons with particle production in other collision systems, such as p-p and p-Pb. Figure 2.4 shows that multiplicity is not merely proportional to the number of participants: in the most central collisions, approximately 10 charged particles are produced for each pair of participants, while for the most peripheral ones, this number drops to about 4 particles.

\begin{figure}[h]
	\centering
	\includegraphics[width=8cm]{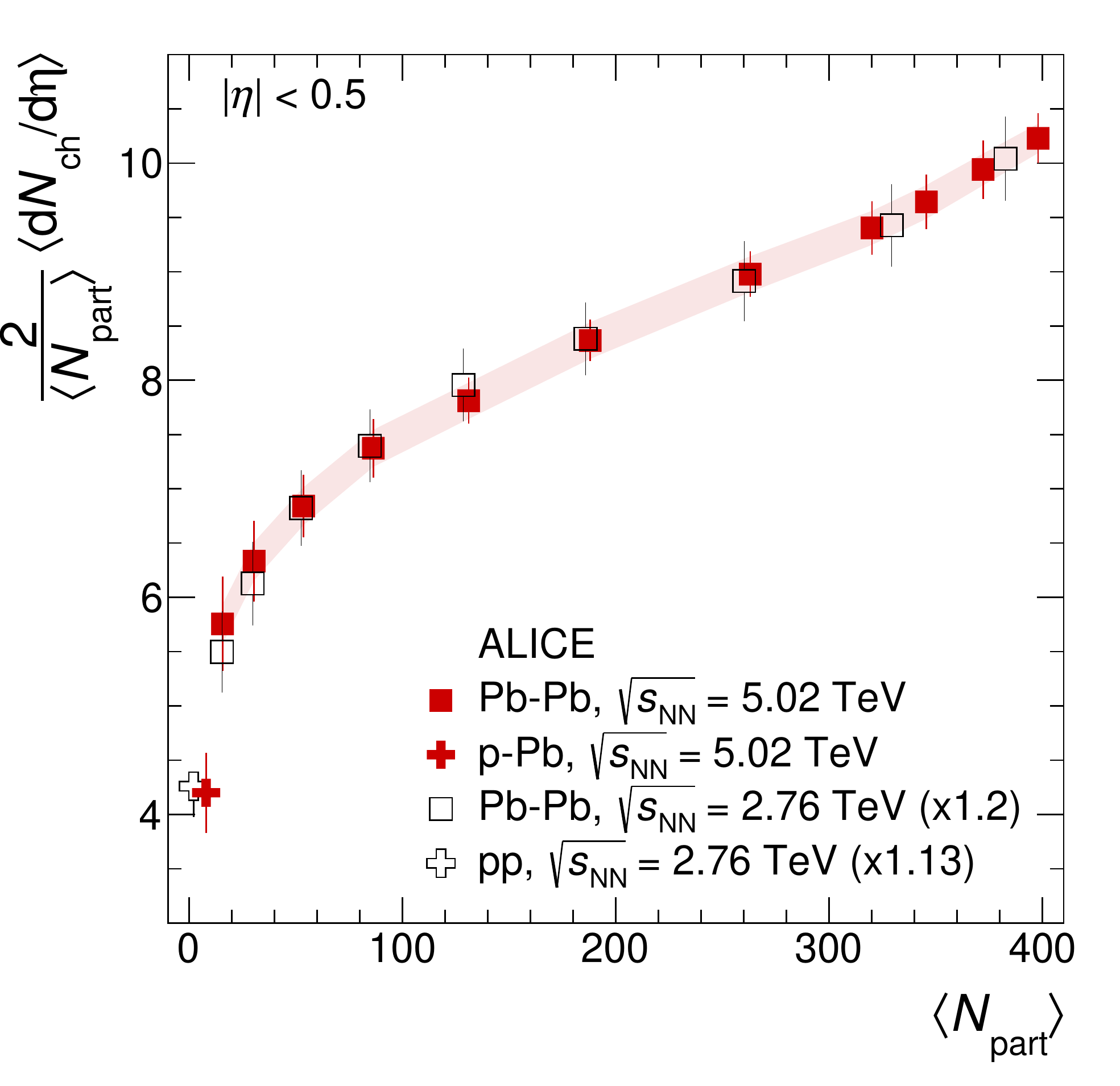}
	\caption{Charged particle multiplicity density in $|\eta| < 0.5$ divided by the number of participant pairs as a function of the number of participant nucleons for different collision systems. From \cite{ALICE_Multiplicity}.}
\end{figure}

Figure 2.4 also shows, as should be expected, that multiplicity increases with increasing collision energy, which can be seen by noticing that the data of Pb-Pb at $\sqrt{s_{\text{NN}}}$ = 2.76 TeV is multiplied by 1.2. From the theoretical point of view, particle production is usually described by two categories of models: two-component models which combine soft interactions and perturbative QCD processes \cite{DPMJET, HIJING} and saturation models \cite{ALbacete:2010ad, Kharzeev:2004if, Armesto:2004ud}. 

\subsection{Transverse momentum distributions}

Transverse momentum distributions ($p_{T}$-spectra) are also among the most commonly measured observables. A transverse momentum distribution is defined simply as a histogram counting the number of particles detected in each $p_{T}$ bin, per unit rapidity. $p_{T}$-spectra contain information about the kinetic properties of the final state particles, and therefore are a powerful tool to study the bulk properties of the QGP. Figure 2.5 shows, for each centrality class, the $p_{T}$-spectra of pions, kaons and protons produced in Pb-Pb collisions at $\sqrt{s_{\text{NN}}}$ = 2.76 TeV, measured by the ALICE Collaboration. 

\begin{figure}[H]
    \centering
    \begin{subfigure}{.3\textwidth}
        \centering
        \includegraphics[width=5cm]{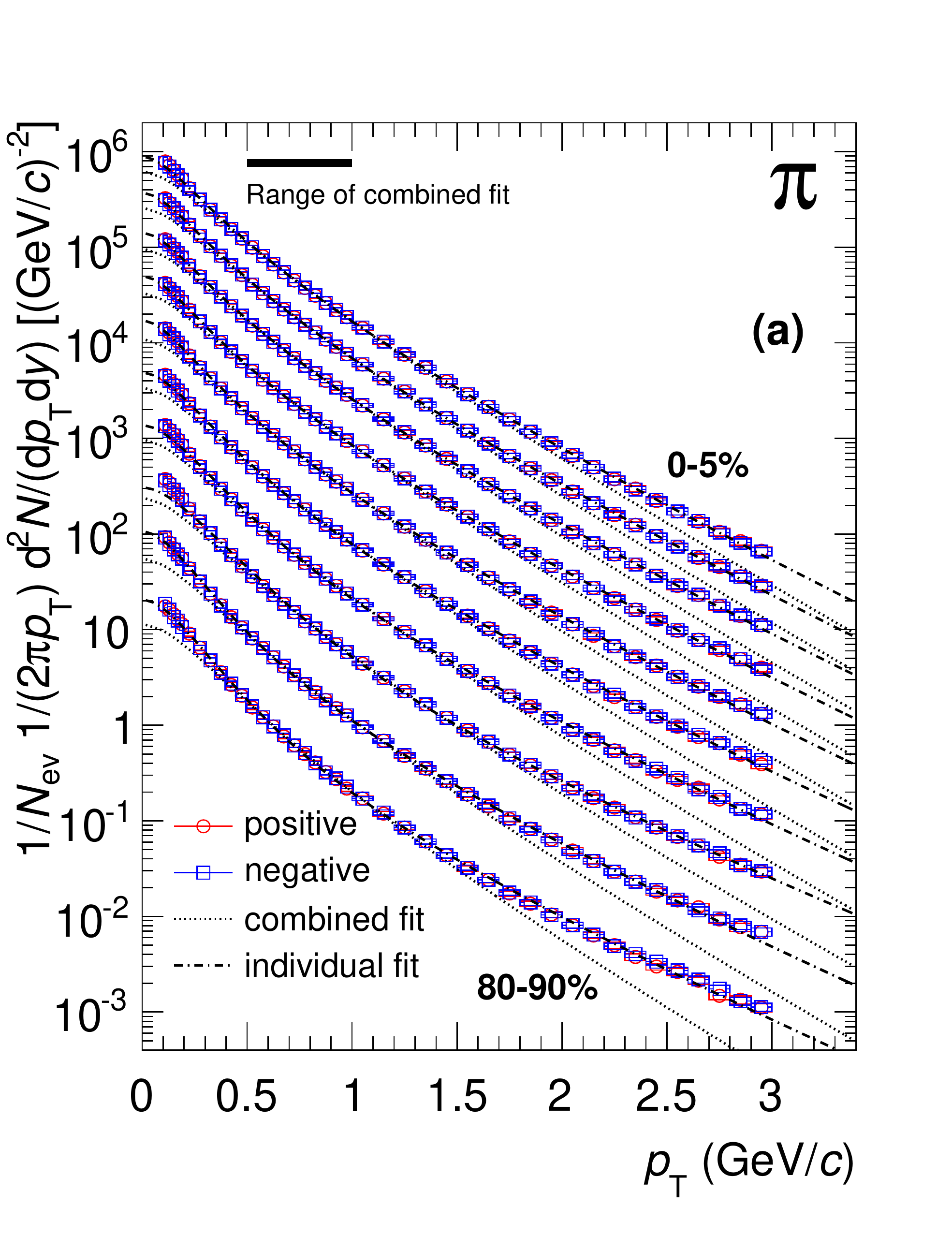}
    \end{subfigure}%
    \begin{subfigure}{.3\textwidth}
        \centering
        \includegraphics[width=5cm]{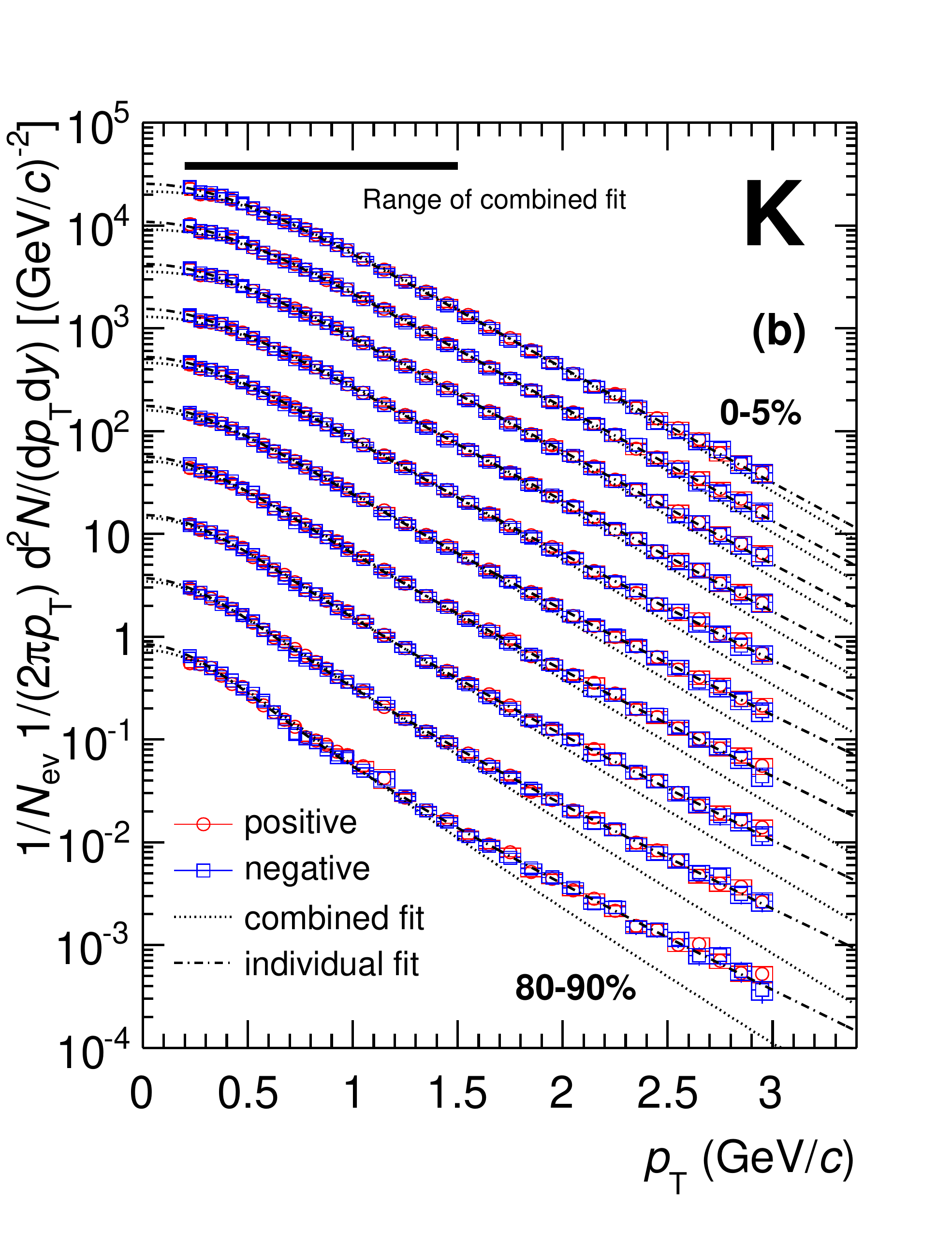}
    \end{subfigure}
    \begin{subfigure}{.3\textwidth}
        \centering
        \includegraphics[width=5cm]{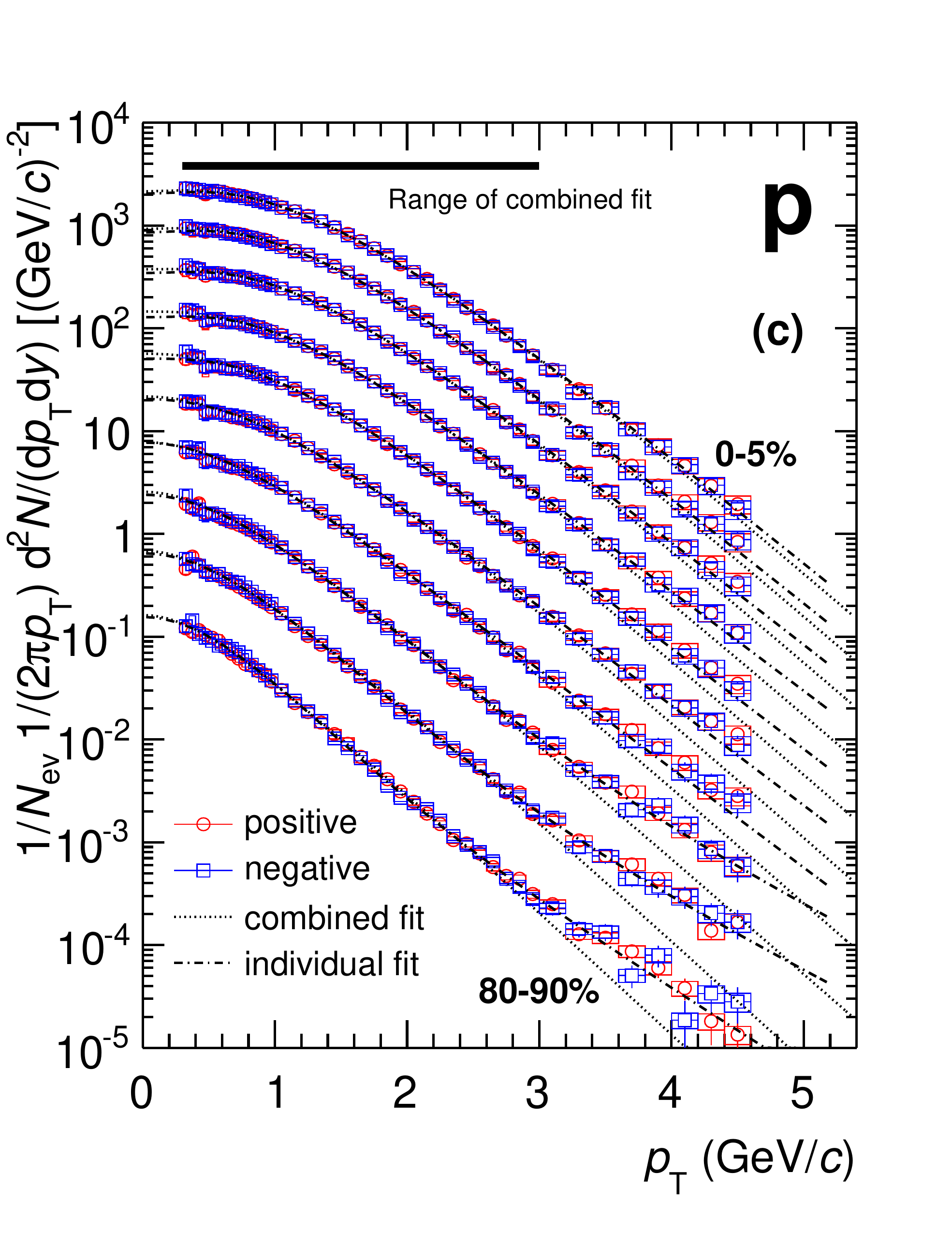}
    \end{subfigure}
    \caption{Transverse momentum distributions of pions, kaons and protons in Pb-Pb collisions at $\sqrt{s_{\text{NN}}}$ = 2.76 TeV measured by ALICE. From \cite{ALICE_Pt}.}
\end{figure}

Most particles are produced at low-$p_{T}$ ($p_{T} \lesssim$ 1 GeV) and the soft region of the spectra, with $p_{T} \lesssim$ 3 GeV, is well described by a thermal distribution. The high-$p_{T}$ (hard) part of the spectra, on the other hand, exhibits a power-law behavior. Many statistical models have been employed to extract physical parameters by fitting $p_{T}$ spectra \cite{Wang:2018hqq, Gupta:2021efj}. These include the non-extensive Tsallis statistics \cite{Tsallis}, the QCD-inspired Hagedorn inverse power law \cite{Hagedorn1, Hagedorn2} and the Pearson distribution \cite{Gupta:2021ksp}. A phenomenological model widely used to characterize $p_{T}$-spectra and obtain information about the kinetic freeze-out is the Blast Wave (BW) model \cite{BlastWave}, which is used in this work (see Section 5.2.4).

\begin{figure}[h]
	\centering
	\begin{subfigure}{.4\textwidth}
	\centering
	\includegraphics[width=7cm]{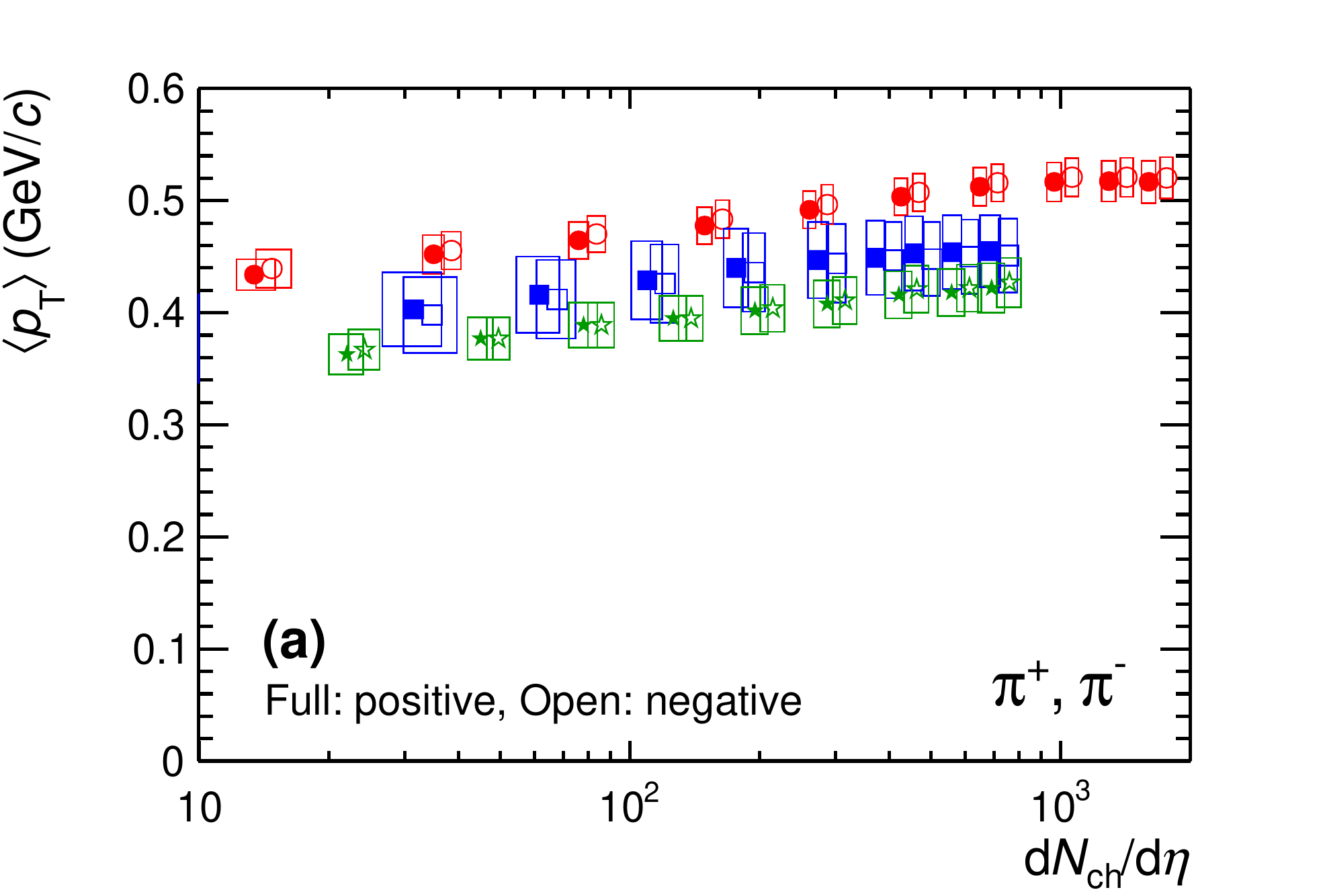}
	\end{subfigure}
	\begin{subfigure}{.4\textwidth}
	\centering
	\includegraphics[width=7cm]{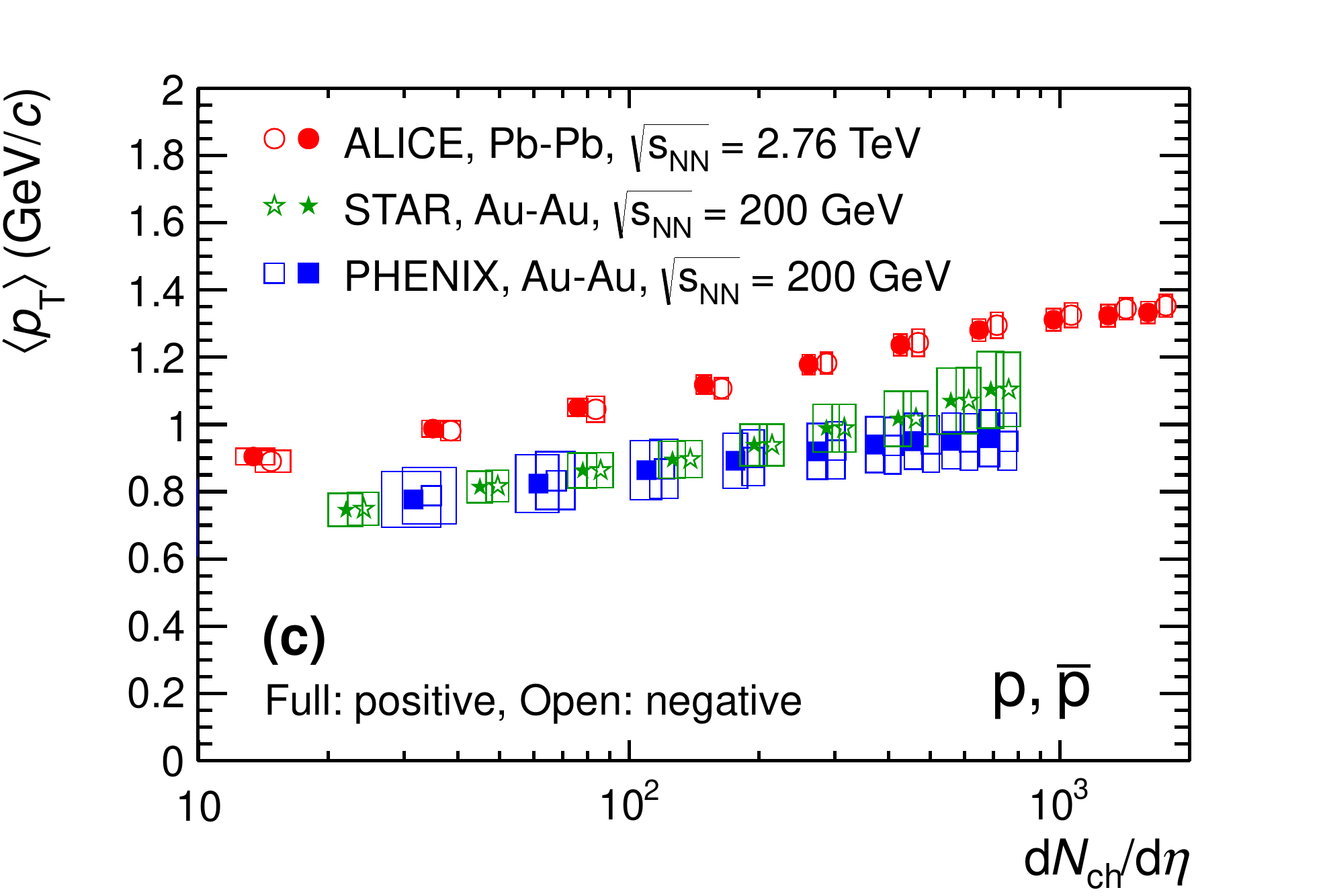}
	\end{subfigure}
	\caption{Mean $p_{T}$ of charged pions (left) and protons (right) in Pb-Pb collisions at $\sqrt{s_{\text{NN}}}$ = 2.76 TeV and Au-Au collisions at $\sqrt{s_{\text{NN}}}$ = 200 GeV, as a function of multiplicity. From \cite{ALICE_Pt}.}
\end{figure}

A straightforward quantity to extract from $p_{T}$-spectra is the mean transverse momentum of the detected particles (mean $p_{T}$):
\begin{equation}
\langle p_{T} \rangle = \frac{\int d^{2}p_{T}p_{T}\frac{dN}{d^{2}p_{T}dy}}{\int d^{2}p_{T}\frac{dN}{d^{2}p_{T}dy}}
\end{equation}
Figure 2.6 shows that the mean $p_{T}$ increases with multiplicity, suggesting that the transverse expansion of the system is somewhat more ``violent'' in central collisions. 

\subsection{Anisotropic flow}

Unlike gases, where particles are far apart from each other and rarely meet, in fluids particles are constantly interacting with their neighbors, so that fluids present \textit{collective behavior}. The experimental observation of collective behavior in relativistic heavy-ion collisions is probably the most compelling evidence that indeed a QGP is formed in such experiments. The QGP itself is never observed: collective behavior manifests itself as anisotropy in the momentum distribution of the final state particles. Due to the fact that in a non-central collision the overlap region of the two nuclei has an approximately elliptical (``almond'') shape, greater pressure gradients develop in the $x$-direction as compared to the $y$-direction. In the hydrodynamic evolution, the QGP flows preferentially in the $x$-direction, as the fluid expands and the elliptical shape of the system becomes more circular. In this process, it is said that the spatial anisotropy is transferred to momentum space. Figure 2.7 illustrates, for a non-central collision, the time evolution of the shape of the system in the transverse plane:

\begin{figure}[h]
	\centering
	\includegraphics[width=14cm]{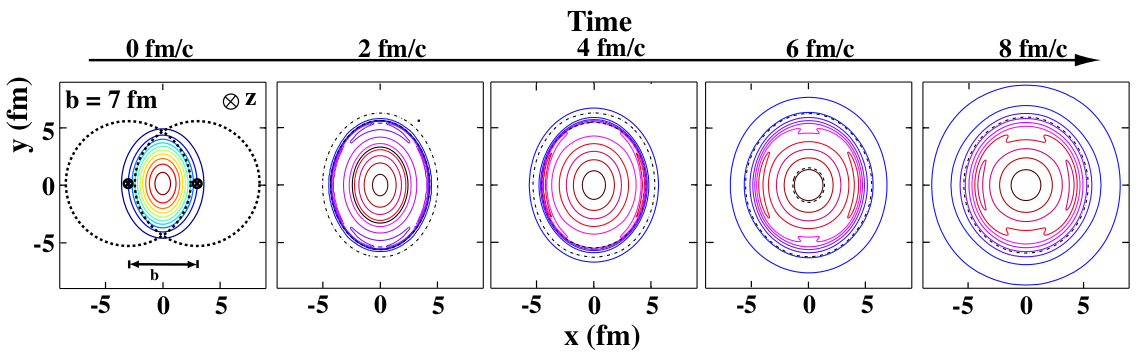}
	\caption{Time evolution of the shape of the system in the transverse plane of the collision in a non-central event. The initially elliptical shape turns more circular as the system expands, and particles flow preferably in the $x$-direction. Figure from \cite{Kolb}.}
\end{figure}

As a result, more particles are detected close to $\phi$ = 0 and $\phi$ = $\pi$ and less particles are detected near $\phi$ = $\pi/2$ and $\phi$ = $3\pi/2$. This momentum anisotropy in the transverse plane can be quantified by expanding the azimuthal distribution of particles in a Fourier series \cite{Voloshin:1994mz}:
\begin{equation}
\frac{dN}{d\phi} \propto 1 + 2 \sum_{n=1}^{\infty} v_{n} \cos[n(\phi - \Psi_{\text{RP}})]
\end{equation}
The coefficients of the series (flow harmonics) are given by:
\begin{equation}
v_{n} = \langle \cos[n(\phi - \Psi_{\text{RP}})] \rangle,
\end{equation}
where the expectation value symbol denotes an average over all particles and $\Psi_{\text{RP}}$ is the reaction plane angle \cite{Voloshin:2008dg}. The series has only the cosine terms, due to the reflection symmetry with respect to the reaction plane. The first coefficient of the series is known as directed flow, while the harmonics $v_{2}$ and $v_{3}$ are called elliptic and triangular flow, respectively.
\begin{figure}[h]
	\centering
	\includegraphics[width=10cm]{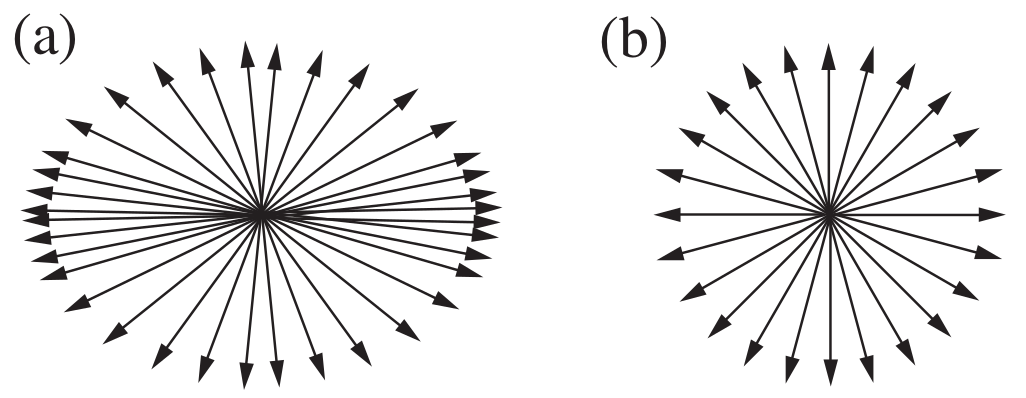}
	\caption{Illustration of particle distributions in the transverse plane of the collision with (a) $v_{2} > 0$ and (b) $v_{2}$ = 0. Figure from \cite{Snellings:2011sz}.}
\end{figure}

Figure 2.9 shows, on the left, the flow harmonics integrated over transverse momentum as a function of centrality, and on the right, the flow harmonics as a function of $p_{T}$ (usually called differential flow) for three centrality classes. 

\begin{figure}[H]
\centering
    \begin{subfigure}{.4\textwidth}
        \centering
        \includegraphics[width=7cm]{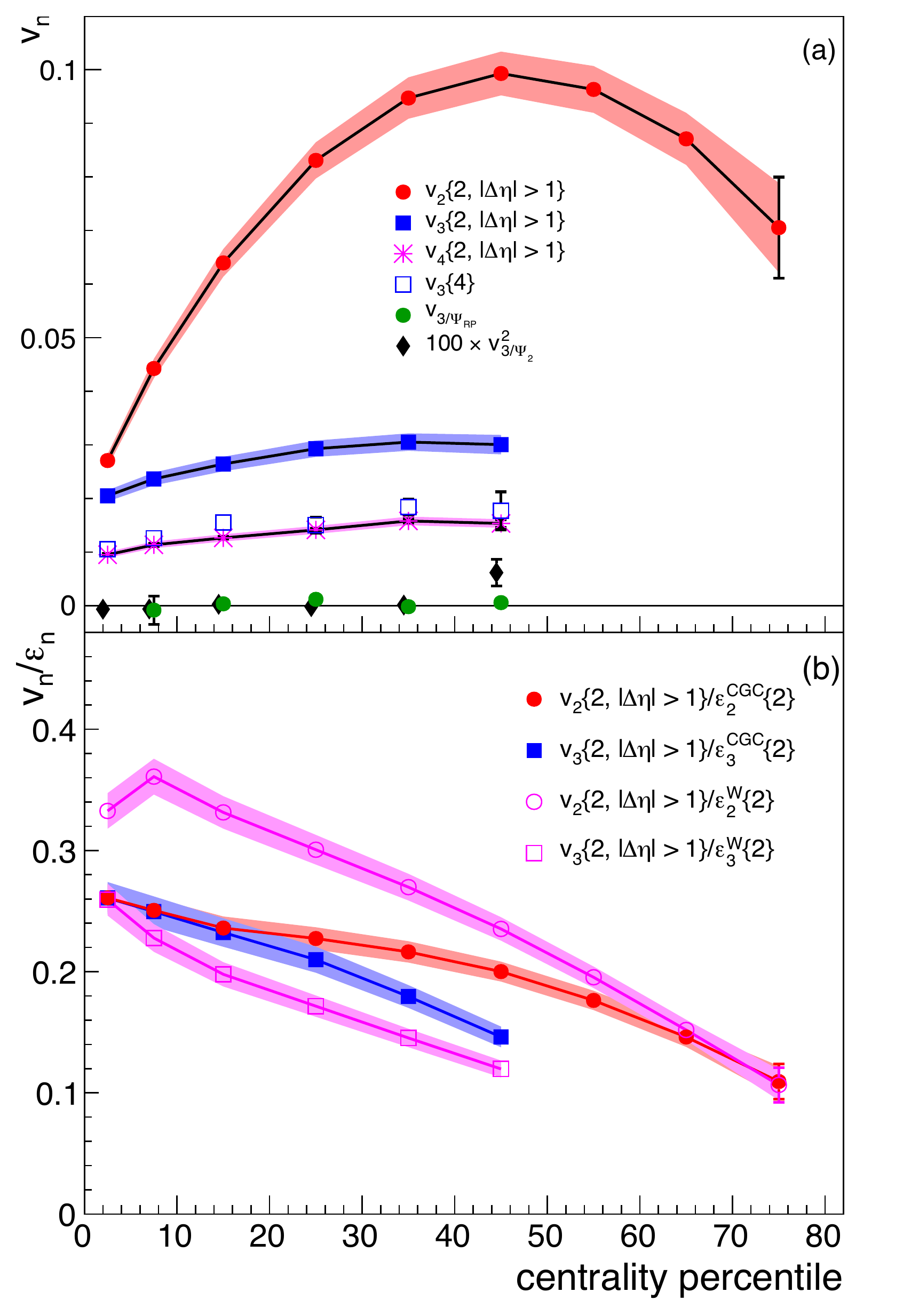}
    \end{subfigure}
    \begin{subfigure}{.4\textwidth}
        \centering
        \includegraphics[width=5cm]{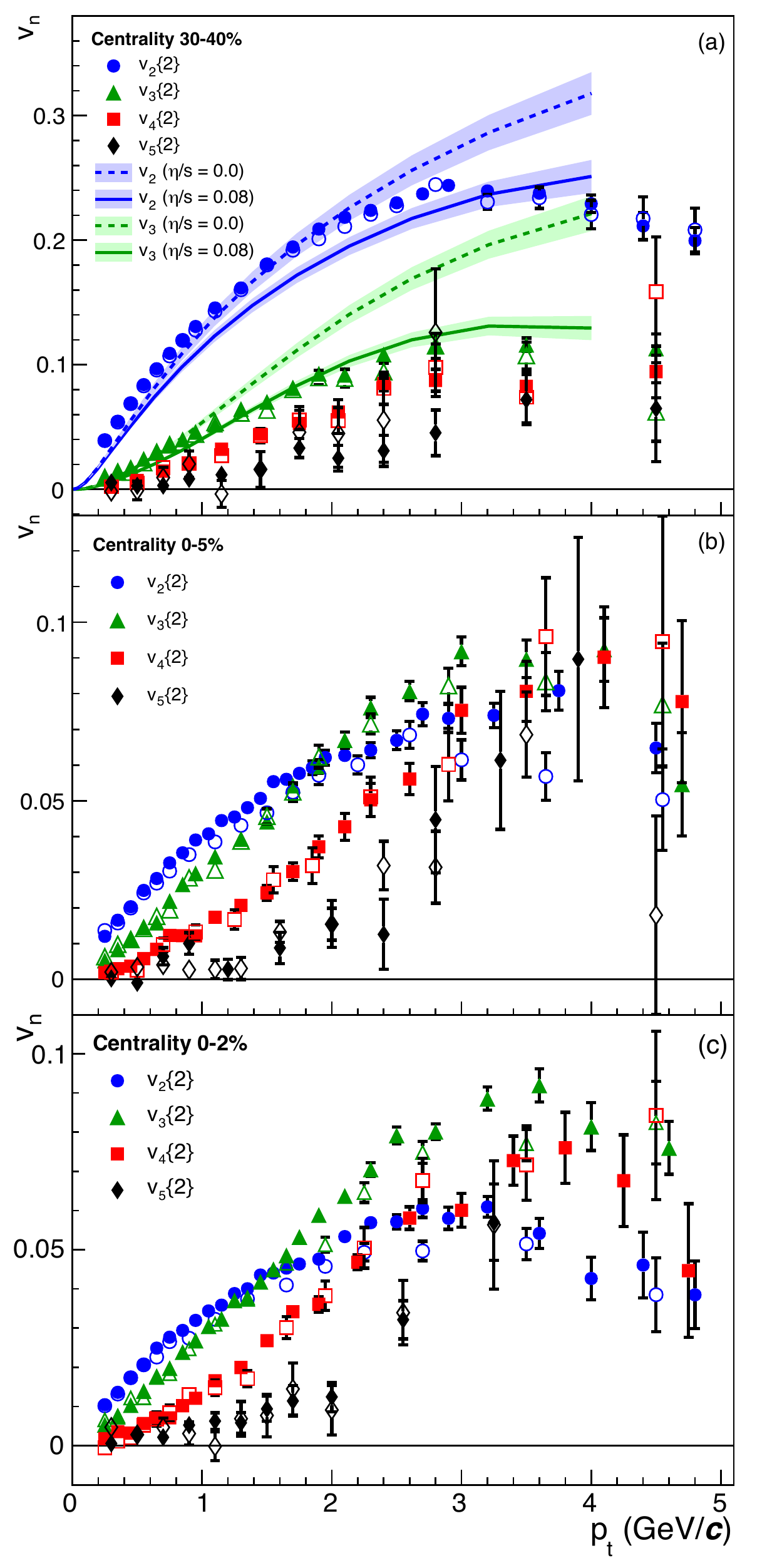}
    \end{subfigure}
    \caption{Flow measurements in Pb-Pb collisions at $\sqrt{s_{\text{NN}}}$ = 2.76 TeV. Left: integrated flow harmonics as a function of centrality. Right: $p_{T}$-differential flow for three centrality classes. From \cite{ALICE_Flow}.}
\end{figure}

There is little flow in central collisions, where the initial spatial anisotropy is small. Flow grows for non-central collisions, as the overlap region between the two colliding nuclei acquires an anisotropic shape. For very peripheral collisions, flow decreases once again. It can also be seen that, inside a given centrality class, elliptic flow increases with transverse momentum. Figure 2.9 also shows, on the bottem left panel, the ratio between the flow harmonics and the corresponding eccentricity harmonics (see Section 2.5.2).

\section{The Initial Condition}
\label{SEC-Phenomenology}

\subsection{The Glauber Model}

In general, the distinction between participants and spectators is not as straightforward as Figure 2.3 might suggest. Moreover, some participants collide only once, while some might suffer multiple collisions, and there is also the possibility of two nucleons ``passing by'' each other without interacting at all. A quantitative estimate of the number of participant nucleons ($N_{\text{part}}$) and nucleon-nucleon (or binary) collisions ($N_{\text{coll}}$) can be provided by the Glauber Model \cite{Miller, dEnterria:2020dwq}. In the 1950's Roy Glauber pioneered the use of quantum scattering theory for composite systems such as heavy nuclei, allowing the calculation of geometric quantities which can be related to the experimental observables. The Glauber formalism treats collisions of composite systems, such as p-A and A-A as a superposition of collisions between their constituent nucleons. The model uses the eikonal approximation for quantum scattering, which assumes that the exchanged momentum between the colliding projectiles is very small compared to their total longitudinal momentum, so that the nucleons travel in straight trajectories.  

There are two most relevant inputs for any Glauber based calculation. The first of them is the nuclear density distribution, for which the most natural choice is a Woods-Saxon distribution:
\begin{equation}
\rho(r) = \rho_{0} \left( 1 + \exp{\left(\frac{r - R}{a}\right)} \right)^{-1}
\end{equation}
In Equation (2.7), $\rho_{0}$ is the nucleon density in the center of the nucleus, $R$ corresponds to the nuclear radius, $a$ is the skin depth and $r$ is the distance from the center of the nucleus. The second one is the inelastic nucleon-nucleon cross section, which is assumed to be dependent only on energy, and not on the collision system, the position of the nucleon inside the nucleus, or any other characteristic of the nuclear environment. Table 2.2 shows some measured values of the nucleon-nucleon inelastic scattering cross-section ($\sigma_{\text{NN}}$) for different collision energies.

\begin{table}
\centering
\begin{tabular}{c | c | c}
$\sqrt{s}$ (TeV) &  $\sigma_{\text{NN}}$ ($\text{fm}^{2}$)   &  Reference \\ \hline
0.2  &  4.23  & \cite{PHENIX:2015tbb}  \\ 
2.76  & 6.28  &  \cite{ALICE:2012fjm} \\ 
5.02  & 7.0  & \cite{ALICE:2012xs} \\ 
7  &  7.32 & \cite{ALICE:2012fjm} \\ 
\end{tabular}
\caption{Nucleon-nucleon inelastic scattering cross-section for different collision energies}
\end{table}

In a collision of two nuclei $A$ and $B$ at impact parameter $b$, the transverse density of each nucleus is represented by the \textit{thickness function}, obtained by integrating the nuclear density in the beam direction:
\begin{equation}
T_{A}(x,y) = \int_{-\infty}^{+\infty} \rho_{A}(x,y,z) dz,
\end{equation}
and analogously for B. The density of binary collisions at point $(x,y)$ is given by the product of the thickness functions of the two nuclei and the inelastic nucleon-nucleon cross-section:
\begin{equation}
n_{\text{BC}}(x,y;b) = \sigma_{\text{NN}} T_{A}(x + b/2,y,z) T_{B}(x - b/2,y,z),
\end{equation}
from which the total number of binary collisions is obtained by integration in the transverse plane:
\begin{equation}
N_{\text{BC}}(b) = \int \sigma_{\text{NN}} T_{A}(x + b/2,y,z) T_{B}(x - b/2,y,z) dx dy.
\end{equation}
The number of participants (or wounded nucleons), on its turn, is given by the integral \cite{Bialas:1976ed}:
\begin{equation}
\begin{split}
N_{\text{WN}}(b) = \int T_{A}(x + b/2,y,z) \left( 1- \left( 1 - \frac{\sigma_{\text{NN}}T_{B}(x - b/2,y,z)}{B}\right)^{B}\right) \\ + T_{B}(x - b/2,y,z) \left( 1- \left( 1 - \frac{\sigma_{\text{NN}}T_{A}(x + b/2,y,z)}{A}\right)^{A}\right) dxdy
\end{split}
\end{equation}

In modern computational simulations, the Glauber Model is usually implemented through Monte Carlo calculations. In the Monte Carlo Glauber approach, the two colliding nuclei are assembled according to a nuclear density distribution on an event-by-event basis. Then, the binary collisions are performed.

\begin{figure}[h]
	\centering
	\includegraphics[width=12cm]{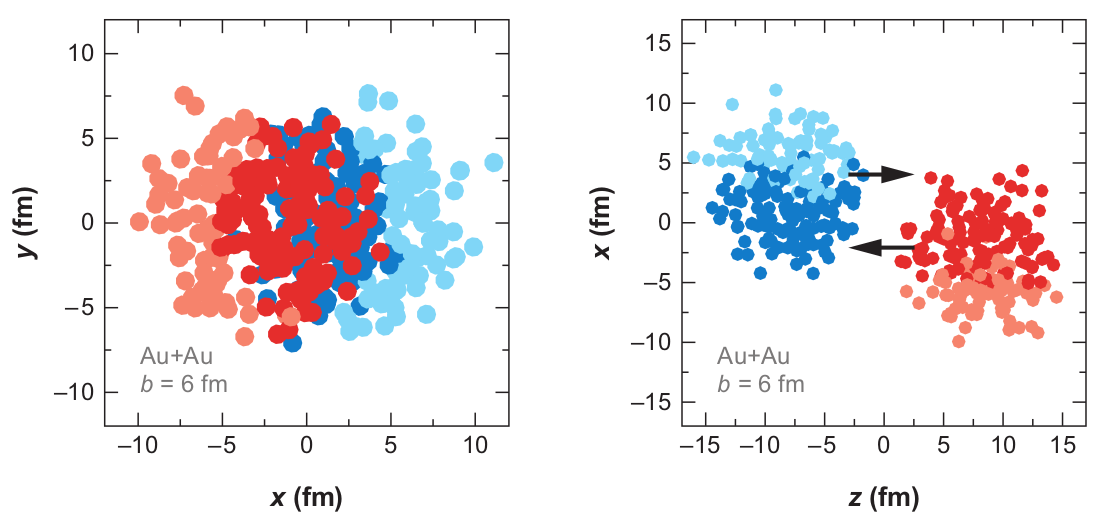}
	\caption{Event generated using a Monte Carlo Glauber approach. Participants are represented by darker circles. Left: transverse plane view. Right: view along the beam axis. Figure from \cite{Miller}.}
\end{figure}

The simplest criterion that can be employed to count the number of binary collisions is simply geometric, and justified by the black disk approximation \cite{Block:2012nj}. It consists in assuming that a nucleon-nucleon collision occurs every time that:
\begin{equation}
d \leq \sqrt{\frac{\sigma_{\text{inel}}^{NN}}{\pi}},
\end{equation}
where $d$ is the distance between two nucleons in the transverse plane. It is well known that the nucleons themselves are composite particles, and it should be noted that there has also been some progress in performing Glauber calculations at the sub-nucleonic level \cite{Loizides}.

\subsection{Eccentricity harmonics}

In a collision with $b = 0$, it is natural to imagine that the transverse geometry of the system should be approximately circular. In a non-central collision, on the other hand, the overlap region of the two colliding nuclei should have a rather elliptic shape, being elongated in the $y$ direction. Besides this evident ``almond'' shape, more complex geometries of the system are generated by event-by-event fluctuations on the spatial distribution of the nucleons inside the nuclei. 

\begin{figure}[h]
	\centering
	\includegraphics[width=12cm]{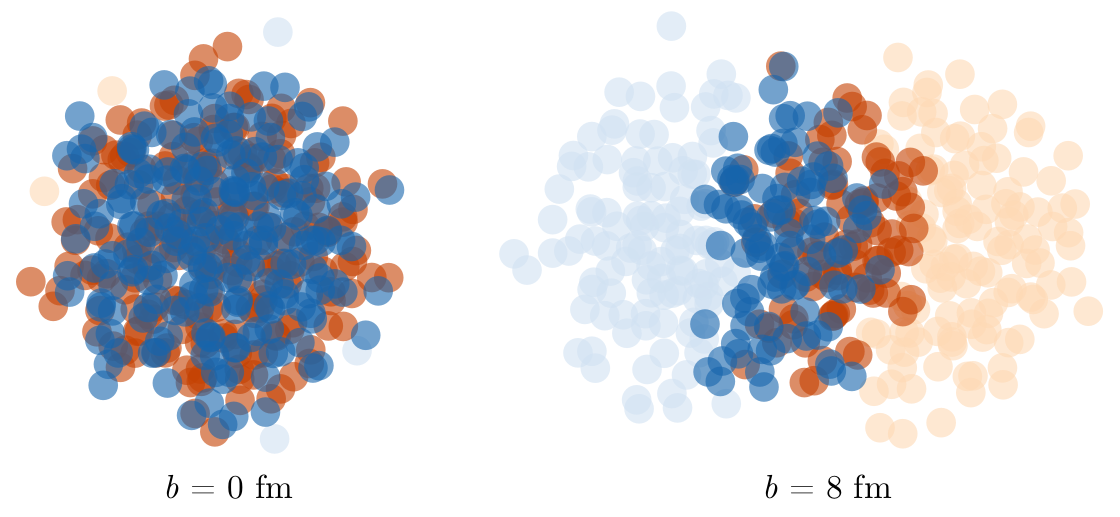}
	\caption{Transverse view of two colliding nuclei. In central events the geometry is approximately circular, and in non-central events it is approximately elliptical, but more complex irregularities are also always presents. Figure from \cite{BernhardThesis}.}
\end{figure}

In practice, the shape of the overlap area of the two colliding nuclei is not perfectly elliptical, and Figure 2.11 shows that even for $b = 0$ the overlap region is not perfectly circular: there is always some spatial anisotropy. This spatial anisotropy can be quantified by the eccentricity harmonics:

\begin{equation}
\varepsilon_{n} = \frac{\int r^{n} e^{in\varphi} s(r, \varphi) r dr d\varphi}{\int r^{n} s(r, \varphi) r dr d\varphi},
\end{equation}
which are calculated using the transverse entropy density distribution of the system $s(x,y)$ as a weight function. The eccentricity harmonics have a geometric interpretation: each harmonic quantifies the spatial anisotropy associated with a particular geometric shape in the initial state (Figure 2.12). In this way, $\varepsilon_{2}$ is appropriately called ellipticity, while $\varepsilon_{3}$ and $\varepsilon_{4}$ are referred to as triangularity and quadrangularity, respectively. 
\begin{figure}[h]
	\centering
	\includegraphics[width=16cm]{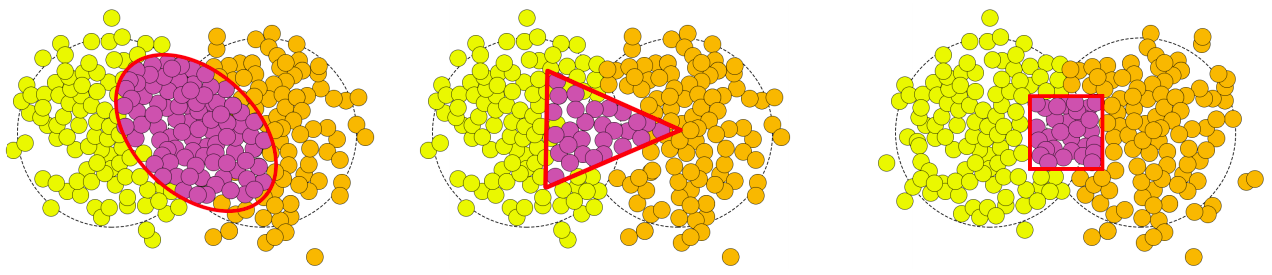}
	\caption{Eccentricity harmonics $\varepsilon_{n}$ for $n$ = 2, 3, 4.}
\end{figure}

On an event-by-event basis, there is a strong relation between the initial state geometry and final state flow observables. The initial geometry of the system dictates the direction of the pressure gradients in the beginning of the hydrodynamic evolution, so that is said that the final state momentum anisotropy is a \textit{hydrodynamic response} to the initial state anisotropy. For $n$ = 2, 3 there is a mapping between the eccentricity harmonics and its corresponding flow harmonics \cite{Noronha-Hostler, Gardim:2011xv, Hippert}, which can be expressed as:
\begin{equation}
v_{n} = f(\varepsilon_{n}) +\delta_{n},
\end{equation}
where $\delta_{n}$ is an error. As a first approximation, which is excellent for central events, a linear mapping can be used:
\begin{equation}
f(\varepsilon_{n}) = \kappa_{n}\varepsilon_{n},
\end{equation}
but for more peripheral events a linear + cubic mapping is a better predictor:
\begin{equation}
f(\varepsilon_{n}) = \kappa_{n}\varepsilon_{n} + \kappa_{n}^{'}|\varepsilon_{n}|^{2}\varepsilon_{n}
\end{equation}
Figure 2.13 shows the mapping between ellipticity and elliptic flow, and between triangularity and triangular flow for Pb-Pb collisions at $\sqrt{s_{\text{NN}}}$ = 2.76 TeV in the 45 - 50 \% centrality class.  

\begin{figure}[H]
    \centering
    \begin{subfigure}{.5\textwidth}
        \centering
        \includegraphics[width=8cm]{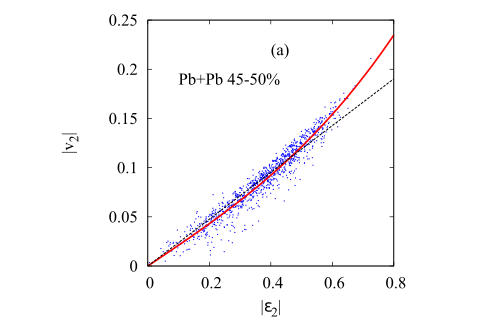}
    \end{subfigure}%
    \begin{subfigure}{.5\textwidth}
        \centering
        \includegraphics[width=8cm]{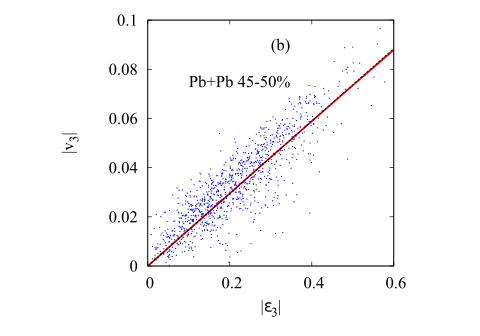}
    \end{subfigure}
    \caption{Mapping between eccentricity harmonics and flow harmonics for $n=2,3$. Each blue dot corresponds to an event. The dotted black line represents the linear estimator, and the full red line represents the cubic estimator. From \cite{Noronha-Hostler}.}
\end{figure}

\chapter{Simulating heavy-ion collisions}
\label{CHAP-Computational}
\noindent\rule{\textwidth}{.1pt}\\[1ex]

Contradicting the expectation that the deconfined state formed in heavy-ion collisions would be similar to a ``gas'' of quarks and gluons, experimental data collected at RHIC showed that instead a strongly interacting state is formed, similar to an almost ideal fluid. This discovery motivated the employment of relativistic hydrodynamics modelling to simulate heavy-ion collisions, which had great success in describing a variety of soft hadronic observables. Hydrodynamics has passed a series of tests and showed to be a good effective description of the bulk evolution of the QGP. More recently, hybrid simulations, where a different model is used to simulate each specific stage of the collision, are considered the most modern way to simulate heavy-ion collisions.

Chapter 3 provides an introduction to the models which constitute the hybrid simulation chain used to generate the data presented in this work. Section 3.1 introduces T$_{\text{R}}$ENTo, the initial condition generator. Section 3.2 and Section 3.3 present K$\o{}$MP$\o{}$ST and MUSIC, which are responsible for simulating the pre-hydrodynamic phase and hydrodynamic phase of the collision, respectively. Section 3.4 and Section 3.5 are devoted to the late stages of the collision: particlization, and final state resonance decays and hadronic interactions. 

\section{The initial condition: T$_{\text{R}}$ENTo}
\label{SEC-TRENTo}

T$_{\text{R}}$ENTo (Reduced Thickness Event-by-event Nuclear Topology) is a Glauber inspired model used to generate an initial entropy density profile of two colliding projectiles (proton-proton, proton-nucleus or nucleus-nucleus) in the transverse plane of the collision \cite{TRENToArticle, TRENToWebsite}. It is an effective model, meaning that it is not based in first principles, and no assumptions are made about the specific mechanisms of entropy production.  

\begin{figure}[h]
	\centering
	\includegraphics[width=14cm]{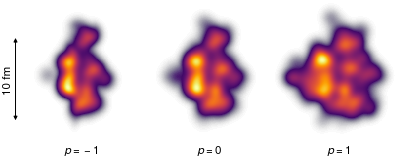}
	\caption{Transverse entropy distributions from T$_{\text{R}}$ENTo for three values of the entropy deposition parameter $p$. Figure from \cite{TRENToWebsite}.}
\end{figure}

In a collision between two protons $A$ and $B$, with impact parameter $b$ in the $x$ direction, their nuclear densities are given by:
\begin{equation}
\rho_{A,B} = \rho_{\text{proton}}(x \pm b/2, y, z).
\end{equation}
In T$_{\text{R}}$ENTo, the thickness function of each nucleon:
\begin{equation}
T_{A,B}(x,y) = \int dz \rho_{A,B}(x,y,z)
\end{equation}
is given by a boosted two-dimensional Gaussian:
\begin{equation}
T_{A,B}(x,y) = \frac{1}{2 \pi w^{2}}\exp\left(-\frac{x^{2}+y^{2}}{2 w^{2}}\right)
\end{equation}
A \textit{fluctuated thickness} is assigned to each one of the protons $A$ and $B$:
\begin{equation}
\tilde{T}_{A,B}(x,y) = w_{A,B} . T_{A,B}(x,y),
\end{equation}
where $w_{A,B}$ are independent random weights sampled from a gamma distribution with unit mean:
\begin{equation}
P_{k}(w) = \frac{k^{k}}{\Gamma(k)}w^{k-1}e^{-kw}.
\end{equation}
These weights introduce additional multiplicity fluctuations, which are necessary to reproduce the large multiplicity fluctuations observed experimentally in proton-proton collisions \cite{ALICE:2017pcy}. 
T$_{\text{R}}$ENTo proposes a function $f(\tilde{T}_{A},\tilde{T}_{B})$ to convert projectile thickness into entropy deposition:
\begin{equation}
f(\tilde{T}_{A},\tilde{T}_{B}) \propto \frac{dS}{d^{2}x_{\perp} d\eta} = s(x,y)
\end{equation}
This function is the \textit{reduced thickness}, which is a generalized mean controlled by a real parameter $p$:
\begin{equation}
f = \tilde{T}_{R}(p;\tilde{T}_{A},\tilde{T}_{B}) = \left (\frac{\tilde{T}_{A}^{p} + \tilde{T}_{B}^{p}}{2} \right)^{1/p}.
\end{equation}
Entropy is determined by the reduced thickness up to an overall normalization factor. For certain values of $p$, the reduced thickness simplifies to the arithmetic, geometric and harmonic means:
\begin{equation}
    \tilde{T}_{R}=
\small
\begin{dcases}
    \text{min}(\tilde{T}_{A},\tilde{T}_{B}) & \quad p \rightarrow -\infty \\
    2\tilde{T}_{A}\tilde{T}_{B}/(\tilde{T}_{A} + \tilde{T}_{B}) &  \quad p = -1 \quad \text{(harmonic)} \\
    \sqrt{\tilde{T}_{A}\tilde{T}_{B}} &  \quad p = 0 \quad \text{(geometric)} \\
    (\tilde{T}_{A} + \tilde{T}_{B})/2 &  \quad p = 1 \quad \text{(arithmetic)} \\
    \text{max}(\tilde{T}_{A},\tilde{T}_{B}) & \quad p \rightarrow +\infty \\
\end{dcases}
\end{equation}

\begin{figure}[h!]
	\centering
	\includegraphics[width=10cm]{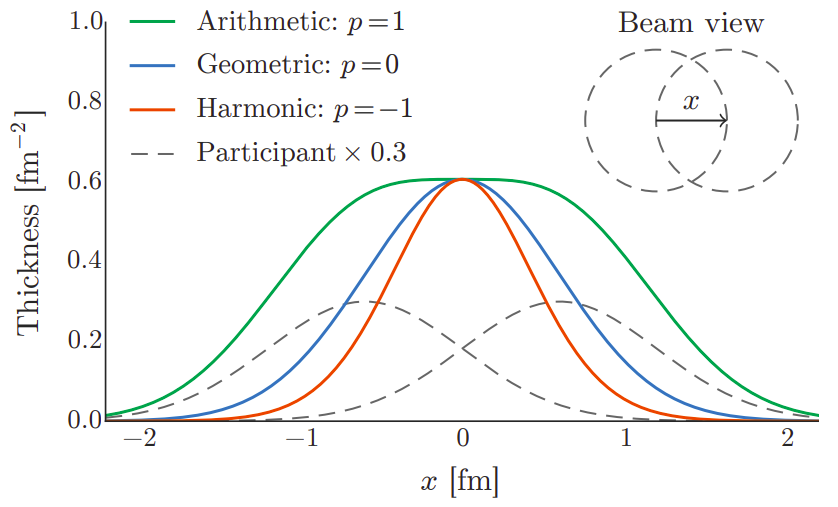}
	\caption{Reduced thickness function for three values of the entropy deposition parameter $p$ in a non-central collision. Figure taken from \cite{TRENToArticle}.}
	\label{FIG-p_parameter}
\end{figure}

Composite systems are treated as superpositions of proton-proton collisions. In a nucleus-nucleus collision, a set of nucleon positions is sampled for each projectile using an uncorrelated Woods-Saxon distribution. In this process, nucleons are forbidden to be placed closer than a distance $d_{\text{min}}$ from another one that was previously positioned. For each pair of nucleons, the collision probability:
\begin{equation}
P_{\text{coll}}(b) = 1 - \exp \left [-\sigma_{gg} T_{AB}(b) \right]
\end{equation}
is sampled once to decide if the nucleons collide. In Equation (3.9), $T_{AB}(b)$ is the overlap integral of the two nucleons $A$ and $B$:
\begin{equation}
T_{AB}(b) = \int dx dy T_{A}(x-b/2,y) T_{B}(x+b/2,y)
\end{equation}
The nucleons which interact at least once are labelled as participants and assigned a fluctuated thickness, and the others are discarded. Then, in a collision between two \textit{nuclei} $A$ and $B$, the fluctuated thickness for nucleus $A$ is the sum of the individual fluctuated thickness of each of its participants:
\begin{equation}
\tilde{T}_{A} = \sum_{i=1}^{N_{\text{part}}} w_{i} \int dz \rho_{\text{proton}}, (x-x_{i},y-y_{i},z-z_{i})
\end{equation}
and analogously for nucleus $B$. $\sigma_{gg}$ is an effective parton-parton cross-section tuned so that the total proton-proton cross section matches the experimental inelastic nucleon-nucleon cross-section:
\begin{equation}
\int 2 \pi b db P_{\text{coll}}(b) = \sigma_{\text{NN}}^{\text{inel.}}
\end{equation}
Following the Glauber model assumptions, the inelastic nucleon-nucleon cross-section remains constant, and the probability of two nucleons interacting depends solely on the impact parameter. It is independent of the number of previous collisions suffered by each nucleon, and the location of the nucleon inside the nucleus. 

\section{Pre-equilibrium dynamics: K$\o{}$MP$\o{}$ST}
\label{SEC-KoMPoST}

The initial state formed just after the collision is far from equilibrium, and complicated. In the time scale of approximately 1 fm/c, this out of equilibrium initial state evolves to a state of local equilibrium, which defines the beginning of the hydrodynamic evolution. K$\o{}$MP$\o{}$ST \cite{KompostArticle} proposes an effective macroscopic description of the dynamics of the system in the early moments of the collision, providing a map of the form:
\begin{equation}
T^{\mu \nu}(\tau_{0}, \mathbf{x})|_{\text{out-of-equilibrium}} \longrightarrow T^{\mu \nu} (\tau_{\text{hydro}}, \mathbf{x}), 
\end{equation}
relating the energy-momentum tensor in an early time $\tau_{0} \sim$ 0.1 fm/c to a later time where hydrodynamics should become applicable, $\tau_{\text{hydro}} \sim$ 1 fm/c.

\begin{figure}[h]
	\centering
	\includegraphics[width=13cm]{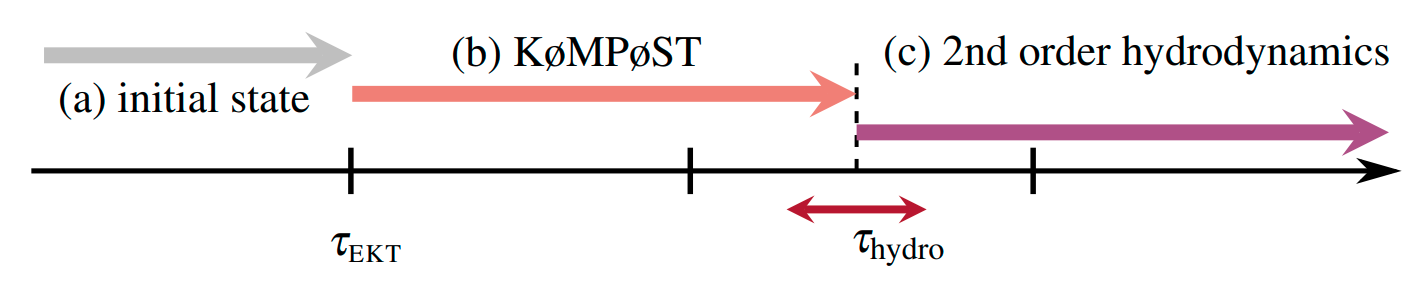}
	\caption{Ilustration of the time evolution of the simulation: an initial condition passes through K$\o{}$MP$\o{}$ST pre-equilibrium before entering relativistic hydrodynamics. Figure from \cite{Kurkela:2018qwx}.}
\end{figure}

K$\o{}$MP$\o{}$ST initially divides the energy-momentum tensor into a local homogeneous background and small perturbations around it:
\begin{equation}
T^{\mu \nu}(\tau_{0},\mathbf{x'}) = \overline{T}^{\mu \nu}(\tau_{0}) + \delta T_{\mu \nu}(\tau_{0},\mathbf{x'})
\end{equation}
As a first approximation, these perturbations can be studied in the framework of linear response theory. Then the energy-momentum tensor at $\tau_{\text{hydro}}$ can be calculated by adding to the evolved background the response to the initial perturbations from equilibrium: 
\begin{equation}
T^{\mu \nu}(\tau_{\text{hydro}},\mathbf{x'}) = \overline{T}_{\mathbf{x}}^{\mu \nu}(\tau_{\text{hydro}}) + \frac{\overline{T}_{\mathbf{x}}^{\tau \tau}(\tau_{\text{hydro}})}{\overline{T}_{\mathbf{x}}^{\tau \tau}(\tau_{0})} \int d^{2}\mathbf{x'} G_{\alpha \beta}^{\mu \nu}(\mathbf{x}, \mathbf{x'}, \tau_{\text{hydro}}, \tau_{0}) \delta T_{\mathbf{x}}^{\alpha \beta}(\tau_{0}, \mathbf{x'})
\end{equation}
The first term is the (non-linear) equilibration of the background and the second term is a convolution of the initial perturbations and the response functions, which is responsible for populating the non-diagonal terms of the energy-momentum tensor. The second term is the convolution of the initial deviations from equilibrium and the response functions, normalized by the background energy density. It is important to note that, due to causality, the energy-momentum tensor at ($\tau_{\text{hydro}}, \mathbf{x}$) is only affected by cells that are within the causal neighborhood of $\mathbf{x}$:
\begin{equation}
|\mathbf{x'} - \mathbf{x}| < c(\tau_{\text{hydro}} - \tau_{0}),
\end{equation}
which is represented by the white circle in Figure 3.4. 

The non-equilibrium evolution of the background and the response functions have to be calculated according to an underlying microscopic description. A microscopic description of the out-of-equilibrium system formed in the collision can be provided by QCD effective kinetic-theory (EKT) \cite{Arnold} :
\begin{equation}
p^{\mu}\partial_{\mu} f(x,p) = \mathcal{C}_{2 \leftrightarrow 2}[f] + \mathcal{C}_{1 \leftrightarrow 2}[f]
\end{equation}
In this relativistic Boltzmann equation, $\mathcal{C}_{2 \leftrightarrow 2}[f]$ is the collision integral for elastic scatterings at leading order and $\mathcal{C}_{1 \leftrightarrow 2}[f]$ is the collision integral for inelastic, particle number changing processes. K$\o{}$MP$\o{}$ST can also be used in a free streaming limit. In the free streaming scenario the partons, which are taken to be massless, don't interact, so that the evolution of the system is dictated by a homogeneous Boltzmann equation:
\begin{equation}
p^{\mu}\partial_{\mu} f(x,p) = 0
\end{equation}
In this case, the perturbations around equilibrium can be obtained analytically and expressed in terms of the Bessel functions \cite{KompostArticle}.   
\begin{figure}[h]
	\centering
	\includegraphics[width=7cm]{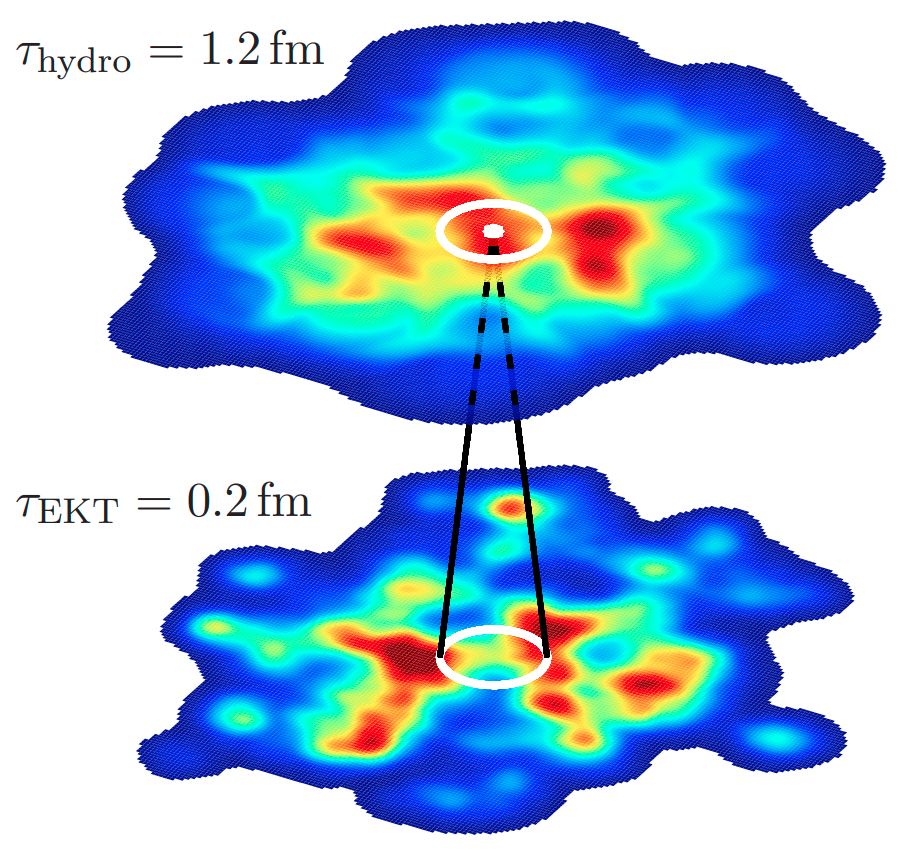}
	\caption{Transverse energy density distribution evolved from $\tau_{\text{EKT}}$ = 0.2 fm/c to $\tau_{\text{hydro}}$ = 1.2 fm/c using K$\o{}$MP$\o{}$ST EKT. The white circle represents the causal circle of the white dot. From \cite{KompostArticle}.}
\end{figure}

The energy-momentum tensor coming from T$_{\text{R}}$ENTo (or from any Glauber-based calculation) initially has the simple, diagonal form:
\begin{center}
$T^{\mu \nu}(\tau_{0}, \mathbf{x}) = \begin{pmatrix}
e(\mathbf{x}) & 0 & 0 & 0 \\
0 & \frac{1}{2}e(\mathbf{x}) & 0 & 0 \\ 
0 & 0 & \frac{1}{2}e(\mathbf{x}) & 0 \\
0 & 0 & 0 & 0
\end{pmatrix}$
\end{center}
At the end of the simulation, the energy-momentum tensor can be decomposed in the Landau frame and written in the usual form used in hydrodynamic simulations:
\begin{equation}
T^{\mu \nu} = e u^{\mu} u^{\nu} + P(e) \Delta^{\mu \nu} + \pi^{\mu \nu},
\end{equation}
where $\Delta^{\mu \nu} = g^{\mu \nu} + u^{\mu} u^{\nu}$, and $P=P(e)$ is the equation of state (see Section 3.3). At this point, the energy density distribution of the system should be smoother, closer to the assumptions of local equilibrium required by hydrodynamics (Figure 3.4).

\section{Hydrodynamic evolution: MUSIC}
\label{SEC-MUSIC}

After pre-equilibrium and thermalization, the dynamics of the QGP is well described by relativistic hydrodynamics \cite{Heinz, Gale}, which is based on \textit{local} conservation laws:
\begin{equation}
\partial_{\mu} T^{\mu \nu} = 0 
\end{equation}
For a relativistic ideal fluid, the energy-momentum tensor takes the form:
\begin{equation}
T^{\mu \nu} = e u^{\mu} u^{\nu} - P(g^{\mu \nu} - u^{\mu}u^{\nu}), 
\end{equation}
where $e$ is the energy density and $P$ is the pressure. $u^{\mu}$ is the fluid velocity field:
\begin{equation}
u^{\mu} = \frac{dx^{\mu}}{d \tau},
\end{equation}
which is normalized so that $u^{\mu}u_{\mu} = 1$, and the derivative is taken with respect to proper time. Including the viscous corrections, the energy-momentum tensor can be written as:
\begin{equation}
T^{\mu \nu} = e u^{\mu} u^{\nu} - (P + \Pi) \Delta^{\mu \nu} + \pi^{\mu \nu}
\end{equation}
In Equation (3.23) $\pi^{\mu \nu}$ is the shear stress tensor, $\Pi$ is the bulk viscous pressure and $\Delta^{\mu \nu} = g^{\mu \nu} - u^{\mu} u^{\nu}$ is called the projection operator. Shear viscosity is a measure of the resistance between layers of a fluid, while the bulk viscosity is associated with the fluid expansion rate (Figure 3.5). Pressure is not independent, it is related to the energy density by the equation of state $P = P(e)$.
\begin{figure}[h]
	\centering
	\includegraphics[width=10cm]{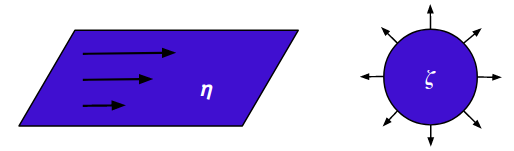}
	\caption{Illustration of shear viscosity (left) and bulk viscosity (right).}
	\label{FIG-Viscosity}
\end{figure}

MUSIC \cite{MUSICWebsite} is a publicly available C++ code which performs 2+1D relativistic hydrodynamics simulations. Given an initial condition (which is the energy density distribution of the system at $\tau_{\text{hydro}}$) and an equation of state, MUSIC evolves the energy-momentum tensor by solving the hydrodynamic equations of motion \cite{Denicol:2012cn, Molnar:2013lta} numerically, while respecting the conservation laws in Equation (3.20). 

MUSIC uses temperature-dependent parametrizations for the shear viscosity to entropy density ratio $\eta/s$ and also for the bulk viscosity to entropy density ratio $\zeta/s$. The entropy density can be used as a substitute for number density, so that these quantities are an attempt of capturing the ``viscosity per unit''. For the shear viscosity, which should reach a minimum near the critical temperature $T_{c}$, a modified linear parametrization is used:
\begin{equation}
(\eta/s)(T) = (\eta/s)_{\text{min}} + (\eta/s)_{\text{slope}} (T-T_{c}) \left(\frac{T}{T_{c}} \right)^{(\eta/s)_{\text{curve}}},
\end{equation}
which contains three free parameters: a minimum value, a slope above $T_{c}$ and a curvature parameter. When the curvature parameter $(\eta/s)_{\text{curve}}$ is set to zero, Equation (3.24) reduces to a simple linear parametrization. For the bulk viscosity, an unnormalized Cauchy distribution is used:
\begin{equation}
(\zeta/s)(T) = \frac{(\zeta/s)_{\text{max}}}{1 + \left(\frac{T - (\zeta/s)T_{0}}{(\zeta/s)_{\text{width}}} \right)^{2}},
\end{equation}
which also has three tunable parameters: a maximum value, the width of the peak and its location $T_{0}$. Figure 3.6 shows the functional form of Equations (3.8) and (3.9) for some values of the free parameters. 

\begin{figure}[h]
	\centering
	\includegraphics[width=14cm]{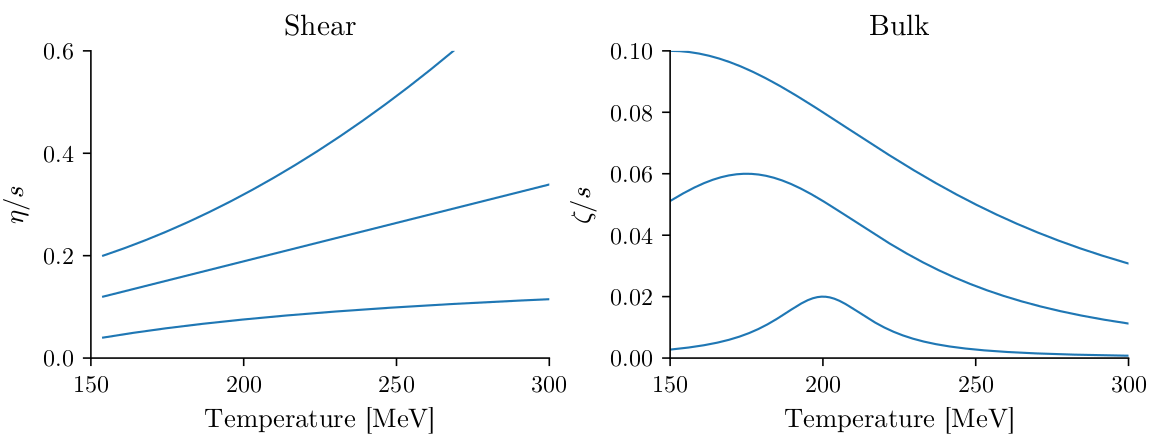}
	\caption{Shear and bulk viscosity parametrizations given in Equations (3.8) and (3.9) for different choices of the free parameters. Figure taken from \cite{BernhardThesis}.}
	\label{FIG-Viscosity}
\end{figure}

In MUSIC, reaching a switching temperature $T_{\text{switch}}$ is the criterion used for stopping hydrodynamics. Connecting all the cells at which the temperature reached the switching temperature forms a 4-dimensional spacetime hypersurface (Figure 3.8). This hypersurface is complete when the temperature of all cells drops to this switching temperature, and hydrodynamic evolution reaches an end.

\section{Particlization: iSS}
\label{SEC-iSS}

Once all points of the grid have cooled down to the switching temperature, it is necessary to change the description of the system from the hydrodynamic picture to a hadron gas description, so that each cell must be converted into discrete particles. This change from hydrodynamics to a microscopic description is known as \textit{particlization} \cite{Huovinen}.

\begin{figure}[h]
	\centering
	\includegraphics[width=7cm]{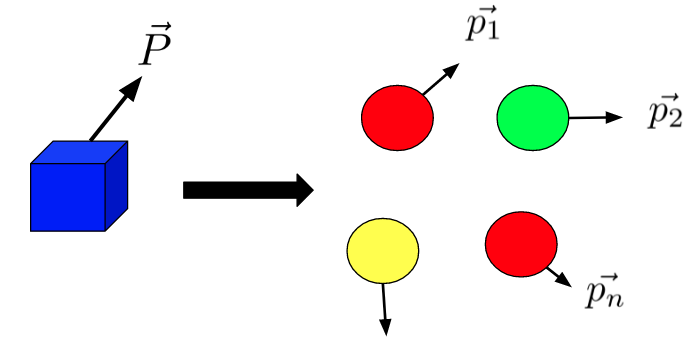}
	\caption{Illustration of the particlization procedure: a fluid cell is converted into hadrons.}
	\label{FIG-Viscosity}
\end{figure}

The momentum distribution of particles of species $i$ (with degeneracy factor $g_{i}$) emerging from particlization can be calculated using the Cooper-Frye prescription \cite{Cooper-Frye}:
\begin{equation}
E \frac{dN_{i}}{dp^{3}} = \frac{g_{i}}{(2\pi)^{3}}\int_{\Sigma} f(p) p^{\mu} d \sigma_{\mu}
\end{equation}
The integral in Equation (3.26) is performed on the isothermal hypersurface $\Sigma$ of temperature $T_{\text{switch}}$ and $d \sigma_{\mu}$ is a volume element of the four-dimensional surface normal to the surface. The integral is calculated using the one-particle distribution function. 
\begin{figure}[h]
	\centering
	\includegraphics[width=7cm]{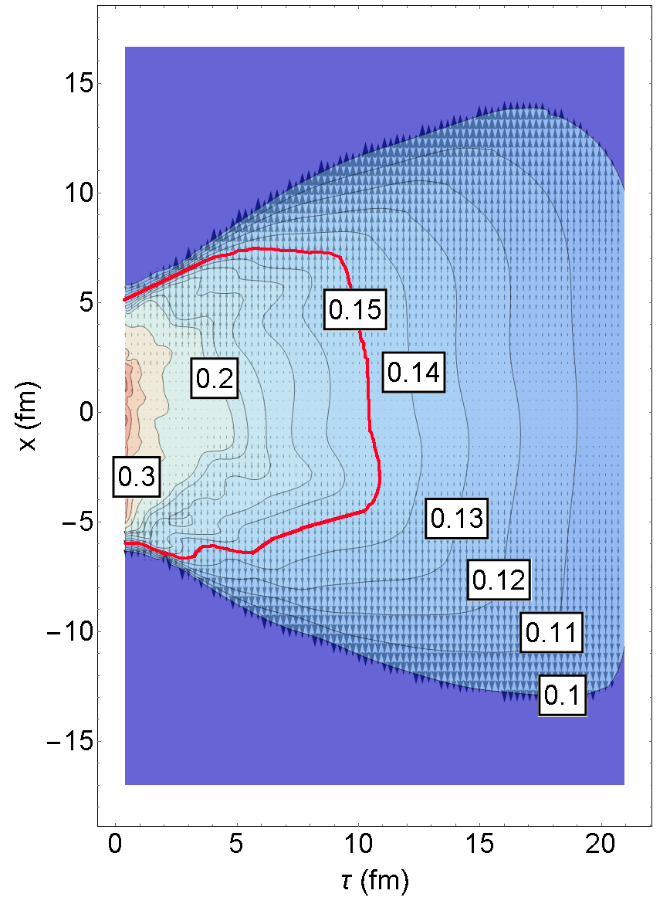}
	\caption{Two-dimensional visualization of a freeze-out hypersurface at $T_{\text{switch}}$ = 150 MeV (red line). Figure from \cite{MUSICWebsite}.}
	\label{FIG-Viscosity}
\end{figure}

If the system was at perfect thermal equilibrium, the distribution function would simply be the Bose-Einstein or Fermi-Dirac distribution:
\begin{equation}
f_{0}(p) = \frac{1}{e^{p \cdot u/T} \mp 1}
\end{equation}
This is, however, not the case and out-of-equilibrium corrections to the distribution function must be made. These corrections can be divided in two main categories. One possibility is to use small linear corrections to the equilibrium distribution function $f_{0}$ \cite{ANDERSON1974466, Chapman}:
\begin{equation}
f = f_{0} + \delta f
\end{equation}
Another strategy is to perform a transformation in the momentum vector inside the distribution function \cite{PrattTorrieri} using a linear transformation matrix $\lambda_{ij}$:
\begin{equation}
p_{i} \rightarrow p_{i}' + \sum_{j} \lambda_{ij} p_{j}
\end{equation}
In this work, we use a linear correction derived from the Boltzmann equation using relaxation time approximation (RTA) \cite{Bozek:2009dw, Dusling:2011fd, Teaney:2003kp}:
\begin{equation}
\delta f = f_{0}(1 \pm f_{0}) \frac{\tau}{ET}\left[\frac{1}{2 \eta} p^{i} p ^{j} \pi_{ij} + \frac{1}{\zeta} \left( \frac{p^{2}}{3} - c_{s}^{2}E^{2} \right) \Pi \right],
\end{equation}
where $\pi_{ij}$ is the shear stress tensor and $\tau$ is a shear and bulk relaxation time, which is taken to be constant and the same for all particle species.

iSS (iSpectraSampler) \cite{iSSarticle, iSSWebsite} is a Monte Carlo particle sampler that uses the Cooper-Frye formula to calculate the momentum distribution of particles coming from a freeze-out hypersurface. iSS generates sets of momenta and positions for particles emitted at each cell at the end of the hydrodynamic simulation. The procedure of doing this many times on the same hypersurface is known as \textit{oversampling}. Oversampling is often used due to the fact that the hydrodynamic part of the simulation is the most expensive one computationally, and one would like to ``reuse'' the same hydrodynamic evolution to generate more events. 

\section{Hadronic phase: UrQMD}
\label{SEC-UrQMD}

Once the hydrodynamic cells have been converted into particles, it is necessary to simulate the last stage of the collision, in which these hadrons travel to the detector. In the way, they interact and resonances might decay into more stable species. A microscopic description of this hadronic phase is provided by the Ultra-Relativistic Quantum Molecular Dynamics Model (UrQMD) \cite{Bass, Bleicher}, which was developed in the late 90's.

UrQMD is a microscopic transport model, in which particles travel in classical trajectories and interact stochastically according to experimental hadron-hadron cross-sections. The UrQMD package \cite{UrQMDWebsite} solves a Boltzmann equation for each hadron species:
\begin{equation}
\frac{df_{i}(x,p)}{dt} = \mathcal{C}_{i}(x,p),
\end{equation}
where the collision term accounts for $2 \leftrightarrow 2$ and $1 \leftrightarrow 2$ processes, such as binary collisions and decays. UrQMD contains 53 baryon species and 24 different meson species. Two particles collide if the distance between them satisfies:
\begin{equation}
d \le d_{0} = \sqrt{\frac{\sigma_{\text{tot}}}{\pi}},
\end{equation}
where the cross-section is taken to be the free cross-section of the interaction in question. The interactions and decays performed by UrQMD constitute the last stage of the hybrid simulation.

\chapter{Motivation}
\label{CHAP-Motivation}
\noindent\rule{\textwidth}{.1pt}\\[1ex]

Chapter 4 presents the motivation of this work: evidences that the size of nucleons, which is a free parameter in the initial condition, has been overestimated in most recent Bayesian Analyses, in which the model is confronted with experimental data to obtain a best-fit value of each parameter. Section 4.1 brings a short review about the current understanding about the nucleon size. Section 4.2 provides an overview of the general structure of Bayesian Analyses used to constrain the parameters of hybrid simulations of heavy-ion collisions. In Section 4.3, the motivation of the work is exposed in detail, which sets the ground for Chapter 5.

\section{The nucleon size}

Experimentally, probing inside hadrons is a huge challenge. From the theoretical point of view, studying the hadronic structure is also a hard task, as it is associated with the non-perturbative part of QCD. Not by accident, details about the structure of hadrons are in general still poorly known, including information about the nucleons. 

The charge radius of the proton can be defined as the slope of its electromagnetic form factor at zero momentum transfer:
\begin{equation}
r_{p}^{2} = 6  \left. \frac{dG_{p}^{E}}{d Q^{2}} \right|_{Q^{2}=0}
\end{equation}
The first indirect measurement of the proton charge radius was made by Hofstader et al. in the famous electron scattering experiment \cite{Hofstadter1, Hofstadter2} which later on won the Nobel Prize in 1961. In this experiment, the proton radius was estimated by fitting the electric form factor with a dipole form. The charge radius of the proton can also be determined by Lamb shift measurements, performed for both electronic and muonic hydrogen atoms \cite{PhysRev.86.288}. Most of the first Lamb shift experiments performed with electronic atoms gave as a result for the proton charge radius roughly the same value as electron scattering experiments, $r_{p}$ = 0.88 fm, which came to be known as the \textit{large radius}. Lamb shift results with muonic atoms, on the other hand, gave the so-called \textit{small radius}, $r_{p}$ = 0.84184(67) fm, which differed by 5$\sigma$ from the large radius. This discrepancy came to be known as the proton radius puzzle \cite{Hammer, Pohl}. As both theory and experiment advanced, electronic and muonic Lamb shift experiments came to an agreement, and the present CODATA value for the proton charge radius is $r_{p}$ = 0.8414(19) fm \cite{CODATA}, closer to the small radius. Table 4.1 shows the results of some of the most modern measurements of the proton charge radius performed with different experimental techniques. The Paris electronic Lamb shift measurement from 2018 \cite{Fleurbaey:2018fih} remains an exception, with an extracted charge radius that is actually closer to the large radius. 

\begin{table}[h]
\centering
\begin{tabular}{c | c | c |c}
$r_{p}$ (fm) &  Year & Method  &  Reference \\ \hline
0.8335(95)  &  2017  & Electronic Lamb shift &  \cite{Beyer:2017gug} \\ 
0.877(13) & 2018  & Electronic Lamb shift & \cite{Fleurbaey:2018fih} \\ 
0.833(10) & 2019  & Electronic Lamb shift & \cite{Bezginov:2019mdi}\\ 
0.831(7)(12) &  2019 & $e^{-}p$ scattering & \cite{Xiong:2019umf} \\ 
\end{tabular}
\caption{Recent extractions of the proton charge radius from electronic Lamb shift experiments and $e^{-}p$ scattering.}
\end{table}

The strong radius of the proton should be smaller than its charge radius, once it is associated with the short-range strong interaction. The electron is electrically charged, and penetrates in the nucleus, which makes it a good probe for the electronic structure of the proton. On the other hand, it is not sensitive to the color fields in the nuclear interior. Low-energy scattering of protons on nuclei provides information about both the electronic and strong interaction. However, unlike electrons, the protons are extended, composite objects, and interact in a much more complicated way as compared to point-like particles. The gluonic structure of nuclei can be studied via $J / \Psi$ scattering. The $J / \Psi$ meson is formed by a charm quark-antiquark pair and has a naturally small dipole, which should scatter on the nucleons individually. The $J / \Psi$ meson interacts mainly by two-gluon exchange, so that scattering off a nuclei should really probe the spatial distribution of the \textit{color} fields of the nucleons. The proton strong radius has been determined from analysis of diffractive $J / \Psi$ photoproduction in $e^{-} p$ collisions at the HERA collider \cite{JPsiScattering}. Results suggest that the transverse strong size of the proton is approximately 0.5 fm.

The mass radius of the proton is a quantity that has not yet been determined experimentally. The gravitational field created by a single proton is extremely weak, so that a direct measurement is currently very difficult. Nevertheless, the proton mass radius was recently estimated by Kharzeev, who defines the mass radius of the proton through the form factor of the trace of the energy-momentum tensor of QCD in the weak gravitational field approximation \cite{Kharzeev}. This form factor is extracted from data on photoproduction of $J / \Psi$ and $\Upsilon$ quarkonia from the GlueX Collaboration \cite{GlueX}. The extracted mass radius of the proton was $R_{\text{m}}$ = 0.55 $\pm$ 0.03 fm.

\section{Constraining the hybrid simulations: Bayesian Analysis}

Hybrid simulations such as the one presented in Chapter 3 are mostly based on effective models, which contain numerous free parameters. Each of these parameters affects every one of the observables and, reciprocally, each observable is in general affected by every one of the parameters. Due to the large amount of free parameters and their complex (usually non-linear) relation to the observables, a robust quantitative tool is necessary to constrain the values of the model parameters and guarantee that they are connectable to actual physical properties of the QGP. Loosely speaking, this procedure should consist in some sort of ``global fit'', in which the model is confronted with experimental data, and all parameters and observables are simultaneously taken into account. 

The framework of Bayesian statistics provides a tool for tackling this kind of problem, and has been employed to constrain the parameters of the models which make up hybrid simulations of heavy-ion collisions. Such analyses are based in a powerful identity, Bayes' theorem, which seeks to quantitatively measure a "degree of belief" about something, given some previously available information. Given a proposition A and some known information B (the evidence), Bayes' theorem states that the probability $P(\text{A}|\text{B})$ of A being true given that evidence B is known (the posterior probability) is given by:
\begin{equation}
P(\text{A}|\text{B}) = \frac{P(\text{B}|\text{A})P(\text{A})}{P(\text{B})}
\end{equation}
$P(\text{A})$ is the prior, and represents the initial degree of belief in A. $P(\text{B}|\text{A})$ is the likelihood for B to be true if the proposition A holds. $P(\text{B})$, usually referred to as Bayes evidence for information B, acts as a normalization factor:
\begin{equation}
P(\text{B}) = \int P(\text{B}|\text{A})P(\text{A}) d \text{A},
\end{equation}
so that the statement of Bayes' theorem can be expressed as:
\begin{equation}
P(\text{A}|\text{B}) \propto P(\text{B}|\text{A}) P(\text{A})
\end{equation}
It states that the posterior probability distribution of A given that B is true is proportional to the product of the previous knowledge about A and the likelihood of B given A. 

How can this formalism be applied to constrain the parameters associated with properties of the QGP? In this case, the hybrid model is the proposition to be tested, and the experimental data collected in colliders constitute the available evidence. In this context, Equation (4.4) can be written as:
\begin{equation}
\mathcal{P}(\mathbf{x}|\mathbf{y}_{\text{exp}}) \propto \mathcal{P}(\mathbf{y}_{\text{exp}}|\mathbf{x})\mathcal{P}(\mathbf{x}),
\end{equation}
where $\mathbf{x}$ is a vector that stores in its entries the model parameters and $\mathbf{y}$ is a vector containing $m$ observation points (sometimes called calibration data). 

For the prior distribution, which expresses the knowledge about the model parameters without any comparison with data, a uniform distribution is chosen for each parameter $x_{i}$:
\begin{equation}
    \mathcal{P}(x_{i}) \propto
\small
\begin{dcases}
  1,   &  \text{if} \quad x_{i,\text{min}} \leq x_{i} \leq x_{i,\text{max}} \\
  0,   &  \text{else} \\
\end{dcases}
\end{equation}
This is also sometimes referred to as the agnostic probability distribution: each parameter is previously believed to be equally likely to assume any value inside an interval, which is chosen based on basic physical constraints. It is assumed that the priors for each parameter are independent, so that the joint prior is the product of the individual priors:
\begin{equation}
\mathcal{P}(\mathbf{x}) \propto \prod_{i} \Theta(x_{i} - x_{i,\text{min}}) \Theta(x_{i} - x_{i,\text{max}})
\end{equation}

The exact form of the likelihood function is rarely known. In the case where uncertainties are normally distributed, the most natural choice is a multivariate Gaussian:
\begin{equation}
\mathcal{P}(\mathbf{y}_{\text{exp}}|\mathbf{x}) = \frac{1}{\sqrt{(2\pi)^{m} \text{det} \Sigma }}\exp\left(- \frac{1}{2} \Delta \mathbf{y}^{T} \Sigma^{-1} \Delta \mathbf{y} \right),
\end{equation}
where:
\begin{equation}
\Delta \mathbf{y} = \mathbf{y}_{\text{model}} - \mathbf{y}_{\text{exp}}
\end{equation}
and $\Sigma$ is the total covariance matrix, which contains the model uncertainties, experimental uncertainties and also correlations between uncertainties. In the case of a multi-system analysis, the joint likelihood function is given by the product of the systems' individual likelihood functions. For example, in an analysis which uses simultaneously data from Pb-Pb and p-Pb collisions:
\begin{equation}
\mathcal{P}(\mathbf{y}_{\text{exp}}|\mathbf{x}) = \mathcal{P}(\mathbf{y}_{\text{exp}}^{\text{Pb-Pb}}|\mathbf{x}) \mathcal{P}(\mathbf{y}_{\text{exp}}^{\text{p-Pb}}|\mathbf{x}) 
\end{equation}

With all the ingredients on the right side of Equation (4.5) in hand, estimation of the posterior is accomplished via Markov Chain Monte Carlo (MCMC) algorithms \cite{MCMC}, which allow estimation of the shape of the posterior distribution. Such methods produce a representative sample of the posterior distribution by performing a random walk in parameter space weighted by the posterior. To speed up the output calculation for each parameter space point, a Gaussian process emulator is used to substitute the model. The set of parameters which maximizes the posterior (the mode of the posterior distribution) is called the \textit{maximum a posteriori} (MAP) set of parameters:
\begin{equation}
\mathbf{x}_{\text{MAP}} = \underset{\mathbf{x}}{\text{arg max}} \: \mathcal{P}(\mathbf{x}|\mathbf{y}_{\text{exp}}).
\end{equation}
Since all the priors are uniform distributions, the set of parameters that maximizes the posterior is the one which maximizes the likelihood function. In other words, the MAP parameters are the ones which best fit the experimental data. The MAP value of each parameter can be visualized as the maximum of its marginalized distribution:
\begin{equation}
\mathcal{P}(x_{i}|\mathbf{y}_{\text{exp}}) = \int dx_{1} ... dx_{i-1} dx_{i+1} ... dx_{n} \mathcal{P}(\mathbf{x}|\mathbf{y}_{\text{exp}}),
\end{equation}
which is obtained by integrating out all the other parameters on the posterior distribution (shown in the diagonal panels of Figure 4.1).
\begin{figure}[h]
	\centering
	\includegraphics[width=10cm]{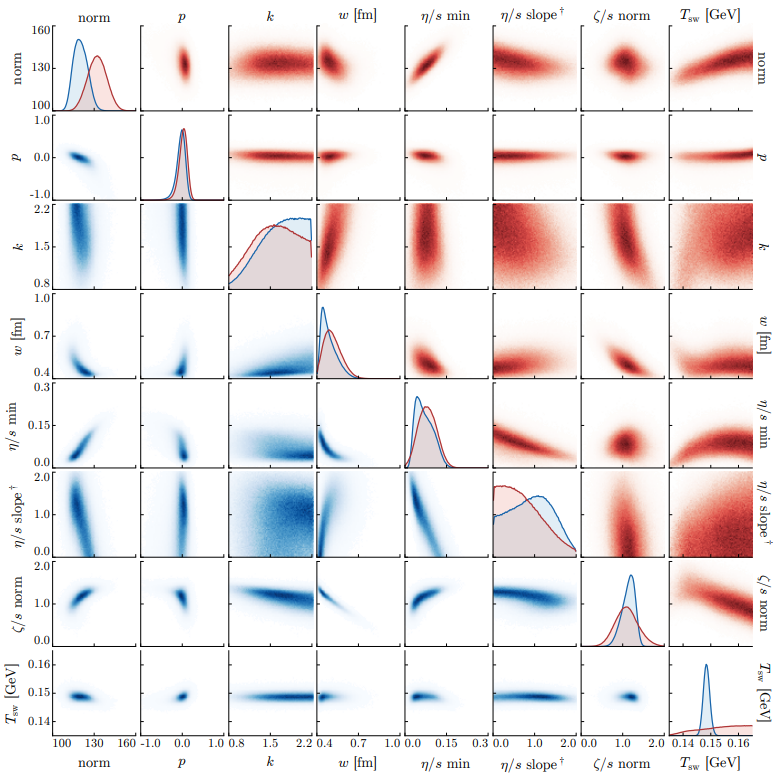}
	\caption{Posterior distributions for the model parameters. In blue, results obtained by calibrating to identified particles data. In red, results obtained by calibrating the model to charged particle yields. In the diagonal, the marginalized posterior for each parameter is shown, and correlations between parameters are shown in the off-diagonal. From \cite{BernhardThesis}.}
\end{figure}

In the numerical setup used for this work (which was presented in Chapter 3), the simulation has a total of 13 free parameters. These are:
\begin{itemize}
    \item The entropy deposition parameter ($p$) in Equation (3.7).
    \item Fluctuation parameter ($k$) in Equation (3.5).
    \item Gaussian nucleon-width ($w$) in Equation (3.3).
    \item Minimum distance between nucleons ($d_{\text{min}}$).
    \item Normalization constant of the entropy density profile, which converts reduced thickness into deposited entropy in Equation (3.6). This normalization factor depends on the energy of the beam.
    \item Pre-equilibrium time. In this work, we use a free-streaming scenario.
    \item Shear viscosity to entropy ratio minimum value $(\eta/s)_{\text{min}}$, slope parameter $(\eta/s)_{\text{slope}}$ and curvature parameter $(\eta/s)_{\text{crv}}$ in Equation (3.24).
    \item Bulk viscosity to entropy ratio maximum value $(\zeta/s)_{\text{max}}$, width $(\zeta/s)_{\text{width}}$ and location $T_{0}$ in Equation (3.25).
    \item Switching temperature between hydrodynamics and the hadronic phase ($T_{\text{switch}}$).
\end{itemize}
Since the main goal of this work was to study the importance of the nucleon-width, we used a set of parameters taken from the Bayesian analysis of the DUKE group in \cite{DUKE2} (see Section 4.3), and changed only the Gaussian nucleon width. 

\section{The nucleon-width parameter}

Event-by-event fluctuations on the spatial distribution of nucleons inside the nuclei in heavy-ion collisions have a large impact on observables \cite{PHOBOS:2006dbo}. Monte Carlo Glauber based approaches incorporate this idea by assembling the nuclei on an event-by-event basis using some nuclear distribution. In the process of assembling two nuclei in order to generate an initial condition of a collision, it is inevitable to somehow make some specification about the \textit{size} of their constituent nucleons. 

The exact form of the transverse distribution of charge inside a nucleon is not well known, and a Gaussian \textit{ansatz} has been successfully implemented as an effective description of the transverse profile of a nucleon moments before the collision: 
\begin{equation}
T_{\text{nucleon}}(x,y) = \frac{1}{2 \pi w^{2}}\exp\left(-\frac{x^{2}+y^{2}}{2 w^{2}}\right)
\end{equation}
The Gaussian shape seems to capture the general idea that the charges of the nucleon (both electric and color charge) are not uniformly distributed: they are more concentrated near its center, although the exact shape of the distribution is not known. 

Since the beginning of the 2000's, hydrodynamic simulations of heavy-ion collisions have used the Gaussian \textit{ansatz} to model the nucleons in the initial condition, with a width of approximately $w$ = 0.4 fm. With this Gaussian width, one can estimate the corresponding ``nucleon radius'' as the root mean square (RMS) transverse radius:
\begin{displaymath}
\sqrt{\langle r^{2} \rangle} = w\sqrt{2} \approx 0.56 \ \text{fm},
\end{displaymath}
which falls in the interval between the proton color radius and its electric charge radius (See Section 4.1). Consistency between the nucleon-width parameter $w$ and experimental estimates of the nucleon size is certainly desirable. At this point, hybrid simulations as sophisticated as the one presented in Chapter 3 did not exist, and the Bayesian formalism discussed in Section 4.2 had never been applied to constrain any of the available models parameters. 

In 2016, the first Bayesian analysis was performed by the Duke group \cite{DUKE1}, using T$_{\text{R}}$ENTo as the initial condition generator, coupled to a hydrodynamic simulation and a hadronic afterburner. The model was compared with Pb-Pb data at $\sqrt{s_{\text{NN}}}$ = 2.76 TeV, and found an optimal value for the nucleon-width parameter of $w$ = 0.43 fm for an analysis using only identified particle yields and $w$ = 0.49 fm when considering only charged particle yields. This results was in agreement with the latest experimental estimates of the proton size. Moreover, hydrodynamic simulations using IP-Glasma initial conditions with a nucleon-width of 0.4 fm had a great success in quantitatively describing experimental data \cite{Schenke:2020mbo}. 

In 2019, the Duke group performed a second global analysis \cite{DUKE2}, with considerable differences from the first one. This time, T$_{\text{R}}$ENTo was used to initialize the \textit{energy} density profile, and a free-streaming pre-equilibrium dynamics was introduced between the initial condition and hydrodynamics. There were new parametrizations for the temperature dependence of specific shear and bulk viscosity (Equations (3.8) and (3.9)), and the model was compared with Pb-Pb data at both $\sqrt{s_{\text{NN}}}$ = 2.76 TeV and 5.02 TeV. This second Bayesian analysis by the Duke group returned a MAP value of 0.956 fm for the nucleon-width parameter, which came as a striking surprise. The corresponding RMS transverse radius is approximately $\sqrt{\langle r^{2} \rangle}$ = 1.352 fm, which exceeds in over 60 \% the present CODATA value for the proton charge radius \cite{CODATA}.
\begin{figure}[h]
	\centering
	\begin{subfigure}{.4\textwidth}
	\centering
	\includegraphics[width=6cm]{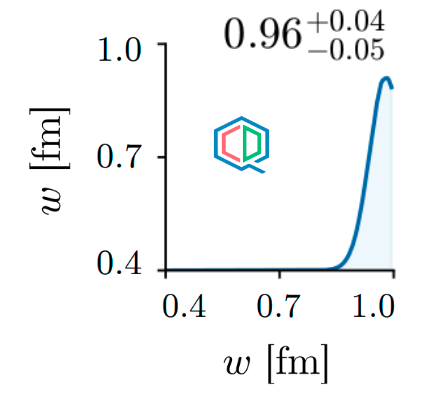}
	\end{subfigure}
	\begin{subfigure}{.4\textwidth}
	\centering
	\includegraphics[width=7cm]{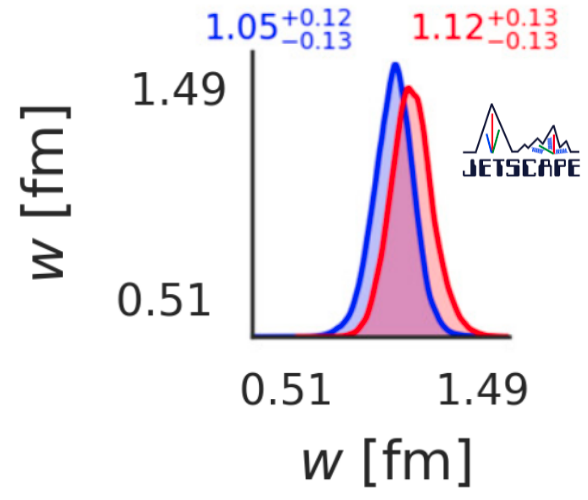}
	\end{subfigure}
	\caption{Marginalized posterior distributions for the nucleon-width parameter from the 2019 DUKE analysis \cite{BernhardThesis, DUKE2} (left) and the JETSCAPE analysis \cite{JETSCAPE:2020mzn} (right). On the right, the result obtained with Grad viscous corrections is shown in blue. In red, the result using Chapman-Enskog corrections.}
\end{figure}
The fact that this result was not expected is visually evident from the marginalized posterior distribution of the nucleon-width parameter (Left side of Figure 4.2) in \cite{BernhardThesis, DUKE2}: the MAP value almost hits the upper bound imposed by the prior (1 fm)!

More recently, an analogous analysis was performed by the JETSCAPE Collaboration in 2020 \cite{JETSCAPE:2020mzn}. The simulation chain used was in general similar to the one employed in the second DUKE analysis. However, different parametrizations were used for the shear and bulk viscosity, and three different out-of-equilibrium corrections to the distribution function in the particlization procedure were considered. The model was simultaneously confronted with Pb-Pb data from the LHC at $\sqrt{s_{\text{NN}}}$ = 2.76 TeV and Au-Au at $\sqrt{s_{\text{NN}}}$ = 200 GeV data from RHIC. Surprisingly, this analysis returned even larger values for the nucleon-width parameter, in the range of 0.9 - 1.2 fm. 

In the latest global analysis by the DUKE group, nucleon sub-structure was enabled in T$_{\text{R}}$ENTo, and the model was compared to data from Pb-Pb and p-Pb collisions at $\sqrt{s_{\text{NN}}}$ = 5.02 TeV. The analysis returned an optimal value of $w$ = 0.88 fm for a nucleon with 6 constituents of width $w_{c}$ = 0.43 fm each. Moreover, other similar analyses seem to always favour such ``diffus'' nucleons, with $w \approx$ 0.8 fm \cite{Nijs:2020ors,Nijs:2020roc, Nijs:2021clz}.

Such findings seem to demand a better understanding of the results of recent Bayesian analyses and of the role played by the nucleon-width parameter inside the simulation as a whole. In this work, we aim not to constrain the value of the nucleon-width, but instead seek a systematic study of its effects in the initial condition characteristics and in final state observables. By using a hybrid simulation chain calibrated with the parameters from \cite{DUKE2} and changing only the value of the nucleon-width parameter, we perform simulations for three values of $w$. How intensely and in which way $w$ affects different observables? Is it possible to obtain a good description of experimental data with the calibrated chain using values of $w$ consistent with estimates of the proton charge radius? If not, which observables(s) are poorly described in this scenario, and why? What led the Bayesian Analyses to favour such large values of the nucleon-width parameter? These are some of the questions that we seek to answer.

\chapter{Results}
\label{CHAP-Results}
\noindent\rule{\textwidth}{.1pt}\\[1ex]

In this work we use the numerical setup presented in Chapter 3 (equivalent to the one used in \cite{DUKE2}) and perform simulations of Pb-Pb collisions at $\sqrt{s_{\text{NN}}}$ = 2.76 TeV using the best-fit (MAP) values obtained in the same reference, with exception of the nucleon-width parameter. For the nucleon-width parameter, we consider three values: 
\begin{itemize}
\item $w$ = 0.5 fm - The ``small nucleon''. This is closer to the values used in simulations before the ``Bayesian Era'', for which the corresponding RMS is consistent with estimates of the proton electric charge radius and strong force radius. 
\item $w$ = 1.0 fm - The ``medium nucleon''. Very close to the result of the Bayesian Analysis considered in this work \cite{DUKE2}. This should be considered to most ``well calibrated'' simulation.
\item $w$ = 1.5 fm - The ``large nucleon''. An exaggeratedly large value for the nucleon-width parameter.  
\end{itemize}

Explicitly, the set of parameters used in this work, which is the same as \cite{DUKE2} except for the nucleon-width, is:

\begin{table}[h!]
\centering
\begin{tabular}{| c | c |}
\hline
Normalization (TeV)  &  286.23   \\ 
$p$  & 0.007  \\ 
$k$  & 0.918  \\ 
$w$ (fm)  &  0.5, 1.0, 1.5  \\ 
$d_{\text{min}}$ (fm)  &  1.27   \\ 
$\tau_{\text{fs}}$ (fm/c)  & 1.2  \\ 
$(\eta/s)_{\text{min}}$  & 0.081  \\ 
$(\eta/s)_{\text{slope}}$ ($\text{GeV}^{-1}$) & 1.11   \\ 
$(\eta/s)_{\text{crv}}$  & -0.48  \\ 
$(\zeta/s)_{\text{max}}$  &  0.052  \\ 
$(\zeta/s)_{\text{width}}$ (GeV) &  0.022  \\ 
$(\zeta/s) T_{0}$ (MeV) &  183  \\ 
$T_{\text{switch}}$ (MeV)  &  151  \\
\hline
\end{tabular}
\caption{Set of parameters used in this work, obtained from the Bayesian analysis in \cite{DUKE2}.}
\end{table}

In Section 5.1, we characterize the effects of changing $w$ in the initial condition. In this analysis, 1.000  T$_{\text{R}}$ENTo initial conditions were generated for 10 values of the impact parameter, with the exception of the ellipticity fluctuations study, for which $10^{6}$ events were generated in 8 impact parameter intervals (centrality classes). In Section 5.2, we present the results of the full simulation and analyze observables. In this analysis, for each value of the nucleon-width parameter, 1.000 minimum-bias (all values of impact parameter mixed together) events were generated. The centrality selection was made after that, based on the total entropy of the initial conditions.

\section{Characterizing the initial condition}
\label{SEC-ICs}

The main objective of this work is to study the impact of the nucleon-width parameter on final observables. We begin this task first by analyzing how the nucleon size affects the general characteristics of the initial condition generated by T$_{\text{R}}$ENTo. There is a strong relation between the initial condition characteristics and observables, so that many of the effects of the nucleon-width parameter on final state observables can be anticipated (at least in approximation) by analyzing its impact on the initial state.

We begin our investigation first by visualizing the initial condition. Figures 5.1 and 5.2 show examples of transverse entropy density profiles $s(x,y)$ generated by T$_{\text{R}}$ENTo in the 0 - 5 \% and 30 - 40 \% centrality classes, respectively:

\begin{figure}[H]
    \centering
    \begin{subfigure}{.3\textwidth}
        \centering
        \includegraphics[width=5cm]{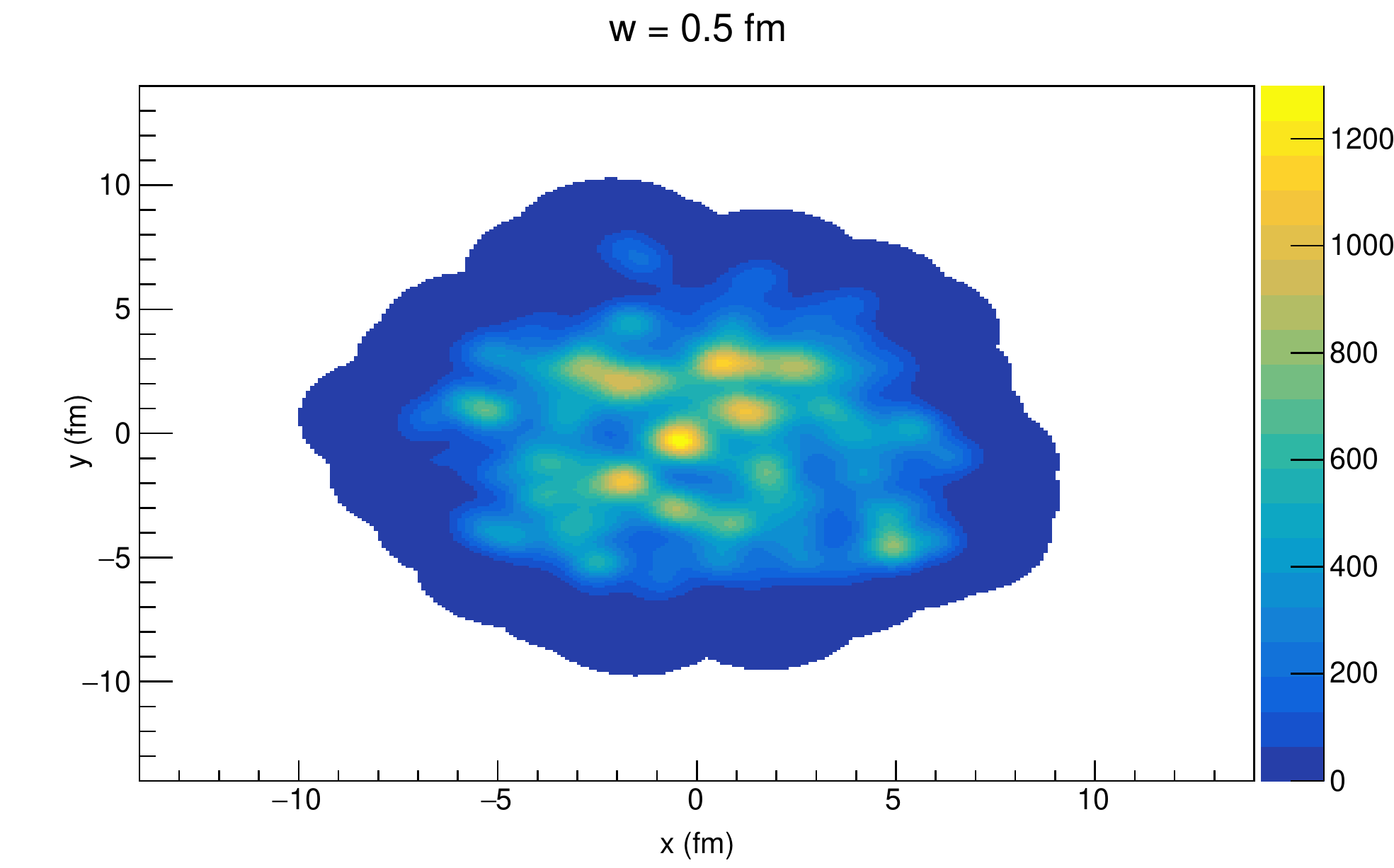}
    \end{subfigure}%
    \begin{subfigure}{.3\textwidth}
        \centering
        \includegraphics[width=5cm]{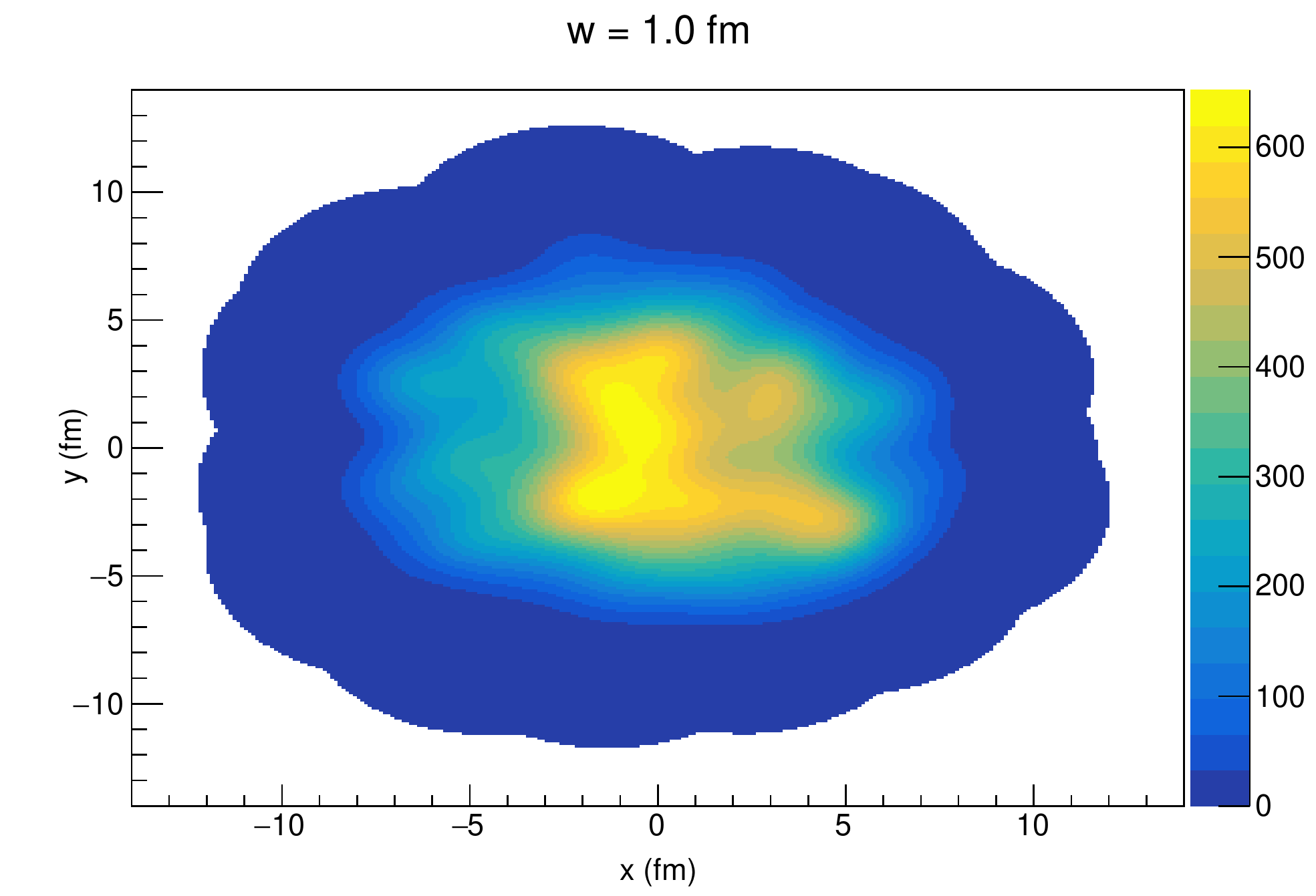}
    \end{subfigure}
        \begin{subfigure}{.3\textwidth}
        \centering
        \includegraphics[width=5cm]{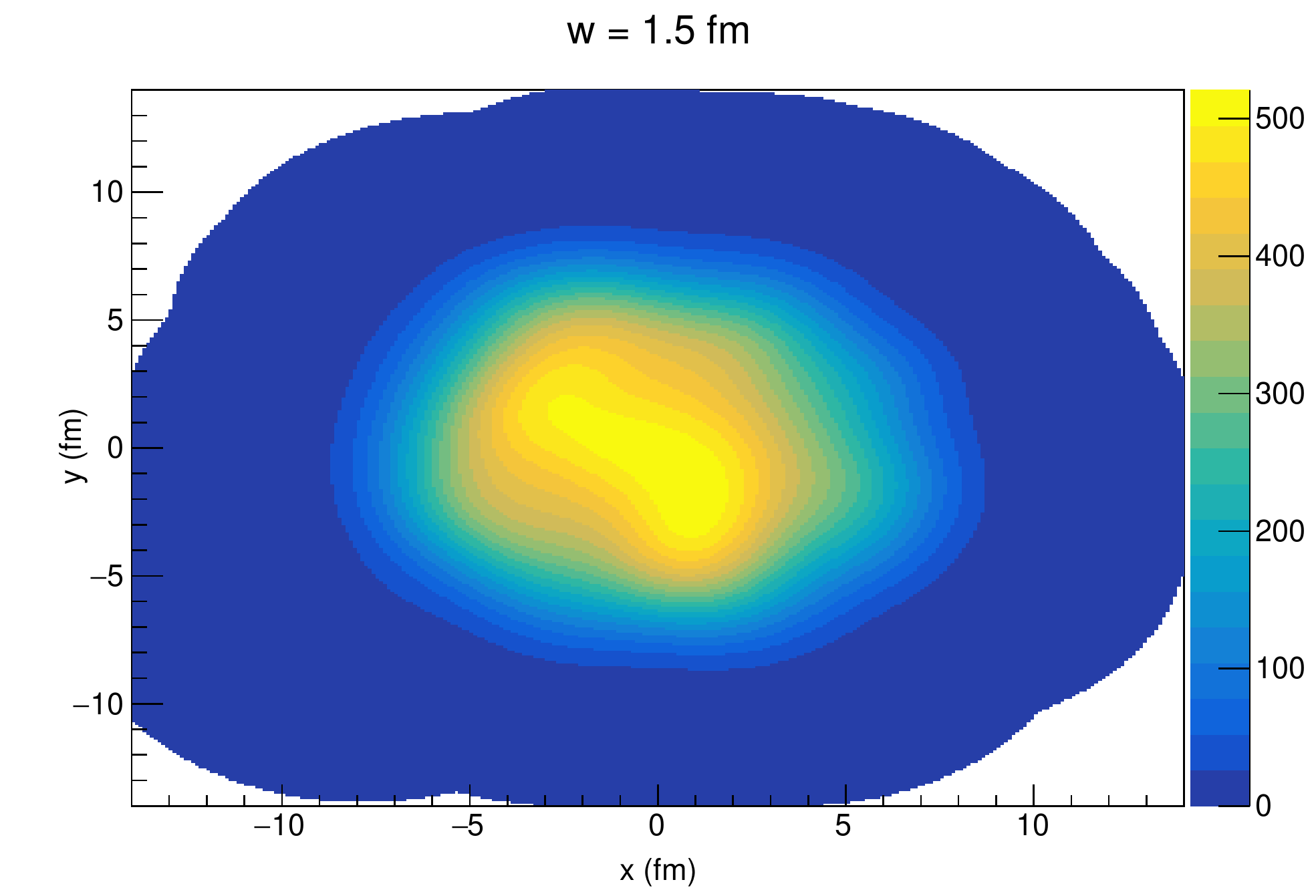}
    \end{subfigure}%
    \caption{Entropy density distribution in the transverse plane of Pb-Pb collisions at $\sqrt{s_{\text{NN}}}$ = 2.76 TeV in the 0 - 5 \% centrality class for: $w = 0.5$ fm (left), $w = 1.0$ fm (center) and $w = 1.5$ fm (right).}
\end{figure}

\begin{figure}[H]
    \centering
    \begin{subfigure}{.3\textwidth}
        \centering
        \includegraphics[width=5cm]{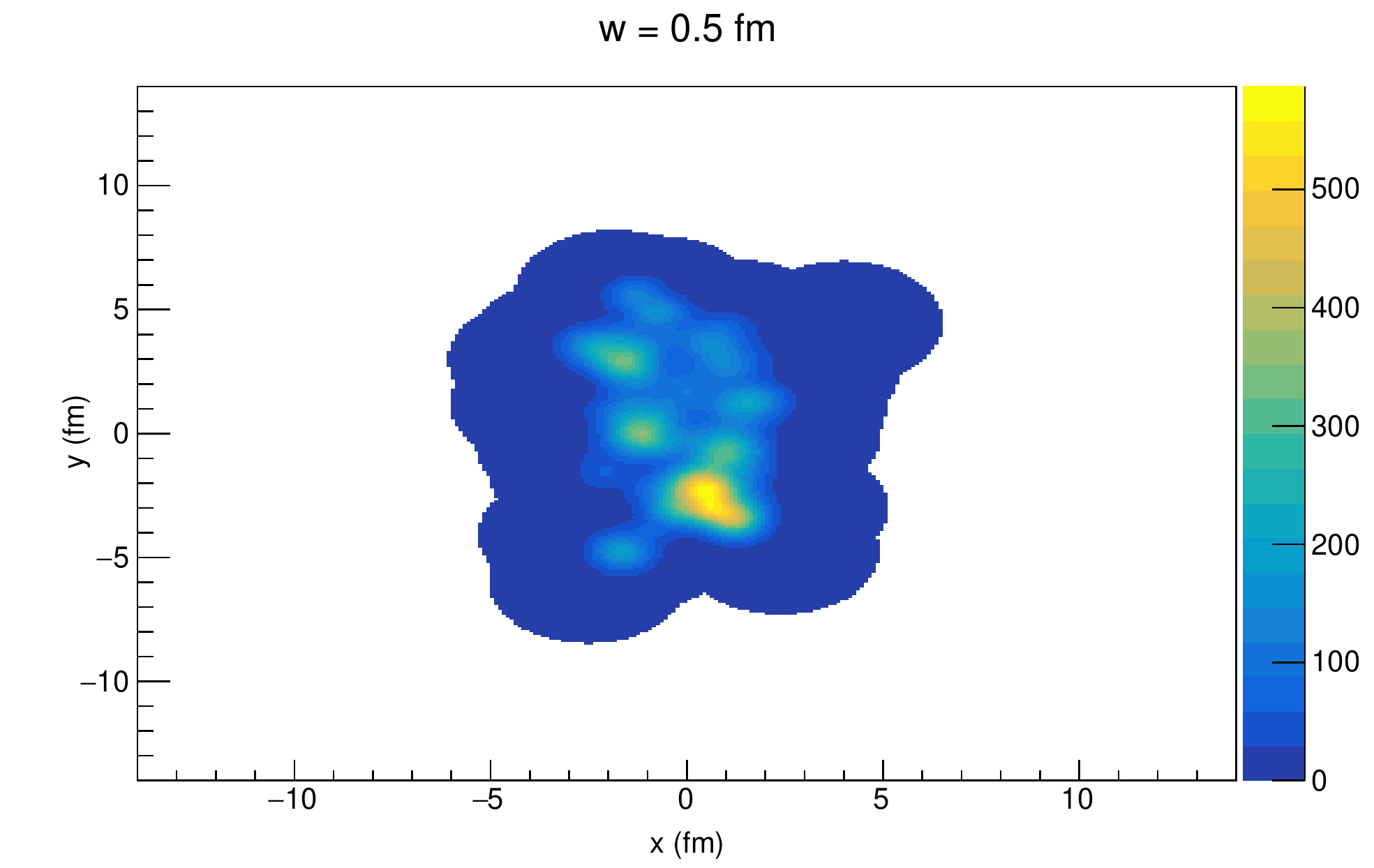}
    \end{subfigure}%
    \begin{subfigure}{.3\textwidth}
        \centering
        \includegraphics[width=5cm]{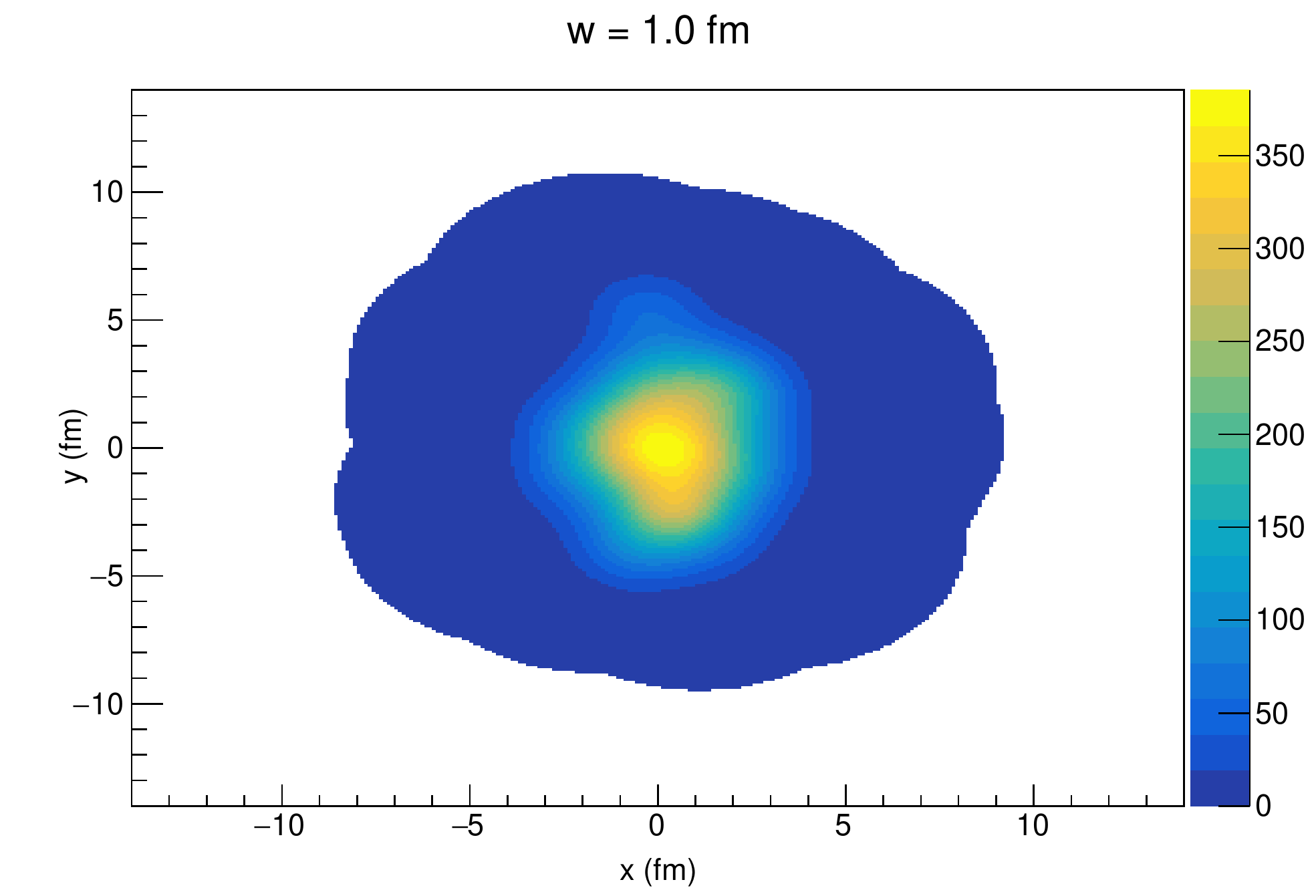}
    \end{subfigure}
        \begin{subfigure}{.3\textwidth}
        \centering
        \includegraphics[width=5cm]{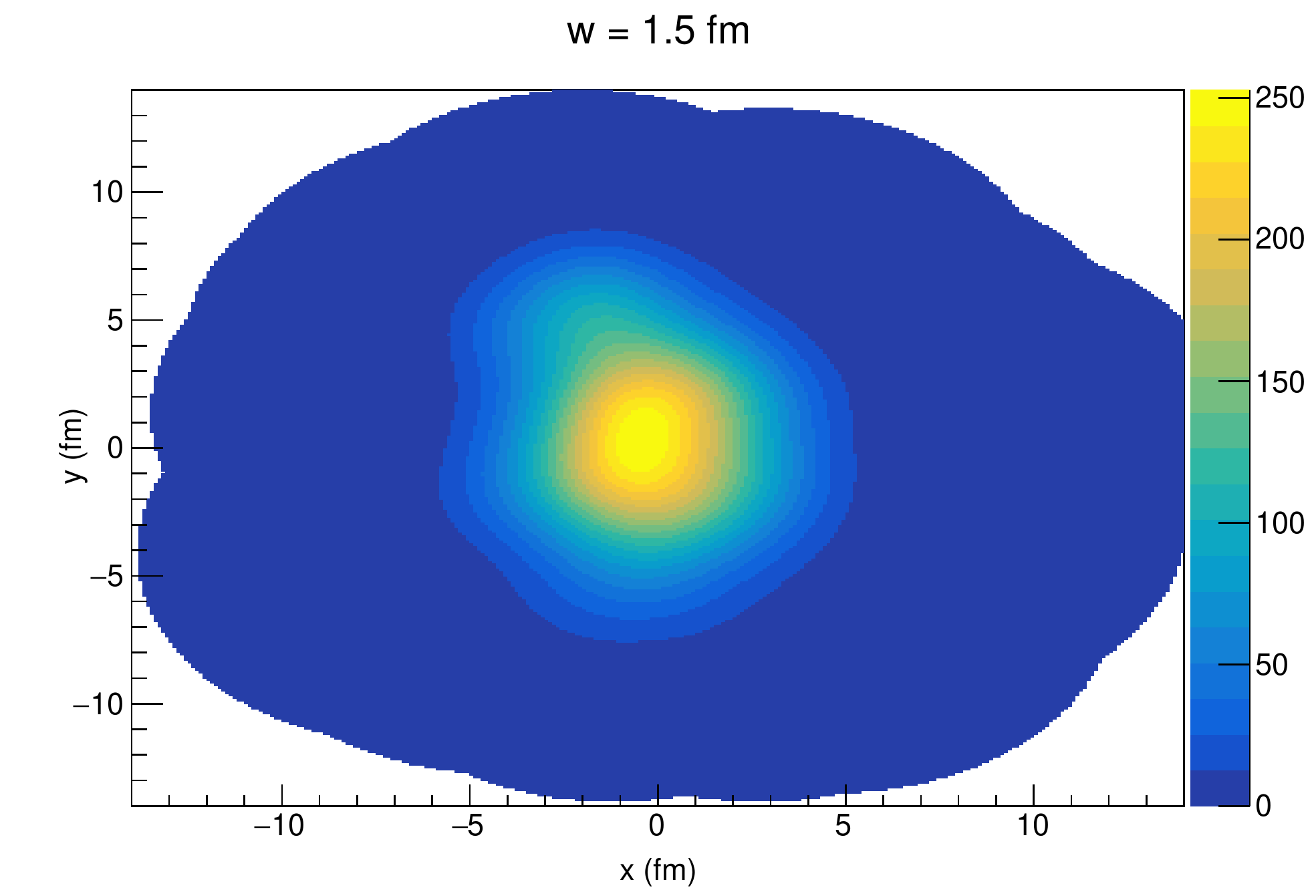}
    \end{subfigure}%
    \caption{Entropy density distribution in the transverse plane of Pb-Pb collisions at $\sqrt{s_{\text{NN}}}$ = 2.76 TeV in the 30 - 40 \% centrality class for: $w = 0.5$ fm (left), $w = 1.0$ fm (center) and $w = 1.5$ fm (right).}
\end{figure}

A straightforward visual analysis of the initial conditions suggests two most prominent effects of changing the nucleon size on the initial state:

\begin{itemize}
    \item The system formed in the collision increases with increasing nucleon size. Furthermore, there is a decrease in system size with increasing centrality (due to the decrease of the overlap area of the two nuclei), which is more pronounced when using smaller nucleons.
    \item The entropy distribution's granularity (``lumpiness'') decreases with increasing nucleon size. In collisions with smaller nucleons, more local high energy density regions (hotspots) can be seen in the initial condition, which gets smoother as the nucleon grows. When using the large nucleon, there is a single peak in the entropy density profile, approximately in the center of the system, which falls smoothly towards the edges of the grid.
\end{itemize}

\begin{figure}[h]
	\centering
	\includegraphics[width=12cm]{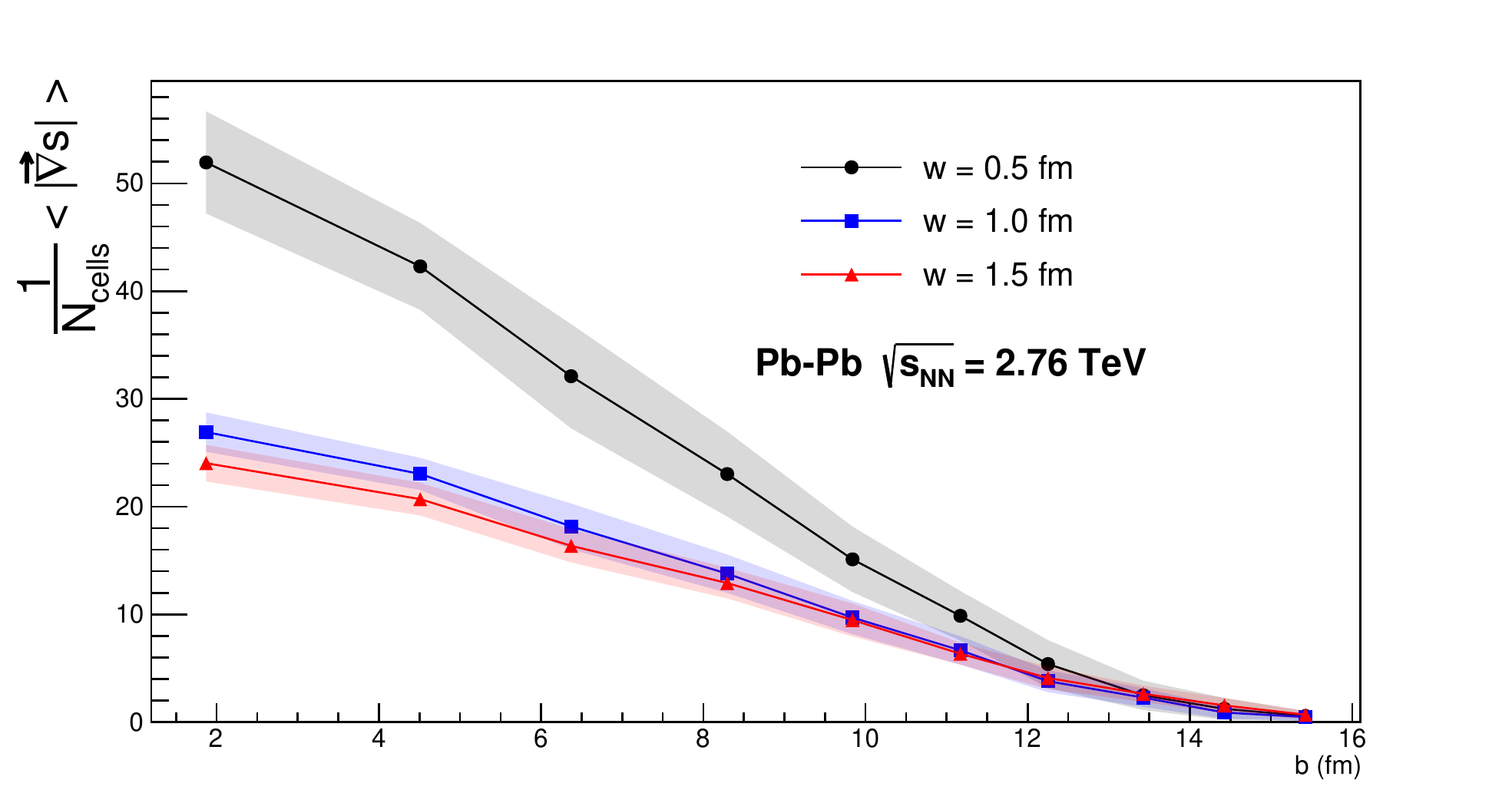}
	\caption{Mean gradient (per cell) of the entropy density profile calculated numerically as a function of impact parameter. Each point is the mean of 1.000 events, and the shaded region represents the event-by-event dispersion around the mean.}
\end{figure}

A visibly higher granularity of the entropy density distribution should mean stronger gradients in the initial condition. This can be verified by sweeping over all cells of the grid an calculating the gradient numerically:
\begin{equation}
|\vec{\nabla} s| = \sqrt{\left (\frac{\partial s}{\partial x} \right)^2 + \left (\frac{\partial s}{\partial y} \right)^2},
\end{equation}
where for each cell $(i,j)$ we have:
\begin{equation}
\frac{\partial s}{\partial x} \approx \frac{s(x_{i+1},y_{j}) - s(x_{i},y_{j})}{x_{i+1} - x_{i}}
\end{equation}
\begin{equation}
\frac{\partial s}{\partial y} \approx \frac{s(x_{i},y_{j+1}) - s(x_{i,}y_{j})}{y_{j+1} - y_{j}}
\end{equation}
Figure 5.3 shows the mean initial condition gradient divided by the number of grid cells as a function of impact parameter, calculated considering 1.000 events generated using T$_{\text{R}}$ENTo for each value of the impact parameter. There is a strong decrease in the gradients from initial conditions generated with $w$ = 0.5 fm to the profiles with $w$ = 1.0 fm. The difference between the medium and the large nucleons, on the other hand, is small. There is a decrease also as a function of impact parameter, simply due to the decrease of the overlap area between the two nuclei, which leads to a decrease of the formed system size. Stronger gradients in the initial condition result in a larger mean of the transverse momentum distribution of the particles detected at the end of the collision \cite{Andrade:2008fa}.

\subsection{Participants and binary collisions}

The number of participant nucleons ($N_{\text{part}}$) and the number of binary collisions ($N_{\text{coll}}$) are the central objects of any Glauber based calculation. How are these affected by changing the nucleon size? Figure 5.4 shows results for $\langle N_{\text{part}} \rangle$ and $\langle N_{\text{coll}} \rangle$ as a function of impact parameter, calculated considering 1.000 events generated using T$_{\text{R}}$ENTo for each value of the impact parameter.
\begin{figure}[H]
    \centering
    \begin{subfigure}{.5\textwidth}
        \centering
        \includegraphics[width=9cm]{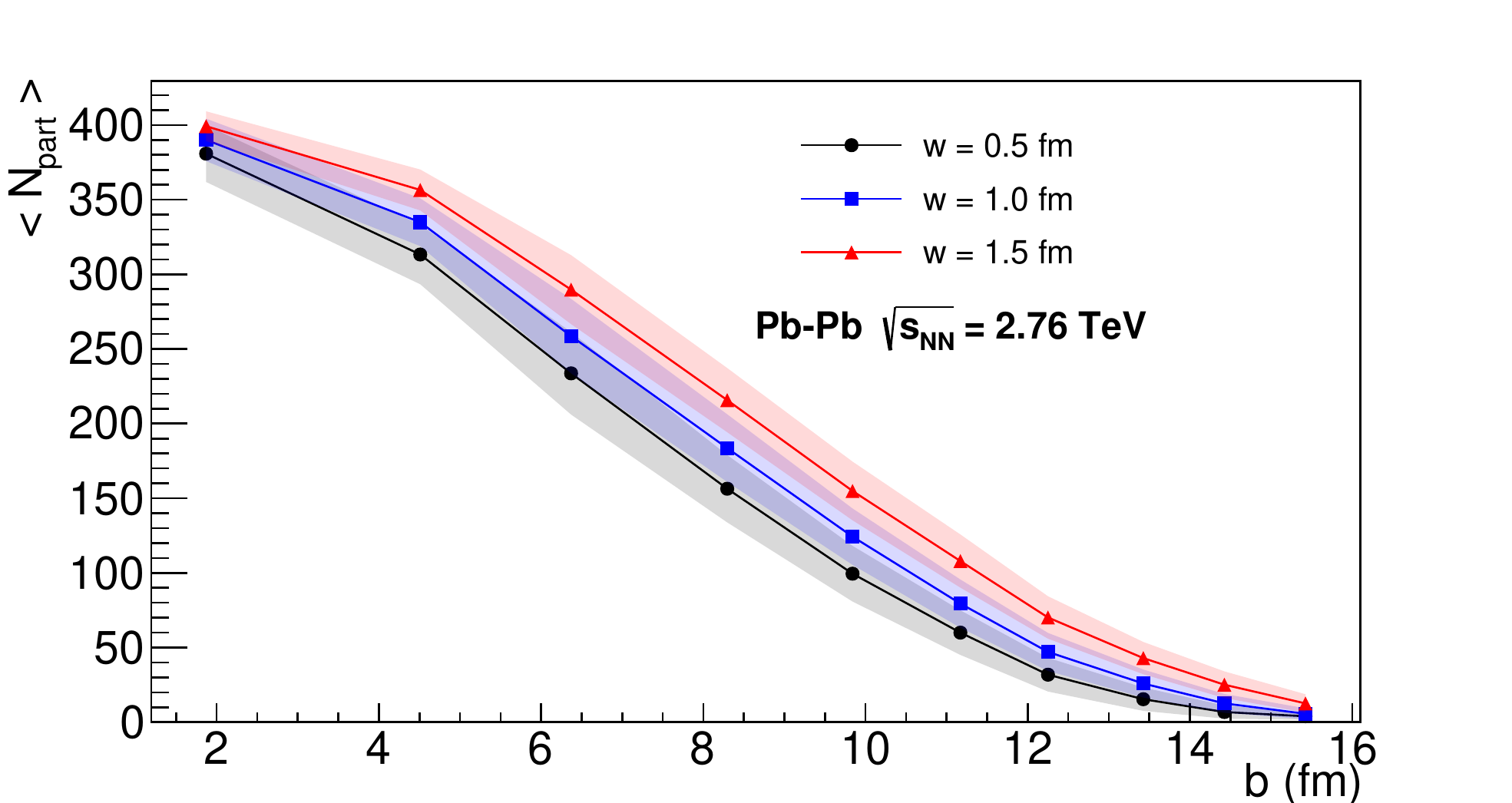}
    \end{subfigure}%
    \begin{subfigure}{.5\textwidth}
        \centering
        \includegraphics[width=9cm]{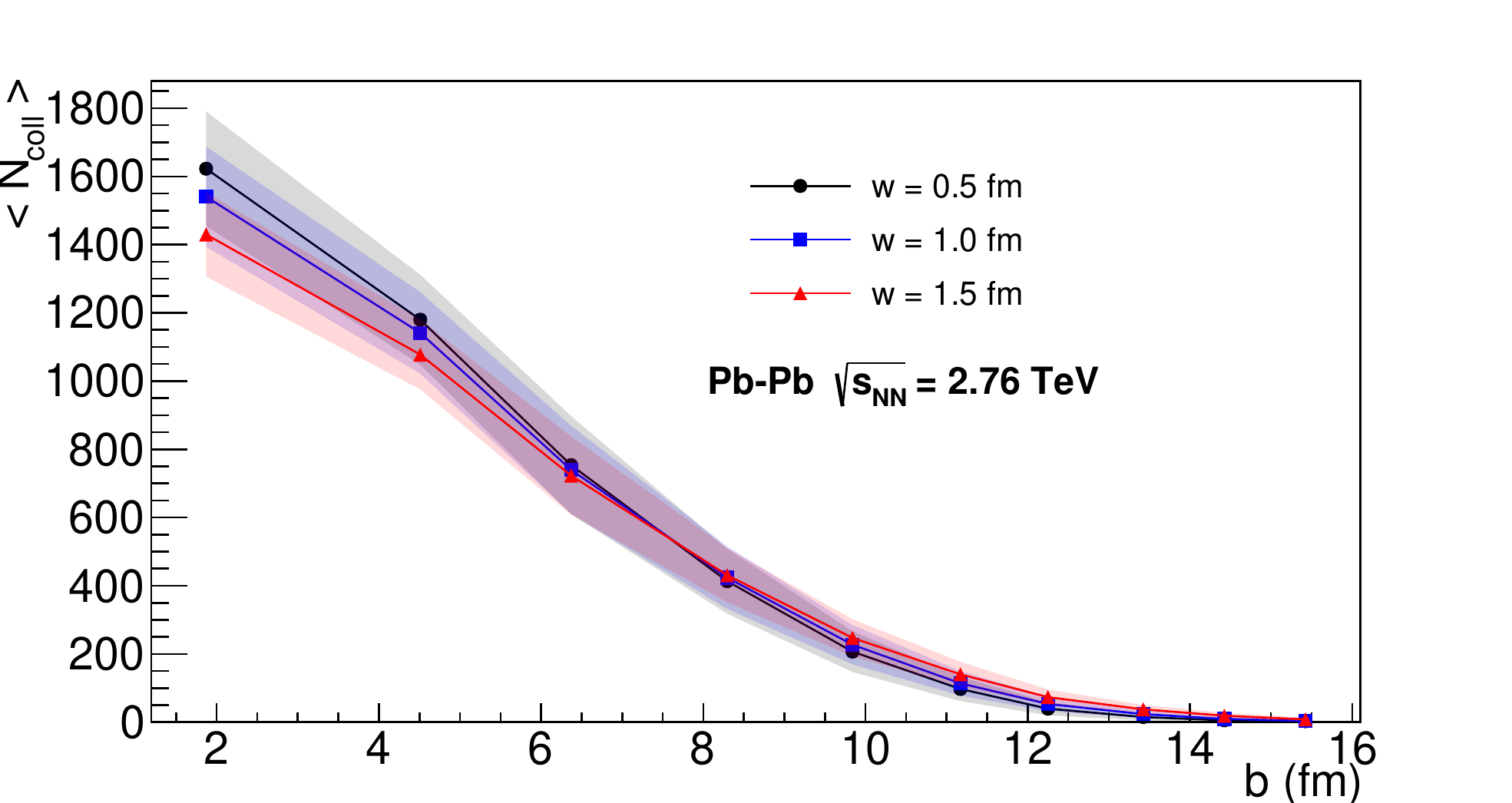}
    \end{subfigure}
    \caption{Mean number of participant nucleons (left) and binary collisions (right) as a function of impact parameter. Each point is the mean of 1.000 events, and the shaded region represents the event-by-event dispersion around the mean.}
\end{figure}
The left side of Figure 5.4 shows that, at the same impact parameter, there is a slight increase in the number of participants for collisions with more diffuse nucleons. In general, the number of binary collisions is simply proportional to the number of participants. On this particular case, however, this is not always true: in central events, the small nucleons interact more often, while in more peripheral events, there are more binary collisions when using the the large nucleons. This is due to the fact that that the collision probability used in T$_{\text{R}}$ENTo:
\begin{equation}
P_{\text{coll}}(b) = 1 - \exp[-\sigma_{gg}T_{AB}(b)]
\end{equation}
depends on the nucleon size. This can be seen by substituting the Gaussian form of the thickness functions in the overlap integral:
\begin{equation}
T_{AB}(b) = \int dx dy \, T_{A}(x-b/2) T_{B}(x+b/2) = \frac{1}{4 \pi w^{2}}\exp\left(- \frac{b^{2}}{4w^{2}}\right).
\end{equation}
Figure 5.5 shows the functional form of Equation (5.4) as a function of the \textit{binary collision} impact parameter, using a constant value for the effective parton-parton cross-section $\sigma_{gg} = 1$.
\begin{figure}[h]
	\centering
	\includegraphics[width=12cm]{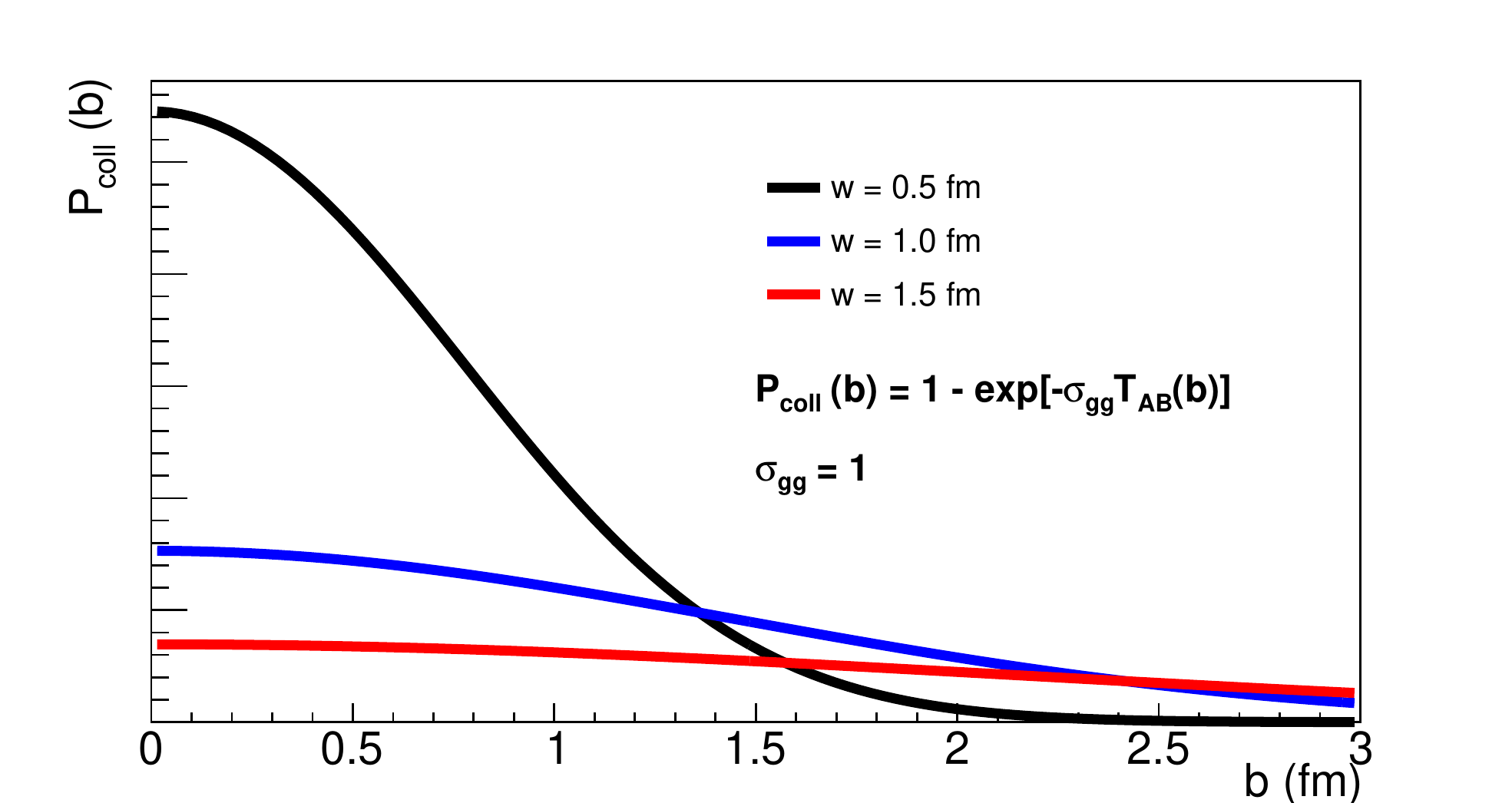}
	\caption{Binary collision probability as a function of impact parameter. The collision probability is more concentrated near $b$ = 0 for the small nucleons. }
\end{figure}
The form of the collision probability reflects the form of the nucleons themselves: when the nucleon-width is small, two nucleons are very likely to interact if they meet at small impact parameter, an this probability falls quickly as the impact parameter grows. As the nucleons get larger, there is a weaker dependence on the impact parameter. This explains why at small impact parameter the small nucleons collide more often, while the opposite happens in more peripheral events: the collision probability is more concentrated near $b$ = 0 for the smaller nucleons. This also explains the slight increase in the number of participants as the nucleon-width increases: for larger nucleons, the collision probability is more spread, so that they are more likely to interact (at least once) when they meet at relatively large distances ($b$ $>$ 2 fm) from each other.

\subsection{Eccentricity harmonics}

A simple visual analysis of Figures 5.1 and 5.2 suggests that the geometry of the system is sensitive to the nucleon size. More quantitatively, this should manifest as a sensitivity of the eccentricity harmonics, as they are calculated using precisely the transverse entropy density distribution of the system as a weight function: 
\begin{equation}
\varepsilon_{n} = \frac{\int r^{n} e^{in\varphi} s(r, \varphi) r dr d\varphi}{\int r^{n} s(r, \varphi) r dr d\varphi}.
\end{equation}
Figure 5.6 shows mean values of ellipticity $\varepsilon_{2}$ (left) and triangularity $\varepsilon_{3}$ (right) of the initials conditions calculated from Equation (5.6) as a function of impact parameter, considering 1.000 events generated using T$_{\text{R}}$ENTo for each value of the impact parameter. 
\begin{figure}[H]
    \centering
    \begin{subfigure}{.5\textwidth}
        \centering
        \includegraphics[width=9cm]{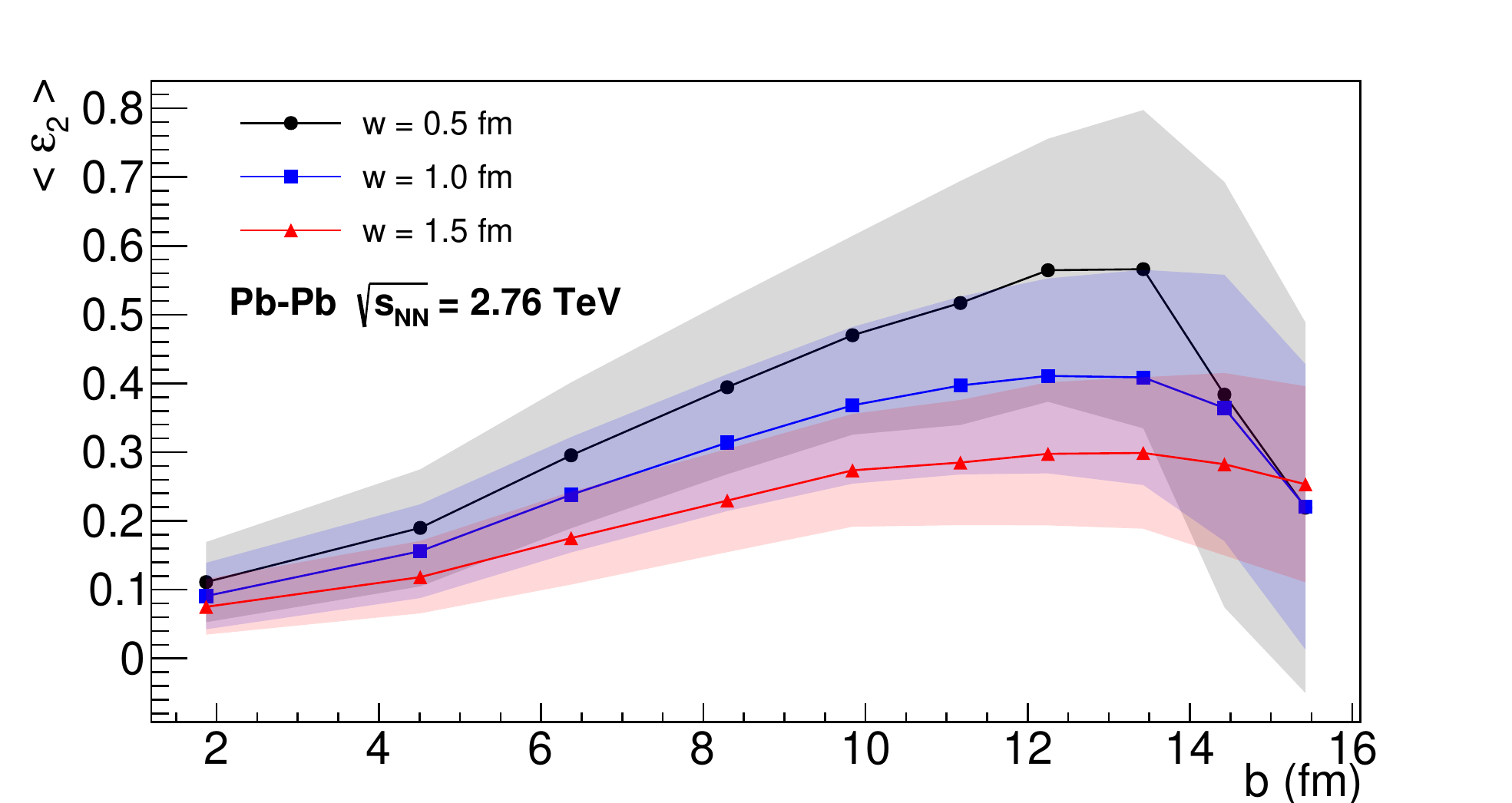}
    \end{subfigure}%
    \begin{subfigure}{.5\textwidth}
        \centering
        \includegraphics[width=9cm]{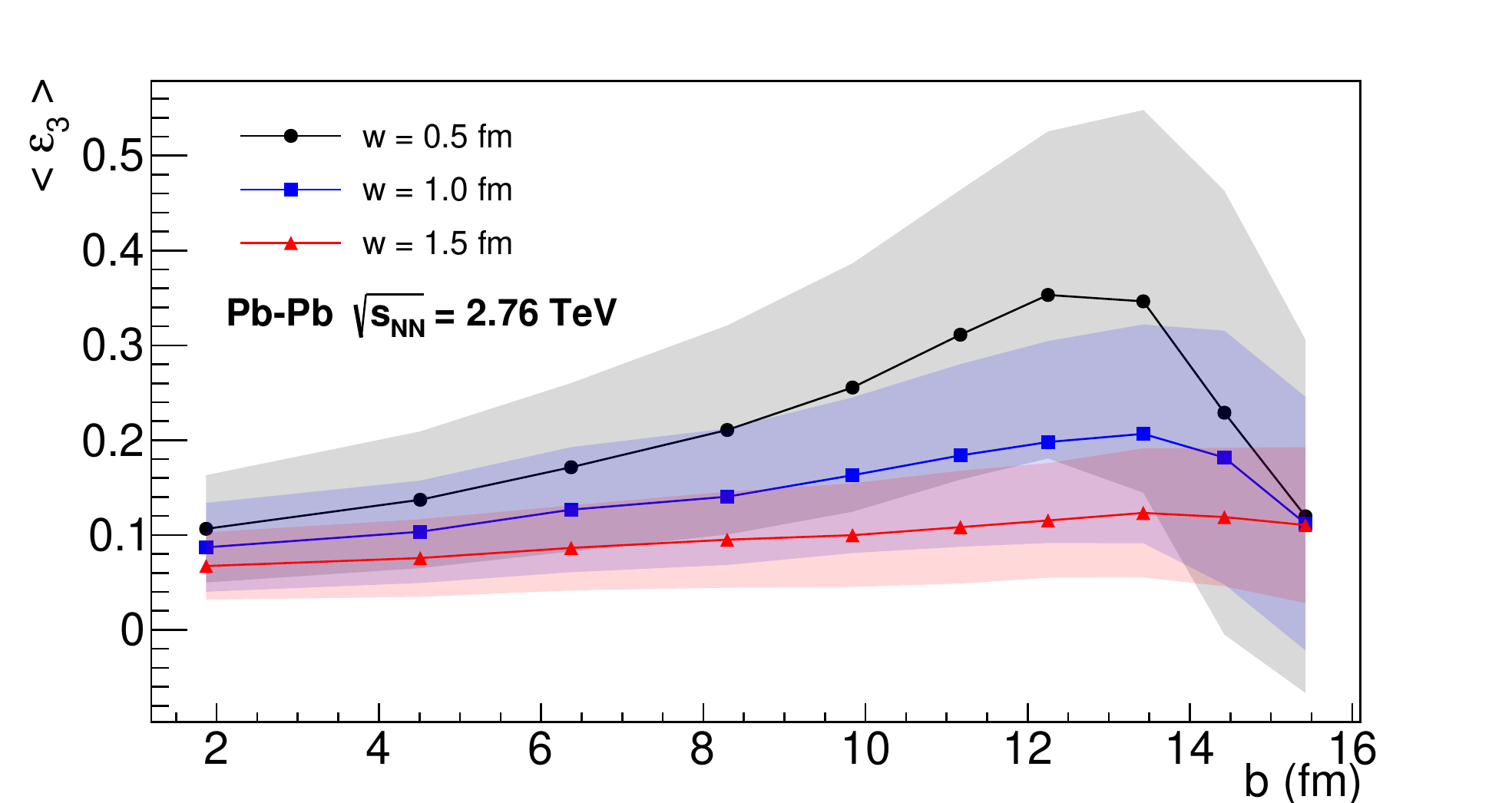}
    \end{subfigure}
    \caption{Mean value of the eccentricity harmonics for $n=2$ (left) and $n=3$ (right) as a function of impact parameter. Each point is the average of 1.000 events, and the shaded region represents the event-by-event dispersion around the mean.}
\end{figure}

In general, $\varepsilon_{2}$ is small in central collisions, where the overlap region between the two colliding nuclei is approximately circular. As the impact parameter grows, the overlap area acquires a more elliptical shape, and the value of $\varepsilon_{2}$ also grows. At some point, for very large values of $b$, the overlap area between the nuclei becomes very small (essentially created by the collision between two nucleons), and ellipticity decreases once again. This behavior as a function of impact parameter is, in general, true for higher order harmonics as well, although there is a weaker dependence with the impact parameter. 

Ellipticity and triangularity are strongly affected by the nucleon size: the mean value of both eccentricity harmonics decreases as the nucleons grow larger. The smoother entropy distributions generated when using larger nucleons are more spatially isotropic. In particular, the triangular pattern is almost not present in the initial conditions generated with $w$ = 1.5 fm, and there is almost no dependence on the impact parameter. 

Borrowing the concept of cumulants from the flow analysis (see Section 5.2.3), we can calculate the analogous estimates for the eccentricity harmonics from cumulants:
\begin{equation}
\varepsilon_{n}\{2\} = \sqrt{\langle |\varepsilon_{n}|^{2} \rangle}
\end{equation}
\begin{equation}
\varepsilon_{n}\{4\} = \sqrt[\leftroot{2}\uproot{2}4]{\langle |\varepsilon_{n}|^{4} \rangle - 2 \langle |\varepsilon_{n}|^{2} \rangle^{2}}
\end{equation}
The ratio $\varepsilon_{n}\{4\}/\varepsilon_{n}\{2\}$ is a standard measure of event-by-event eccentricity fluctuations. Such calculations demand great statistics: in Figure 5.7, which shows ellipticity fluctuations as a function of centrality, each point was calculated using $10^{6}$ events. 
\begin{figure}[h]
	\centering
	\includegraphics[width=12cm]{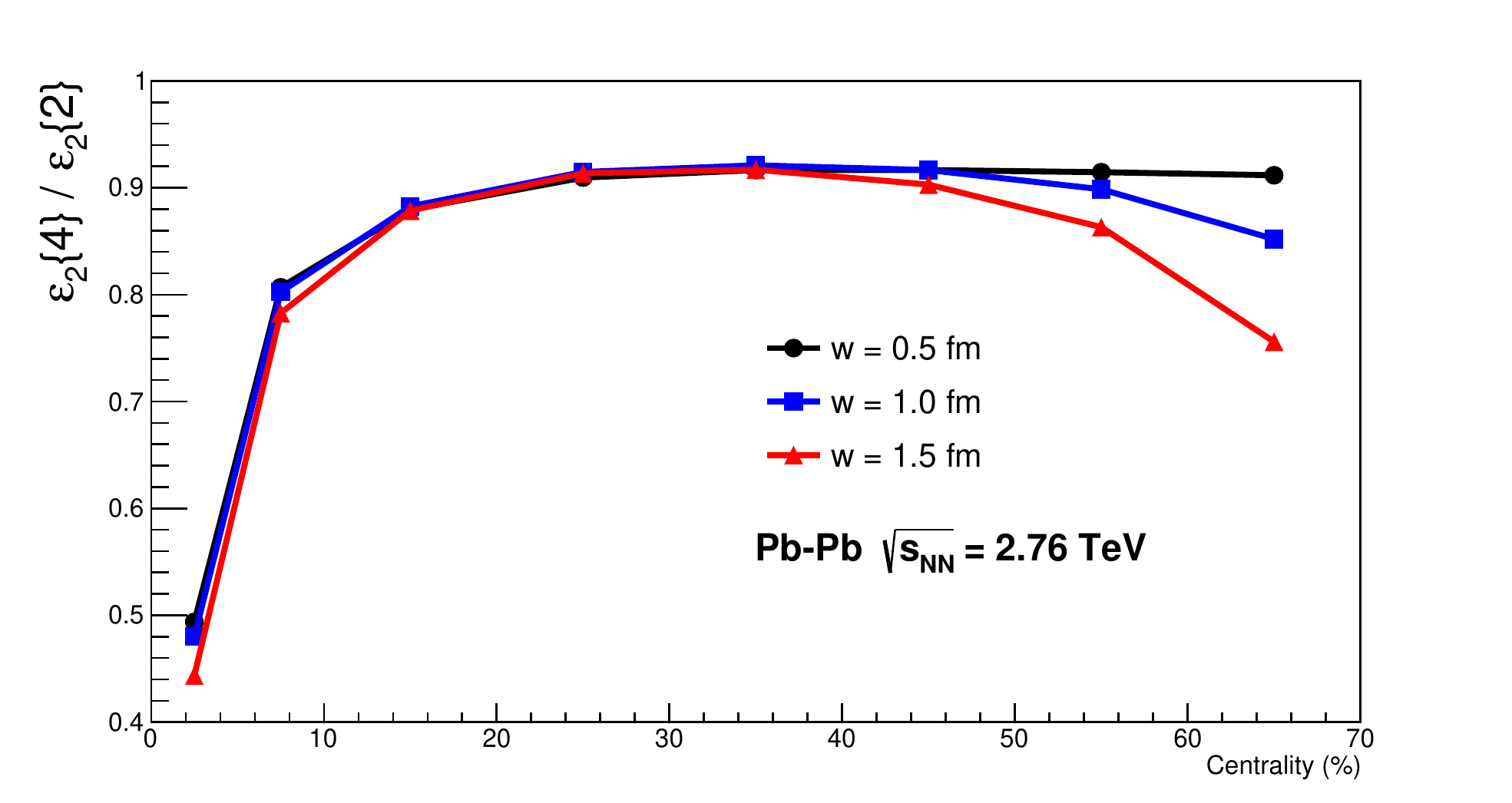}
	\caption{Event-by-event ellipticity fluctuations as a function of centrality. Each dot was calculated using $10^{6}$ T$_{\text{R}}$ENTo initial conditions.}
\end{figure}
Figure 5.7 shows that although the nucleon-width strongly affects the average value of ellipticity $\langle \varepsilon_{2} \rangle$, there is not a strong impact on event-by-event fluctuations around the mean for more central events. For more peripheral events (from 40 - 50 \% on), there is a decrease in fluctuations as the nucleon size increases. It seems intuitive that the degree of fluctuation decreases for larger values of $w$ (as the initial conditions are smoother), and that this effect is more pronounced in peripheral events. In central events, where there are lots of binary collisions, initial conditions are more likely to ``look alike'' regardless of the nucleon size, while in peripheral collisions, where the system size is smaller, the effect of the nucleon size is visible.

Figure 5.8 shows the results for higher order harmonics ($n=4$ and $n=5$), calculated considering 1.000 events generated using T$_{\text{R}}$ENTo for each value of impact parameter. It is possible to see that the effects of changing the nucleon-width are even more pronounced for these higher order harmonics, for which the corresponding geometric patterns are ``sharper'' when compared to ellipticity and triangularity.
\begin{figure}[H]
    \centering
    \begin{subfigure}{.5\textwidth}
        \centering
        \includegraphics[width=9cm]{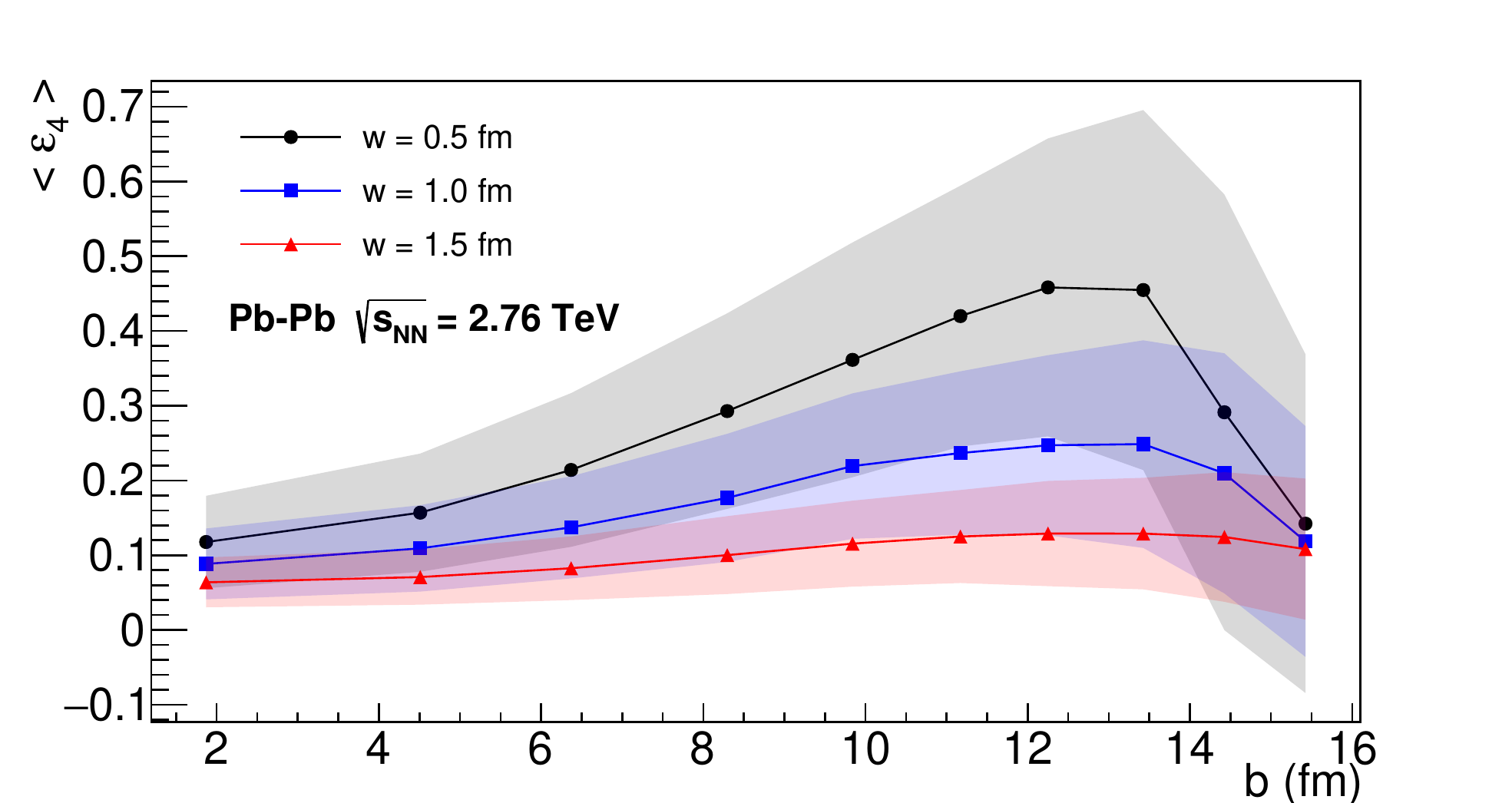}
    \end{subfigure}%
    \begin{subfigure}{.5\textwidth}
        \centering
        \includegraphics[width=9cm]{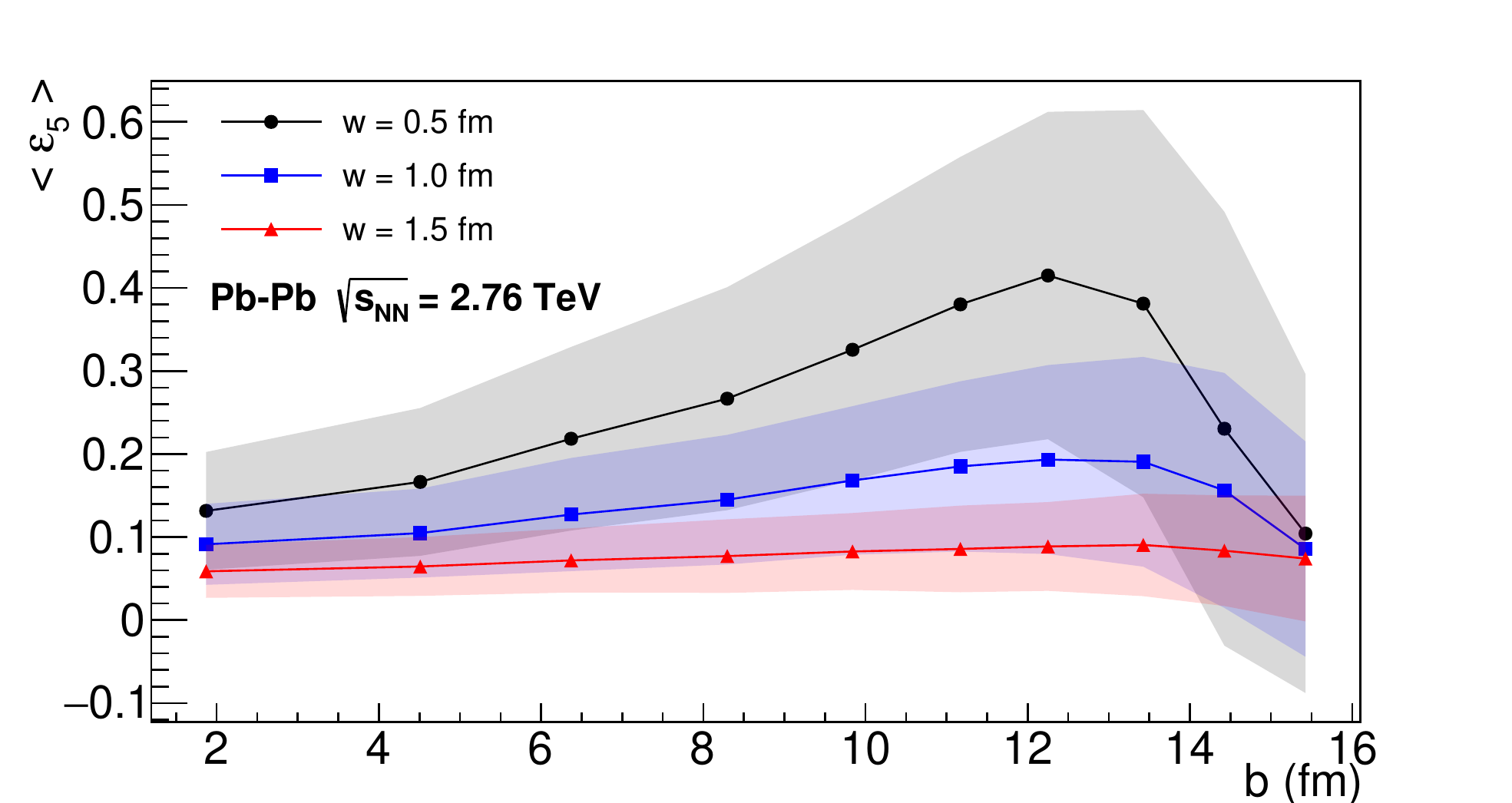}
    \end{subfigure}
    \caption{Mean value of the eccentricity harmonics for $n=4$ (left) and $n=5$ (right) as a function of impact parameter. Each point is the mean of 1.000 events, and the shaded region represents the event-by-event dispersion around the mean.}
\end{figure}
The results for these higher order harmonics show that indeed initial conditions generated using smaller nucleons have a more detailed and complex geometric structure, as Figures 2.1 and 2.2 suggest.

\section{Final state observables}
\label{SEC-Observables}

While Section 5.1 presented results obtained by analyzing T$_{\text{R}}$ENTo initial conditions, this section presents results obtained after the complete simulation, as discussed in Chapter 3. For each value of the nucleon-width parameter, 1.000 minimum-bias (all values of impact parameter mixed together) events were generated. The centrality selection was made after that, based on the total entropy of the initial conditions (in the usual way, by ordering them from the lowest to the highest entropy values, and separating them in percentiles), which is a good predictor for the final multiplicity. 

\subsection{Charged particle multiplicity density at mid-rapidity}

We begin the analysis of final state observables by calculating the charged particle multiplicity density in the mid pseudo-rapidity region ($|\eta|<0.5$). Figure 5.9 shows charged particle multiplicity density in the region $|\eta|<0.5$ as a function of centrality.

\begin{figure}[h]
	\centering
	\includegraphics[width=14cm]{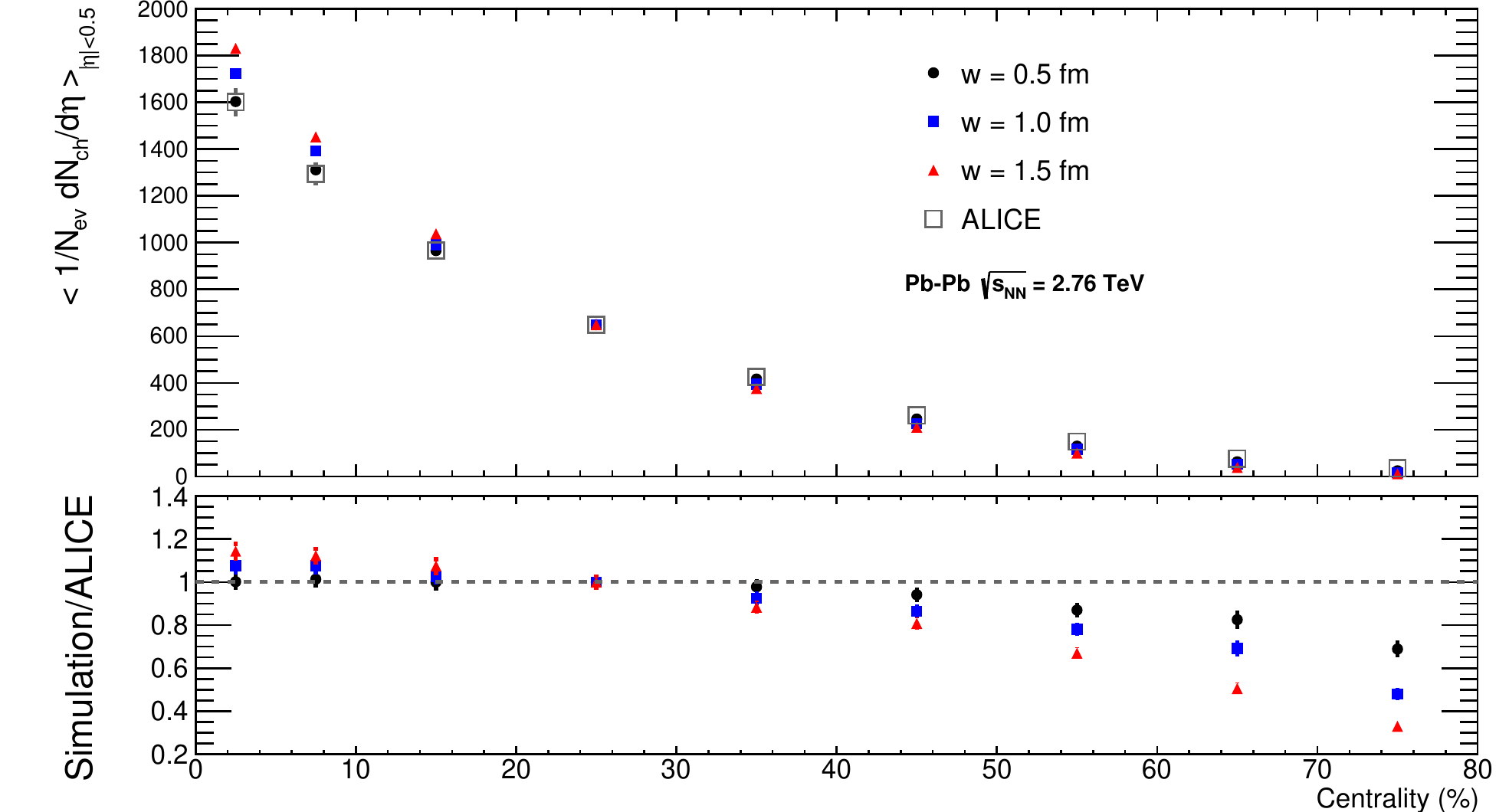}
	\caption{Charged particle multiplicity density in the central pseudo-rapidity region as a function of centrality. Experimental data from the ALICE Collaboration\cite{ALICE_Mult}. The error bars represent the event-by-event dispersion around the mean.}
\end{figure}

As seen in Figure 5.9, our results show that, in more central events, collisions with larger nucleons produce more particles, while the opposite happens for the more peripheral classes. Although $w=1.0$ fm is essentially the MAP value of the nucleon-width in \cite{DUKE2}, a better description of the centrality dependence of charged particle multiplicity is actually obtained using $w=0.5$ fm. In fact, until the 40 - 50 \% class, the simulation using the small nucleon provides a good description (within the 10 \% range) of experimental data. The simulated points were separately normalized by a constant factor so that the simulations agree exactly with experiment (and with each other) in the 20 - 30 \% centrality interval.

\subsection{Mean transverse momentum}

The stronger gradients in the initial energy density profile of the system when using smaller nucleons should have a great effect on raising the mean transverse momentum of particles in the final state of the collision. Figure 5.10 shows the mean transverse momentum of charged pions in the mid-rapidity region for Pb-Pb collisions at $\sqrt{s_{\text{NN}}}$ = 2.76 TeV, compared with data from ALICE. The mean transverse momentum was calculated considering the charged pions spectra from our simulation, in the $|y|<$0.5 rapidity interval, and with 0.1 $<p_{T}<$ 3 GeV/c (to be in agreement with the experimental acceptance).

\begin{figure}[h]
	\centering
	\includegraphics[width=14cm]{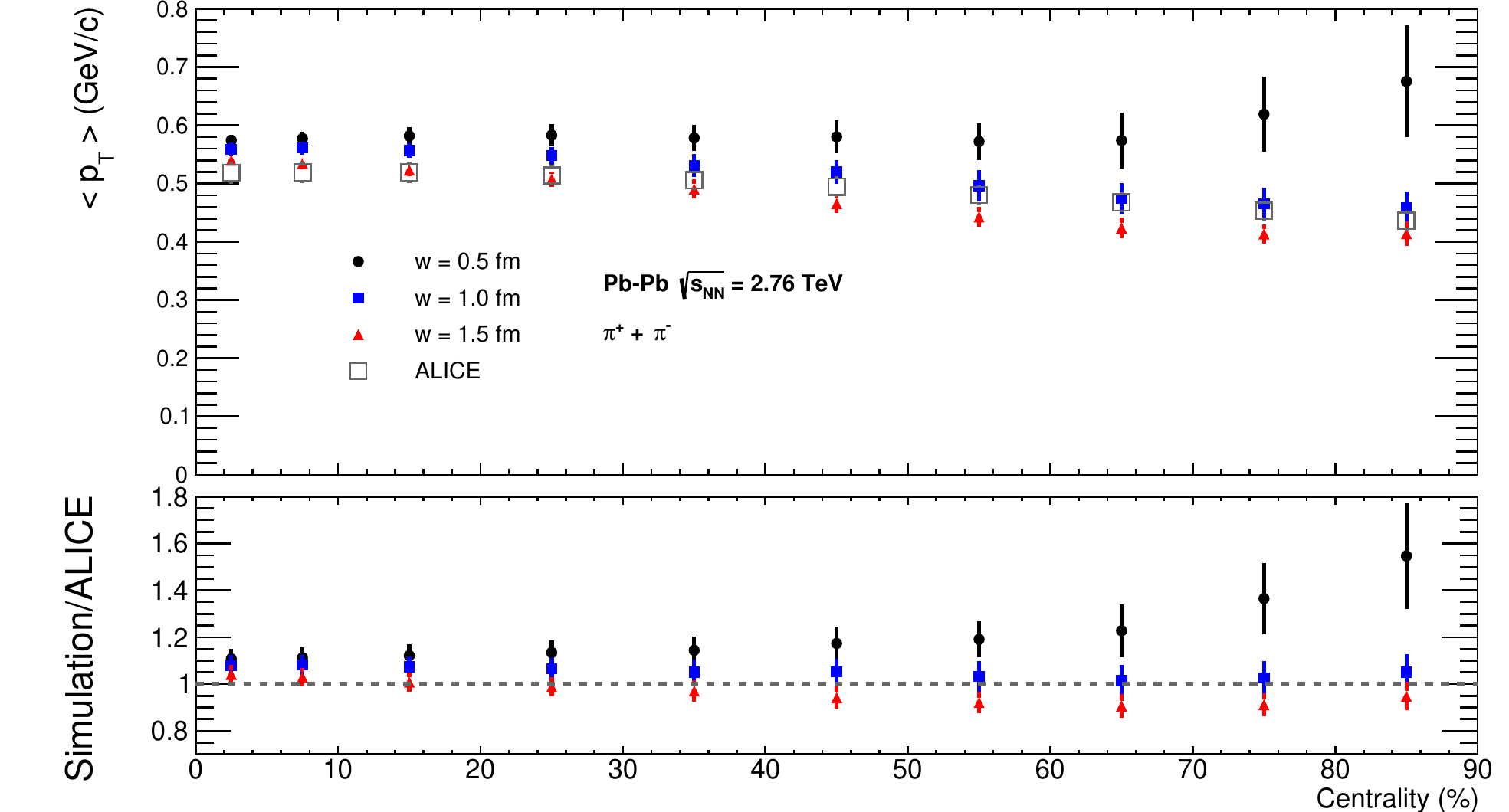}
	\caption{Mean transverse momentum of charged pions as a function of centrality. Data from the ALICE Collaboration \cite{ALICE_Pt}, measured in Pb-Pb collisions at $\sqrt{s_{\text{NN}}}$ = 2.76 TeV. The error bars represent the event-by-event dispersion around the mean.}
\end{figure}

For the 0 - 5 \% centrality class, the mean transverse momentum is roughly the same for the three simulations, and slightly above the experimental data (around 10 \%). The simulations using the two larger nucleons provide a good description of data (within the 10 \% range) across all centralities, but as events get more peripheral, the simulation with $w=0.5$ fm generates particles with too large transverse momentum. In fact, there is an increase with centrality, which is known not to be the case.

Raising the $\langle p_{T} \rangle$ as a consequence of increasing the granularity of the initial condition has been well established for almost 15 years \cite{Andrade:2008fa}. Certainly this effect is present when the nucleon size is changed, but there also seems to be more to it. To investigate this, we note there were two main simplifications in the pre-equilibrium phase considered in \cite{DUKE2}:

\begin{itemize}
\item In the pre-equilibrium phase, partons are assumed to be massless. Therefore, free-streaming takes place with $v=c$. This is known to result in an exaggerated large out-of-equilibrium bulk pressure when switching to hydrodynamics, which, on its turn, is responsible for an \textit{artificially} large value of the mean transverse momentum in the final state \cite{NunesdaSilva:2020bfs}. 
\item The pre-equilibrium phase lasts the same time for all centrality classes. As the system size decreases fore more peripheral collisions, it is reasonable to expect that the pre-equilibrium phase should last less when compared to central events (as is the case for the hydrodynamic evolution, for example). The use of a constant pre-equilibrium time results in an artificial increase of the mean transverse momentum with centrality \cite{Gale:2021emg}, as the violent pre-hydrodynamic expansion lasts longer than it should. 
\end{itemize}

The simulation using the small nucleon is the most sensitive to this second simplification in the free-streaming phase, due to the fact that the system formed is considerably smaller when compared to using the two larger nucleon-width values (Figures 5.1 and 5.2). The first simplification, on the other hand, affects all centrality classes. Figure 5.10 shows that, \textit{in this scenario}, where the effects of the stronger gradients due to using smaller nucleons are \textit{combined} with the effects of these two simplifications made in the pre-hydrodyamic stage of the simulation, the simulation with $w$ = 0.5 fm produces wrong results (exceeding up to 60 \% the experimental data). 

\subsection{Anisotropic flow}

Calculating the flow harmonics explicitly as an expectation value like in Equation (2.6) requires event-by-event knowledge of the reaction plane angle $\Psi_{\text{RP}}$, which is not accessible experimentally. As an alternative, flow harmonics are usually calculated using multiparticle azimuthal correlations \cite{Borghini:2000sa, Borghini:2001vi}, sometimes also referred to as the \textit{cumulants method}. The idea behind the method is that, as collectivity is related to correlations between particles in momentum space, it should be possible to extract flow from correlation functions.

Let $\langle k \rangle$ denote the azimuthal correlation function of $k$ particles in a single event. In this notation, the 2-particle and 4-particle correlation functions are given by \cite{Bilandzic:2010jr}:
\begin{equation}
\langle 2 \rangle = e^{in(\phi_{1} - \phi_{2})} = \frac{1}{P_{M,2}}\sum_{i \neq j}^{M}e^{in(\phi_{i} - \phi_{j})}
\end{equation}
\begin{equation}
\langle 4 \rangle = e^{in(\phi_{1} + \phi_{2} - \phi_{3} - \phi_{4})} = \frac{1}{P_{M,4}}\sum_{i \neq j \neq k \neq l}^{M}e^{in(\phi_{i} + \phi_{j} - \phi{k} - \phi_{l})},
\end{equation}
where $M$ is the number of charged particles in the event, $n$ is the order of the harmonic, $\phi$ is the azimuthal angle, and:
\begin{equation}
P_{M,k} = \frac{M!}{(M-k)!}
\end{equation}
is the number of permutations of $k$ particles in an event of multiplicity $M$, so that:
\begin{equation}
P_{M,2} = M(M-1)
\end{equation}
\begin{equation}
P_{M,4} = M(M-1)(M-2)(M-3).
\end{equation}
Inside a given centrality bin, the $k$-particle correlation function can be written as $\langle \langle k \rangle \rangle$, where the outer brackets denote an average over all events inside the centrality class. 

How can the flow coefficients be estimated from these correlation functions? We begin by rewriting $\langle \langle 2 \rangle \rangle$, adding an subtracting $\Psi_{\text{RP}}$ on the exponential:
\begin{equation}
\langle \langle 2 \rangle \rangle = \langle \langle e^{in[(\phi_{1} - \Psi_{\text{RP}}) - (\phi_{2} - \Psi_{\text{RP}})]} \rangle \rangle.
\end{equation}
Now, assuming that the only correlation between $\phi_{1}$ and $\phi_{2}$ is due to collective flow, the inner average can be factorized in a product:
\begin{equation}
\langle \langle 2 \rangle \rangle \approx \langle \langle e^{in(\phi_{1} - \Psi_{\text{RP}})} \rangle \langle e^{-in(\phi_{2} - \Psi_{\text{RP}})} \rangle  \rangle = \langle v_{n}^{2} \rangle, 
\end{equation}
and analogously, $\langle \langle 4 \rangle \rangle \approx \langle v_{n}^{4} \rangle$. This factorization is not exact, so that the equality is only approximate. Denoting by $c_{n}\{k\}$ the $n$th-order cumulant from correlations of $k$ particles:
\begin{equation}
c_{n}\{2\} = \langle \langle 2 \rangle \rangle
\end{equation}
\begin{equation}
c_{n}\{2\} = \langle \langle 4 \rangle \rangle - 2 \langle \langle 2 \rangle \rangle,
\end{equation}
so that the flow coefficients $v_{n}$ estimated from the 2 and 4-particle cumulants are given by:
\begin{equation}
v_{n}\{2\} = \sqrt{c_{n}\{2\}}
\end{equation}
\begin{equation}
v_{n}\{4\} = \sqrt[\leftroot{2}\uproot{2}4]{-c_{n}\{4\}}
\end{equation}

As the calculation of correlation functions is computationally expensive, in practice the flow harmonics are usually extracted using the $Q$-vectors, which are defined as \cite{Bilandzic:2010jr}:
\begin{equation}
Q_{n} = \sum_{i=1}^{M} e^{in\phi_{i}}.
\end{equation}
Using the $Q$-vector, the single-event correlation functions can be expressed analytically. For example:
\begin{equation}
|Q_{n}|^{2} = \sum_{i,j=1}^{M} e^{in(\phi_{i} - \phi_{j})} = M + \sum_{i \neq j}^{M}e^{in(\phi_{i} - \phi_{j})},
\end{equation}
so that:
\begin{equation}
\langle 2 \rangle = \frac{|Q_{n}|^{2} - M}{M(M-1)},
\end{equation}
and more complicated, yet analogous expressions can be derived for higher order correlations. 

Figure 5.11 shows the $p_{T}$-integrated elliptic flow calculated using two-particle correlations as a function of centrality, compared with data from the ALICE Collaboration \cite{ALICE_Flow}. Each dot is the mean value of the events in that centrality class, and the error bars represent the event-by-event dispersion around the mean.

\begin{figure}[h!]
    \centering
    \includegraphics[width=14cm]{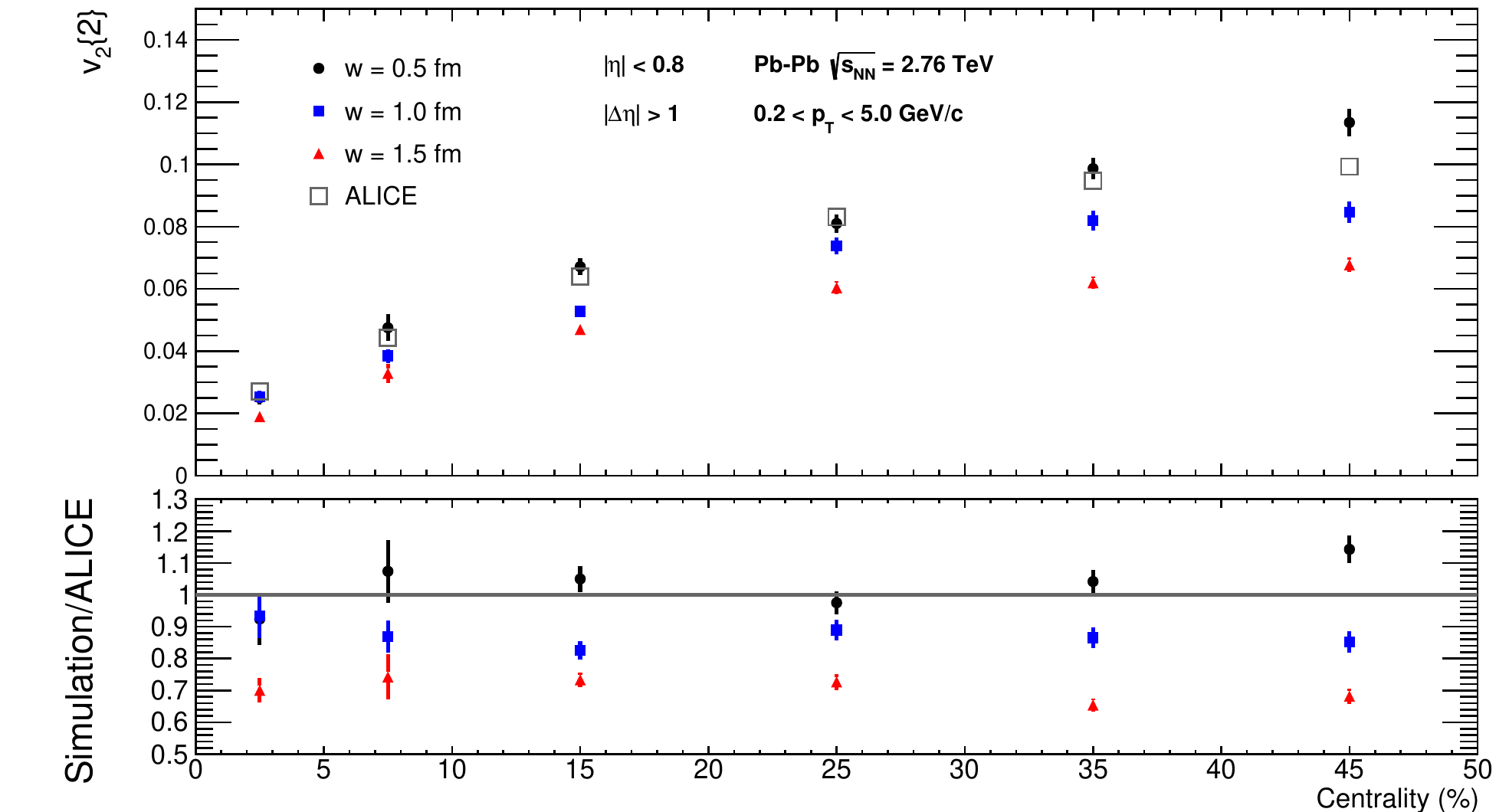}
    \caption{Integrated elliptic flow calculated using two-particle correlations as a function of centrality. Data from \cite{ALICE_Flow}. The error bars represent the event-by-event dispersion around the mean.}
\end{figure}
There is a pronounced effect of the nucleon-width parameter on integrated flow, which decreases as the nucleon size increases. This follows the same trend as ellipticity in the initial condition, and is simply a consequence of the mapping between initial state geometry and final state momentum anisotropy mentioned in Section 2.5.2:
\begin{equation}
v_{n} = f(\varepsilon_{n}) + \delta_{n}
\end{equation}
Using $w=0.5$ fm provides a better description of data description of $v_{2}$ data (within the 10 \% range), while the simulations with the larger nucleons do not produce enough elliptic flow.

\begin{figure}[h!]
    \centering
    \includegraphics[width=14cm]{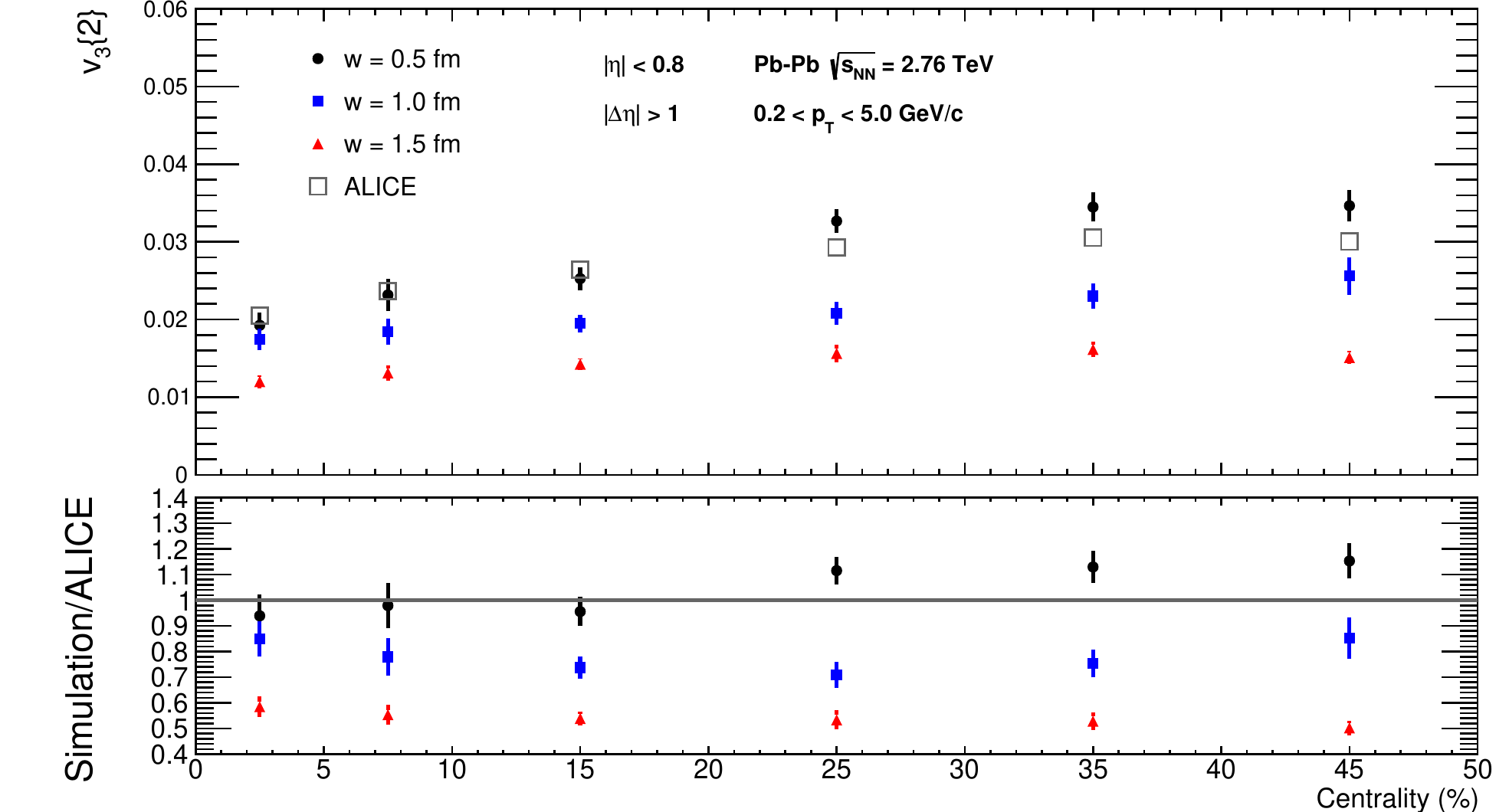}
    \caption{Integrated triangular flow calculated using two-particle correlations as a function of centrality. Data from \cite{ALICE_Flow}. The error bars represent the event-by-event dispersion around the mean.}
\end{figure}

In Figure 5.12, which shows the integrated triangular flow, this effect is even more pronounced: the simulations with the larger nucleons do not produce enough triangular flow, while the simulation with the small nucleons is able to reproduce experimental data within the 10 \% range. Once again, the mapping between initial state eccentricity and anisotropic flow is present. 

\subsection{Blast Wave analysis}

So far the only information we have extracted from the $p_{T}$-spectra was the mean value. What else can we learn about the effects of the nucleon-width parameter in the transverse momentum distributions? A hydrodynamic-inspired phenomenological model widely used to characterize $p_{T}$-spectra is the Blast Wave (BW) model \cite{BlastWave}. In the BW model, particles are emitted from the ``source'' with an approximately thermal distribution, at the same time that they are boosted in the radial direction by the transverse expansion of the system. Starting from this picture, the following analytical expression is obtained:
\begin{equation}
\frac{1}{p_{T}} \frac{d N}{d p_{T}} \propto \int_{0}^{R} r d r m_{T} I_{0}\left(\frac{p_{T} \sinh \rho}{T_{\text{kin}}}\right) K_{1}\left(\frac{m_{T} \cosh \rho}{T_{\text{kin}}}\right),
\end{equation}
where:
\begin{equation}
\rho=\tanh ^{-1} \beta_{T}=\tanh ^{-1}\left(\left(\frac{r}{R}\right)^{n} \beta_{s}\right),
\end{equation}
and $m_{T} = \sqrt{m^{2} + p_{T}^{2}}$ is the transverse mass. $R$ is the system radius, and $I_{0}$ and $K_{1}$ are modified Bessel functions. This expression contains three free parameters:
\begin{itemize}
    \item $T_{\text{kin}}$ - Kinetic freeze-out temperature
    \item $\beta_{T}$ - Transverse expansion velocity ($\beta_{s}$ is the surface velocity)
    \item $n$ - Velocity profile exponent
\end{itemize}
which can be extracted by fitting $p_{T}$-spectra with the expression in Equation (5.24). This is usually done in the form of a \textit{combined fit}, in which the $p_{T}$-spectra of three particle species are fitted simultaneously, providing a more stable fit and more robust insights. We follow the prescription used by the ALICE Collaboration \cite{ALICE_Pt}, in which the $p_{T}$-spectra of charged pions, kaons and protons were fitted in the following $p_{T}$ intervals:
\begin{itemize}
\item $\pi^{+}$ + $\pi^{-}$ : 0.5 - 1.0 GeV/c
\item $K^{+}$ + $K^{-}$ : 0.2 - 1.5 GeV/c
\item $p$ + $\Bar{p}$ : 0.3 - 3 GeV/c
\end{itemize}

\begin{figure}[h]
	\centering
	\includegraphics[width=14cm]{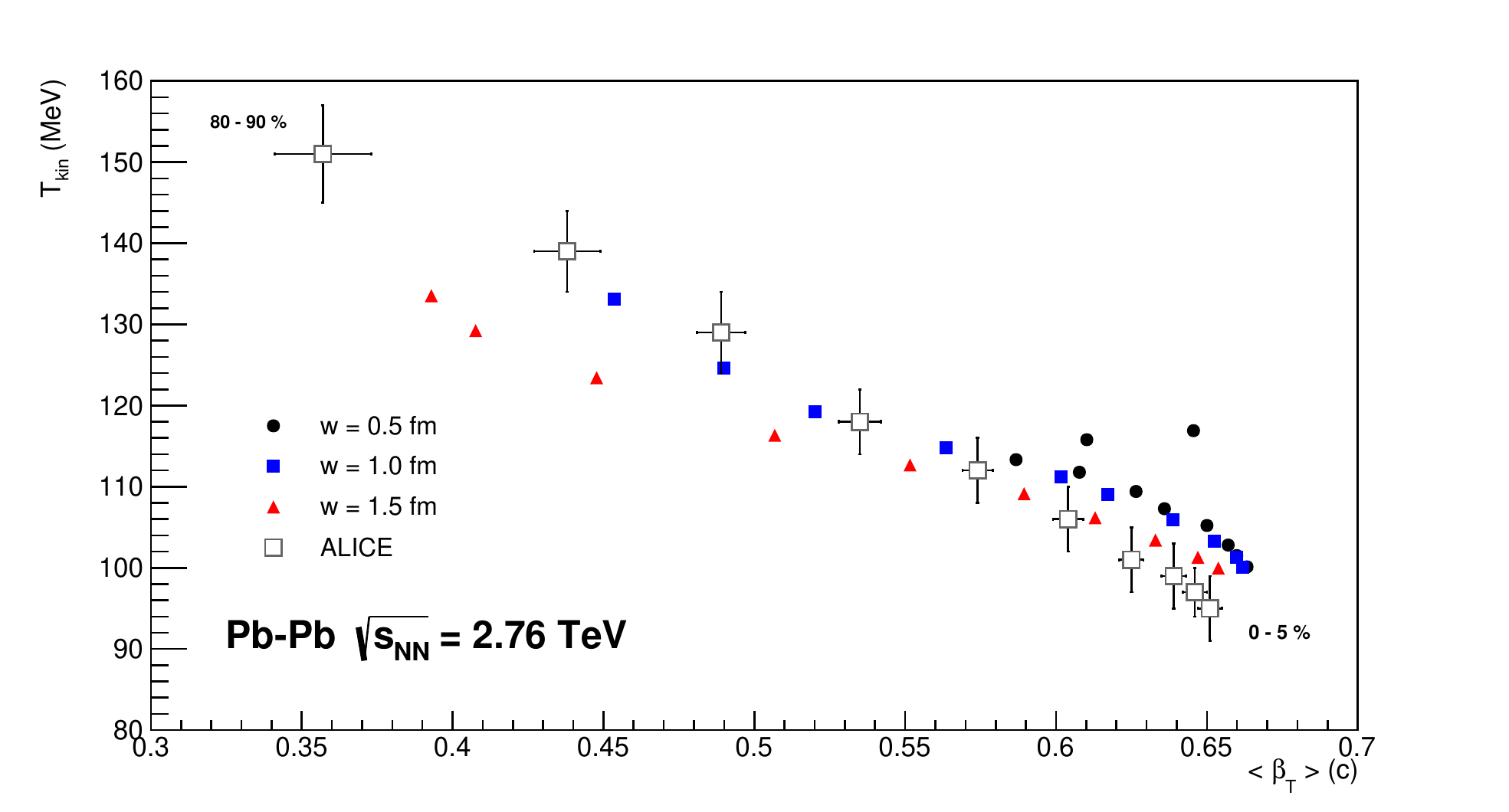}
	\caption{Blast Wave combined fit parameters. Each point corresponds to a centrality class. Central events are on the right, and peripheral events on the left. Data from \cite{ALICE_Pt}.}
\end{figure}

The standard visualization of the results of a BW analysis is such as in Figure 5.13: the transverse expansion velocity is displayed in the horizontal axis and the kinetic freeze-out temperature in the vertical axis. It is clear that the simulation with $w=0.5$ fm provides an insufficient description of experimental data, occupying a small region on the right side of Figure 5.13, while the experimental data is well spread across the plane. This suggests that, besides the mean transverse momentum, the simulation with the smaller nucleon size seems to be problematic in what concerns the description of the shape of $p_{T}$-spectra as a whole. Also, we note that for very peripheral collisions, the simulation using the small nucleons changes the trend of the transverse expansion velocity, following what was already observed in the centrality dependence of the mean transverse momentum of charged pions. 

Based on our results, we conclude that the only experimental observable which constitutes the core of the Bayesian Analyses that is actually poorly described when using the small nucleons is the mean transverse momentum, which is too large. We associate this to two simplifications made in the pre-hydrodynamic stage of the simulation. In this sense, the large values of the nucleon-width parameter might have been an artefact of the Bayesian Analyses, necessary to lower the mean transverse momentum, which was being artificially raised. In addition, the combined BW fit suggests that other issues are present in the transverse momentum distributions.


\chapter{Conclusions}
\label{CHAP-Conclusions}
\noindent\rule{\textwidth}{.1pt}\\[1ex]

This work was motivated by recent results in Bayesian Analyses used to constrain the parameters of simulations of heavy-ion collisions, in which the nucleon-width parameter assumed surprisingly large values, inconsistent with current measurements of the proton size. In this work, we have performed full hybrid simulations of Pb-Pb collisions at $\sqrt{s_{\text{NN}}}$ = 2.76 TeV using a state-of-the-art simulation chain for three different values of the nucleon-width parameter, and systematically analyzed its effects on the initial state characteristics and observables. 

The nucleon size strongly affects the initial condition characteristics: collisions generated using smaller nucleons have stronger entropy gradients and are more spatially anisotropic.

Now we are ready to answer the questions that were asked in Section 4.3:

\begin{itemize}
\item \textbf{How intensely and in which way $w$ affects different observables? :} The nucleon-width visibly affects all the considered observables. As the nucleon size increases, there is a decrease in the mean transverse momentum of final particles (due to the weaker gradients in the initial conditions) and a decrease on elliptic and triangular flow (due to the decrease in the initial condition ellipticity and triangularity). 
\item \textbf{Is it possible to obtain a good description of experimental data with the calibrated chain using values of $w$ consistent with estimates of the proton charge radius? If not, which observables(s) are poorly described in this scenario, and why? :} Overall, it is possible to obtain a good description (within the 10 \% range) of experimental data using $w=0.5$ fm, with the exception of the mean transverse momentum. In this case, the simulation with the smaller nucleons result in values of $\langle p_{T} \rangle$ above data, and a wrong centrality dependence. This was linked to simplifications made in the pre-equilibrium phase which artificially increase the mean transverse momentum.
\item \textbf{What led the Bayesian Analyses to favour such large values of the nucleon-width parameter? :} Our results suggest that the large value of the nucleon-width parameter returned by the Bayesian Analysis in \cite{DUKE2} might have been an ``attempt'' to lower the $\langle p_{T} \rangle$, which was being artificially raised. However, it is not possible to state that this is the only effect present, and totally accounts for the enlargement of the nucleon-width parameter MAP value. 
\end{itemize}
A combined Blast Wave analysis suggests that the shape of the $p_{T}$-spectra as a whole is poorly described by the simulation with $w=$ 0.5 fm.

This work highlights the importance of performing computational simulations based on physical models in order to extract information about the collision and allow for comparisons with experimental data collected in heavy-ion colliders. These simulations are computationally demanding and present a complex relation between its parameters and observables, so that systematic studies such as the ones presented in this work are necessary.

Finally, this work complements the recent findings in \cite{Giacalone:2021clp, Nijs:2022rme}, which show that indeed the nucleon-width parameter should assume values smaller than 0.7 fm.

\newpage


\printbibliography

@misc{SM_Website,
    author    = "MissMJ",
    title     = "Standard Model of Elementary Particles",
    note       = "\url{https://commons.wikimedia.org/wiki/File:Standard_Model_of_Elementary_Particles.svg}"
}

@article{20121,
title = {Observation of a new particle in the search for the Standard Model Higgs boson with the ATLAS detector at the LHC},
journal = {Physics Letters B},
volume = {716},
number = {1},
pages = {1-29},
year = {2012},
issn = {0370-2693},
doi = {https://doi.org/10.1016/j.physletb.2012.08.020},
url = {https://www.sciencedirect.com/science/article/pii/S037026931200857X},
collaboration = {ATLAS Collaboration}
}

@Book{MarkThomson,
   title =     {Modern Particle Physics},
   author =    {Mark Thomson},
   publisher = {Cambridge University Press},
   isbn =      {978-1-107-03426-6},
   year =      {2013}
}

@article{Higgs,
  title = {Broken Symmetries and the Masses of Gauge Bosons},
  author = {Higgs, Peter W.},
  journal = {Phys. Rev. Lett.},
  volume = {13},
  issue = {16},
  pages = {508--509},
  numpages = {0},
  year = {1964},
  month = {Oct},
  publisher = {American Physical Society},
  doi = {10.1103/PhysRevLett.13.508},
  url = {https://link.aps.org/doi/10.1103/PhysRevLett.13.508}
}

@book{Schwartz,
    author = "Schwartz, Matthew D.",
    title = "{Quantum Field Theory and the Standard Model}",
    isbn = "978-1-107-03473-0, 978-1-107-03473-0",
    publisher = "Cambridge University Press",
    month = "3",
    year = "2014"
}

@article{QED1,
  title = {Quantum Electrodynamics. I. A Covariant Formulation},
  author = {Schwinger, Julian},
  journal = {Phys. Rev.},
  volume = {74},
  issue = {10},
  pages = {1439--1461},
  numpages = {0},
  year = {1948},
  month = {Nov},
  publisher = {American Physical Society},
  doi = {10.1103/PhysRev.74.1439},
  url = {https://link.aps.org/doi/10.1103/PhysRev.74.1439}
}

@article{Fritzsch_Gell-Mann,
    author = "Fritzsch, H. and Gell-Mann, Murray and Leutwyler, H.",
    title = "{Advantages of the Color Octet Gluon Picture}",
    reportNumber = "CALT-68-409",
    doi = "10.1016/0370-2693(73)90625-4",
    journal = "Phys. Lett. B",
    volume = "47",
    pages = "365--368",
    year = "1973"
}

@article{Maire,
author = {Maire, Antonin},
year = {2015},
month = {06},
pages = {},
title = {Introduction to the Quark-Gluon Plasma session in RJC 2014}
}

@inproceedings{Tlusty,
    author = "Tlusty, David",
    title = "{The RHIC Beam Energy Scan Phase II: Physics and Upgrades}",
    booktitle = "{13th Conference on the Intersections of Particle and Nuclear Physics}",
    eprint = "1810.04767",
    archivePrefix = "arXiv",
    primaryClass = "nucl-ex",
    month = "10",
    year = "2018"
}

@article{PDG,
    author = {Particle Data Group},
    title = "{Review of Particle Physics}",
    journal = {Progress of Theoretical and Experimental Physics},
    volume = {2020},
    number = {8},
    year = {2020},
    month = {08},
    issn = {2050-3911},
    doi = {10.1093/ptep/ptaa104},
    url = {https://doi.org/10.1093/ptep/ptaa104},
    note = {083C01},
    eprint = {https://academic.oup.com/ptep/article-pdf/2020/8/083C01/34673722/ptaa104.pdf},
}

@article{QCDReview1,
author = {Wilczek, F},
title = {Quantum Chromodynamics: The Modern Theory of the Strong Interaction},
journal = {Annual Review of Nuclear and Particle Science},
volume = {32},
number = {1},
pages = {177-209},
year = {1982},
doi = {10.1146/annurev.ns.32.120182.001141},
URL = {https://doi.org/10.1146/annurev.ns.32.120182.001141},
eprint = { https://doi.org/10.1146/annurev.ns.32.120182.00114}
}

@inproceedings{QCDReview2,
    author = "Ecker, Gerhard",
    title = "{Quantum chromodynamics}",
    booktitle = "{2005 European School of High-Energy Physics}",
    eprint = "hep-ph/0604165",
    archivePrefix = "arXiv",
    reportNumber = "UWTHPH-2006-9",
    month = "4",
    year = "2006"
}

@article{Deur,
title = {The QCD running coupling},
journal = {Progress in Particle and Nuclear Physics},
volume = {90},
pages = {1-74},
year = {2016},
issn = {0146-6410},
doi = {https://doi.org/10.1016/j.ppnp.2016.04.003},
url = {https://www.sciencedirect.com/science/article/pii/S0146641016300035},
author = {Alexandre Deur and Stanley J. Brodsky and Guy F. {de Téramond}}
}

@article{Aoki,
    author = "Aoki, Y. and Endrodi, G. and Fodor, Z. and Katz, S. D. and Szabo, K. K.",
    title = "{The Order of the quantum chromodynamics transition predicted by the standard model of particle physics}",
    eprint = "hep-lat/0611014",
    archivePrefix = "arXiv",
    doi = "10.1038/nature05120",
    journal = "Nature",
    volume = "443",
    pages = "675--678",
    year = "2006"
}

@article{Parisi,
title = {Exponential hadronic spectrum and quark liberation},
journal = {Physics Letters B},
volume = {59},
number = {1},
pages = {67-69},
year = {1975},
issn = {0370-2693},
doi = {https://doi.org/10.1016/0370-2693(75)90158-6},
url = {https://www.sciencedirect.com/science/article/pii/0370269375901586},
author = {N. Cabibbo and G. Parisi},
}

@article{Ding,
author = {Ding, Heng-Tong and Karsch, Frithjof and Mukherjee, Swagato},
year = {2015},
month = {04},
pages = {},
title = {Thermodynamics of strong-interaction matter from Lattice QCD},
volume = {24},
journal = {International Journal of Modern Physics E},
doi = {10.1142/S0218301315300076}
}

@article{GELLMANN1964214,
title = {A schematic model of baryons and mesons},
journal = {Physics Letters},
volume = {8},
number = {3},
pages = {214-215},
year = {1964},
issn = {0031-9163},
doi = {https://doi.org/10.1016/S0031-9163(64)92001-3},
url = {https://www.sciencedirect.com/science/article/pii/S0031916364920013},
author = {M. Gell-Mann}
}

@article{Feynman:1969wa,
    author = "Feynman, R. P.",
    title = "{The behavior of hadron collisions at extreme energies}",
    journal = "Conf. Proc. C",
    volume = "690905",
    pages = "237--258",
    year = "1969"
}

@article{Alford:1997zt,
    author = "Alford, Mark G. and Rajagopal, Krishna and Wilczek, Frank",
    title = "{QCD at finite baryon density: Nucleon droplets and color superconductivity}",
    eprint = "hep-ph/9711395",
    archivePrefix = "arXiv",
    reportNumber = "IASSNS-HEP-97-119, MIT-CTP-2695",
    doi = "10.1016/S0370-2693(98)00051-3",
    journal = "Phys. Lett. B",
    volume = "422",
    pages = "247--256",
    year = "1998"
}

@article{Stock:2004iim,
    author = "Stock, Reinhard",
    editor = "Ritter, Hans Georg and Wang, Xin-Nian",
    title = "{Relativistic nucleus nucleus collisions: From the BEVALAC to RHIC.}",
    eprint = "nucl-ex/0405007",
    archivePrefix = "arXiv",
    doi = "10.1088/0954-3899/30/8/001",
    journal = "J. Phys. G",
    volume = "30",
    pages = "S633--S648",
    year = "2004"
}

@article{Voloshin:1994mz,
    author = "Voloshin, S. and Zhang, Y.",
    title = "{Flow study in relativistic nuclear collisions by Fourier expansion of Azimuthal particle distributions}",
    eprint = "hep-ph/9407282",
    archivePrefix = "arXiv",
    doi = "10.1007/s002880050141",
    journal = "Z. Phys. C",
    volume = "70",
    pages = "665--672",
    year = "1996"
}

@article{Voloshin:2008dg,
    author = "Voloshin, Sergei A. and Poskanzer, Arthur M. and Snellings, Raimond",
    editor = "Stock, R.",
    title = "{Collective phenomena in non-central nuclear collisions}",
    eprint = "0809.2949",
    archivePrefix = "arXiv",
    primaryClass = "nucl-ex",
    doi = "10.1007/978-3-642-01539-7_10",
    journal = "Landolt-Bornstein",
    volume = "23",
    pages = "293--333",
    year = "2010"
}

@article{Bialas:1976ed,
    author = "Bialas, A. and Bleszynski, M. and Czyz, W.",
    title = "{Multiplicity Distributions in Nucleus-Nucleus Collisions at High-Energies}",
    reportNumber = "TPJU-9/76",
    doi = "10.1016/0550-3213(76)90329-1",
    journal = "Nucl. Phys. B",
    volume = "111",
    pages = "461--476",
    year = "1976"
}

@article{Bozek:2009dw,
    author = "Bozek, Piotr",
    title = "{Bulk and shear viscosities of matter created in relativistic heavy-ion collisions}",
    eprint = "0911.2397",
    archivePrefix = "arXiv",
    primaryClass = "nucl-th",
    doi = "10.1103/PhysRevC.81.034909",
    journal = "Phys. Rev. C",
    volume = "81",
    pages = "034909",
    year = "2010"
}

@article{CMS:2012qbp,
    author = "Chatrchyan, Serguei and others",
    collaboration = "CMS",
    title = "{Observation of a New Boson at a Mass of 125 GeV with the CMS Experiment at the LHC}",
    eprint = "1207.7235",
    archivePrefix = "arXiv",
    primaryClass = "hep-ex",
    reportNumber = "CMS-HIG-12-028, CERN-PH-EP-2012-220",
    doi = "10.1016/j.physletb.2012.08.021",
    journal = "Phys. Lett. B",
    volume = "716",
    pages = "30--61",
    year = "2012"
}

@article{CMS:2013btf,
    author = "Chatrchyan, Serguei and others",
    collaboration = "CMS",
    title = "{Observation of a New Boson with Mass Near 125 GeV in $pp$ Collisions at $\sqrt{s}$ = 7 and 8 TeV}",
    eprint = "1303.4571",
    archivePrefix = "arXiv",
    primaryClass = "hep-ex",
    reportNumber = "CMS-HIG-12-036, CERN-PH-EP-2013-035",
    doi = "10.1007/JHEP06(2013)081",
    journal = "JHEP",
    volume = "06",
    pages = "081",
    year = "2013"
}

@article{PhysRevLett.30.1343,
  title = {Ultraviolet Behavior of Non-Abelian Gauge Theories},
  author = {Gross, David J. and Wilczek, Frank},
  journal = {Phys. Rev. Lett.},
  volume = {30},
  issue = {26},
  pages = {1343--1346},
  numpages = {0},
  year = {1973},
  month = {Jun},
  publisher = {American Physical Society},
  doi = {10.1103/PhysRevLett.30.1343},
  url = {https://link.aps.org/doi/10.1103/PhysRevLett.30.1343}
}

@article{DAVIDPOLITZER1974129,
title = {Asymptotic freedom: An approach to strong interactions},
journal = {Physics Reports},
volume = {14},
number = {4},
pages = {129-180},
year = {1974},
issn = {0370-1573},
doi = {https://doi.org/10.1016/0370-1573(74)90014-3},
url = {https://www.sciencedirect.com/science/article/pii/0370157374900143},
author = {H {David Politzer}},
}

@article{Dusling:2011fd,
    author = {Dusling, Kevin and Sch\"afer, Thomas},
    title = "{Bulk viscosity, particle spectra and flow in heavy-ion collisions}",
    eprint = "1109.5181",
    archivePrefix = "arXiv",
    primaryClass = "hep-ph",
    doi = "10.1103/PhysRevC.85.044909",
    journal = "Phys. Rev. C",
    volume = "85",
    pages = "044909",
    year = "2012"
}

@article{Teaney:2003kp,
    author = "Teaney, Derek",
    title = "{The Effects of viscosity on spectra, elliptic flow, and HBT radii}",
    eprint = "nucl-th/0301099",
    archivePrefix = "arXiv",
    doi = "10.1103/PhysRevC.68.034913",
    journal = "Phys. Rev. C",
    volume = "68",
    pages = "034913",
    year = "2003"
}

@article{Baldin_1983,
	doi = {10.1088/0031-8949/1983/t3/009},
	url = {https://doi.org/10.1088/0031-8949/1983/t3/009},
	year = 1983,
	month = {jan},
	publisher = {{IOP} Publishing},
	volume = {T3},
	pages = {43--44},
	author = {A M Baldin and E D Donets and I B Issinsky and A D Kirillov and I N Semenyushkin},
	title = {Beams of Highly Charged Ions at the Dubna Syncrophasotron},
	journal = {Physica Scripta},
}

@article{Baym:2001in,
    author = "Baym, Gordon",
    editor = "Hallman, T. J. and Kharzeev, D. E. and Mitchell, J. T. and Ullrich, T. S.",
    title = "{RHIC: From dreams to beams in two decades}",
    eprint = "hep-ph/0104138",
    archivePrefix = "arXiv",
    doi = "10.1016/S0375-9474(01)01342-2",
    journal = "Nucl. Phys. A",
    volume = "698",
    pages = "XXIII--XXXII",
    year = "2002"
}

@article{PHOBOS:2004zne,
    author = "Back, B. B. and others",
    collaboration = "PHOBOS",
    title = "{The PHOBOS perspective on discoveries at RHIC}",
    eprint = "nucl-ex/0410022",
    archivePrefix = "arXiv",
    doi = "10.1016/j.nuclphysa.2005.03.084",
    journal = "Nucl. Phys. A",
    volume = "757",
    pages = "28--101",
    year = "2005"
}

@article{BRAHMS:2004adc,
    author = "Arsene, I. and others",
    collaboration = "BRAHMS",
    title = "{Quark gluon plasma and color glass condensate at RHIC? The Perspective from the BRAHMS experiment}",
    eprint = "nucl-ex/0410020",
    archivePrefix = "arXiv",
    doi = "10.1016/j.nuclphysa.2005.02.130",
    journal = "Nucl. Phys. A",
    volume = "757",
    pages = "1--27",
    year = "2005"
}

@article{STAR:2005gfr,
    author = "Adams, John and others",
    collaboration = "STAR",
    title = "{Experimental and theoretical challenges in the search for the quark gluon plasma: The STAR Collaboration's critical assessment of the evidence from RHIC collisions}",
    eprint = "nucl-ex/0501009",
    archivePrefix = "arXiv",
    doi = "10.1016/j.nuclphysa.2005.03.085",
    journal = "Nucl. Phys. A",
    volume = "757",
    pages = "102--183",
    year = "2005"
}

@article{PHENIX:2004vcz,
    author = "Adcox, K. and others",
    collaboration = "PHENIX",
    title = "{Formation of dense partonic matter in relativistic nucleus-nucleus collisions at RHIC: Experimental evaluation by the PHENIX collaboration}",
    eprint = "nucl-ex/0410003",
    archivePrefix = "arXiv",
    doi = "10.1016/j.nuclphysa.2005.03.086",
    journal = "Nucl. Phys. A",
    volume = "757",
    pages = "184--283",
    year = "2005"
}

@article{CHEN201438,
title = {Studying the Early Universe via Quark-Gluon Plasma},
journal = {Nuclear Physics B - Proceedings Supplements},
volume = {246-247},
pages = {38-41},
year = {2014},
note = {Proceedings of the 9th International Symposium on Cosmology and Particle Astrophysics},
issn = {0920-5632},
doi = {https://doi.org/10.1016/j.nuclphysbps.2013.10.063},
url = {https://www.sciencedirect.com/science/article/pii/S092056321300652X},
author = {Chin-Hao Chen},
}

@article{ALICE:2019mmy,
    author = "Acharya, Shreyasi and others",
    collaboration = "ALICE",
    title = "{Underlying Event properties in pp collisions at $\sqrt{s}$ = 13 TeV}",
    eprint = "1910.14400",
    archivePrefix = "arXiv",
    primaryClass = "nucl-ex",
    reportNumber = "CERN-EP-2019-235",
    doi = "10.1007/JHEP04(2020)192",
    journal = "JHEP",
    volume = "04",
    pages = "192",
    year = "2020"
}

@article{Busza,
author = {Busza, Wit and Rajagopal, Krishna and van der Schee, Wilke},
title = {Heavy Ion Collisions: The Big Picture and the Big Questions},
journal = {Annual Review of Nuclear and Particle Science},
volume = {68},
number = {1},
pages = {339-376},
year = {2018},
doi = {10.1146/annurev-nucl-101917-020852},
URL = { https://doi.org/10.1146/annurev-nucl-101917-020852}
}

@article{Connors:2017ptx,
    author = "Connors, Megan and Nattrass, Christine and Reed, Rosi and Salur, Sevil",
    title = "{Jet measurements in heavy ion physics}",
    eprint = "1705.01974",
    archivePrefix = "arXiv",
    primaryClass = "nucl-ex",
    doi = "10.1103/RevModPhys.90.025005",
    journal = "Rev. Mod. Phys.",
    volume = "90",
    pages = "025005",
    year = "2018"
}

@article{Brewer:2020tbb,
    author = "Brewer, Jasmine",
    title = "{Jets as a probe of the quark-gluon plasma}",
    eprint = "2012.14457",
    archivePrefix = "arXiv",
    primaryClass = "nucl-th",
    doi = "10.22323/1.387.0012",
    journal = "PoS",
    volume = "HardProbes2020",
    pages = "012",
    year = "2021"
}

@phdthesis{Costanza,
author = {Cavicchioli, Costanza},
year = {2010},
month = {01},
pages = {},
title = {Development and Commissioning of the Pixel Trigger System for the ALICE Experiment at the CERN Large Hadron Collider}
}

@article{DPMJET,
    author = "Bopp, Fritz W. and Engel, R. and Ranft, J. and Roesler, S.",
    editor = "Armesto, N. and Borghini, N. and Jeon, S. and Wiedemann, U. A.",
    title = "{Inclusive distributions at the LHC as predicted from the DPMJET-III model with chain fusion}",
    eprint = "0706.3875",
    archivePrefix = "arXiv",
    primaryClass = "hep-ph",
    reportNumber = "SI-HEP-2007-10",
    journal = "J. Phys. G",
    volume = "35",
    number = "5",
    pages = "054001.20",
    year = "2008"
}

@article{HIJING,
    author = "Deng, Wei-Tian and Wang, Xin-Nian and Xu, Rong",
    title = "{Hadron production in p+p, p+Pb, and Pb+Pb collisions with the HIJING 2.0 model at energies available at the CERN Large Hadron Collider}",
    eprint = "1008.1841",
    archivePrefix = "arXiv",
    primaryClass = "hep-ph",
    doi = "10.1103/PhysRevC.83.014915",
    journal = "Phys. Rev. C",
    volume = "83",
    pages = "014915",
    year = "2011"
}

@article{ALbacete:2010ad,
    author = "ALbacete, Javier L. and Dumitru, Adrian",
    title = "{A model for gluon production in heavy-ion collisions at the LHC with rcBK unintegrated gluon densities}",
    eprint = "1011.5161",
    archivePrefix = "arXiv",
    primaryClass = "hep-ph",
    month = "11",
    year = "2010"
}

@article{Kharzeev:2004if,
    author = "Kharzeev, Dmitri and Levin, Eugene and Nardi, Marzia",
    title = "{Color glass condensate at the LHC: Hadron multiplicities in pp, pA and AA collisions}",
    eprint = "hep-ph/0408050",
    archivePrefix = "arXiv",
    reportNumber = "BNL-NT-04-26, BNL---NT---04-26",
    doi = "10.1016/j.nuclphysa.2004.10.018",
    journal = "Nucl. Phys. A",
    volume = "747",
    pages = "609--629",
    year = "2005"
}

@article{Armesto:2004ud,
    author = "Armesto, Nestor and Salgado, Carlos A. and Wiedemann, Urs Achim",
    title = "{Relating high-energy lepton-hadron, proton-nucleus and nucleus-nucleus collisions through geometric scaling}",
    eprint = "hep-ph/0407018",
    archivePrefix = "arXiv",
    reportNumber = "CERN-PH-TH-2004-121",
    doi = "10.1103/PhysRevLett.94.022002",
    journal = "Phys. Rev. Lett.",
    volume = "94",
    pages = "022002",
    year = "2005"
}

@article{Wang:2018hqq,
    author = "Wang, Qi and Yang, Pei-Pin and Liu, Fu-Hu",
    title = "{Comparing a few distributions of transverse momenta in high energy collisions}",
    eprint = "1811.09883",
    archivePrefix = "arXiv",
    primaryClass = "hep-ph",
    doi = "10.1016/j.rinp.2018.11.067",
    journal = "Results Phys.",
    volume = "12",
    pages = "259--267",
    year = "2019"
}

@article{Gupta:2021efj,
    author = "Gupta, Rohit and Jena, Satyajit",
    title = "{Model comparison of the transverse momentum spectra of charged hadrons produced in $PbPb$ collision at $\sqrt{s_{NN}} = 5.02$ TeV}",
    eprint = "2103.13104",
    archivePrefix = "arXiv",
    primaryClass = "hep-ph",
    month = "3",
    year = "2021"
}

@article{Tsallis,
author = {Tsallis, Constantino},
year = {1988},
month = {07},
pages = {479-487},
title = {Possible generalization of Boltzmann-Gibbs statistics},
volume = {52},
journal = {Journal of Statistical Physics},
doi = {10.1007/BF01016429}
}

@article{Hagedorn1,
    author = "Hagedorn, R.",
    title = "{Statistical thermodynamics of strong interactions at high-energies}",
    reportNumber = "CERN-TH-520",
    journal = "Nuovo Cim. Suppl.",
    volume = "3",
    pages = "147--186",
    year = "1965"
}

@article{Hagedorn2,
    author = "Hagedorn, R. and Ranft, J.",
    title = "{Statistical thermodynamics of strong interactions at high-energies. 2. Momentum spectra of particles produced in pp-collisions}",
    reportNumber = "CERN-TH-851",
    journal = "Nuovo Cim. Suppl.",
    volume = "6",
    pages = "169--354",
    year = "1968"
}

@inproceedings{Gupta:2021ksp,
    author = "Gupta, Rohit and Jena, Satyajit",
    title = "{A unified formalism to study $soft$ as well as $hard$ part of the transverse momentum spectra}",
    booktitle = "{24th DAE-BRNS High Energy Physics Symposium}",
    eprint = "2103.13896",
    archivePrefix = "arXiv",
    primaryClass = "hep-ph",
    month = "3",
    year = "2021"
}

@article{BlastWave,
  title = {Evidence for a Blast Wave from Compressed Nuclear Matter},
  author = {Siemens, Philip J. and Rasmussen, John O.},
  journal = {Phys. Rev. Lett.},
  volume = {42},
  issue = {14},
  pages = {880--883},
  numpages = {0},
  year = {1979},
  month = {Apr},
  publisher = {American Physical Society},
  doi = {10.1103/PhysRevLett.42.880},
  url = {https://link.aps.org/doi/10.1103/PhysRevLett.42.880}
}

@article{Heinz:2013th,
    author = "Heinz, Ulrich and Snellings, Raimond",
    title = "{Collective flow and viscosity in relativistic heavy-ion collisions}",
    eprint = "1301.2826",
    archivePrefix = "arXiv",
    primaryClass = "nucl-th",
    doi = "10.1146/annurev-nucl-102212-170540",
    journal = "Ann. Rev. Nucl. Part. Sci.",
    volume = "63",
    pages = "123--151",
    year = "2013"
}

@article{Betz,
author = {Betz, Barbara},
year = {2009},
month = {10},
pages = {},
title = {Jet Propagation and Mach-Cone Formation in (3+1)-dimensional Ideal Hydrodynamics}
}

@article{LbL,
collaboration = "ATLAS Collaboration",
year = {2017},
month = {09},
pages = {852-858},
title = {Evidence for light-by-light scattering in heavy-ion collisions with the ATLAS detector at the LHC},
volume = {13},
journal = {Nature Physics},
doi = {https://doi.org/10.1038/nphys4208}
}

@article{ALICE_Pt,
    author = "Abelev, Betty and others",
    collaboration = "ALICE",
    title = "{Centrality dependence of $\pi$, K, p production in Pb-Pb collisions at $\sqrt{s_{NN}}$ = 2.76 TeV}",
    eprint = "1303.0737",
    archivePrefix = "arXiv",
    primaryClass = "hep-ex",
    reportNumber = "CERN-PH-EP-2013-019",
    doi = "10.1103/PhysRevC.88.044910",
    journal = "Phys. Rev. C",
    volume = "88",
    pages = "044910",
    year = "2013"
}

@article{Andrade:2008fa,
    author = "Andrade, R. P. G. and dos Reis, A. L. V. R. and Grassi, F. and Hama, Yogiro and Qian, W. L. and Kodama, T. and Ollitrault, J. -Y.",
    title = "{Fluctuations and initial state granularity in heavy ion collisions and their effects on observables from hydrodynamics}",
    eprint = "0812.4143",
    archivePrefix = "arXiv",
    primaryClass = "nucl-th",
    journal = "Acta Phys. Polon. B",
    volume = "40",
    pages = "993--998",
    year = "2009"
}

@article{ALICE_Multiplicity,
    author = "Adam, Jaroslav and others",
    collaboration = "ALICE",
    title = "{Centrality dependence of the charged-particle multiplicity density at midrapidity in Pb-Pb collisions at $\sqrt{s_{\rm NN}}$ = 5.02 TeV}",
    eprint = "1512.06104",
    archivePrefix = "arXiv",
    primaryClass = "nucl-ex",
    reportNumber = "CERN-PH-EP-2015-324, ALICE-PUBLIC-2015-008",
    doi = "10.1103/PhysRevLett.116.222302",
    journal = "Phys. Rev. Lett.",
    volume = "116",
    number = "22",
    pages = "222302",
    year = "2016"
}

@article{ALICE_Mult,
    author = "Aamodt, Kenneth and others",
    collaboration = "ALICE",
    title = "{Centrality dependence of the charged-particle multiplicity density at mid-rapidity in Pb-Pb collisions at $\sqrt{s_{NN}}=2.76$ TeV}",
    eprint = "1012.1657",
    archivePrefix = "arXiv",
    primaryClass = "nucl-ex",
    reportNumber = "CERN-PH-EP-2010-071",
    doi = "10.1103/PhysRevLett.106.032301",
    journal = "Phys. Rev. Lett.",
    volume = "106",
    pages = "032301",
    year = "2011"
}

@article{Kolb,
    author = "Kolb, Peter F. and Heinz, Ulrich W.",
    editor = "Hwa, Rudolph C. and Wang, Xin-Nian",
    title = "{Hydrodynamic description of ultrarelativistic heavy ion collisions}",
    eprint = "nucl-th/0305084",
    archivePrefix = "arXiv",
    reportNumber = "SUNY-NTG-03-06",
    pages = "634--714",
    month = "5",
    year = "2003"
}

@article{ALICE_Flow,
    author = "Aamodt, K. and others",
    collaboration = "ALICE",
    title = "{Higher harmonic anisotropic flow measurements of charged particles in Pb-Pb collisions at $\sqrt{s_{NN}}$=2.76 TeV}",
    eprint = "1105.3865",
    archivePrefix = "arXiv",
    primaryClass = "nucl-ex",
    reportNumber = "CERN-PH-EP-2011-073",
    doi = "10.1103/PhysRevLett.107.032301",
    journal = "Phys. Rev. Lett.",
    volume = "107",
    pages = "032301",
    year = "2011"
}

@article{PHENIX:2015tbb,
    author = "Adare, A. and others",
    collaboration = "PHENIX",
    title = "{Transverse energy production and charged-particle multiplicity at midrapidity in various systems from $\sqrt{s_{NN}}=7.7$ to 200 GeV}",
    eprint = "1509.06727",
    archivePrefix = "arXiv",
    primaryClass = "nucl-ex",
    doi = "10.1103/PhysRevC.93.024901",
    journal = "Phys. Rev. C",
    volume = "93",
    number = "2",
    pages = "024901",
    year = "2016"
}

@article{ALICE:2012fjm,
    author = "Abelev, Betty and others",
    collaboration = "ALICE",
    title = "{Measurement of inelastic, single- and double-diffraction cross sections in proton--proton collisions at the LHC with ALICE}",
    eprint = "1208.4968",
    archivePrefix = "arXiv",
    primaryClass = "hep-ex",
    reportNumber = "CERN-PH-EP-2012-238",
    doi = "10.1140/epjc/s10052-013-2456-0",
    journal = "Eur. Phys. J. C",
    volume = "73",
    number = "6",
    pages = "2456",
    year = "2013"
}

@article{ALICE:2012xs,
    author = "Abelev, Betty and others",
    collaboration = "ALICE",
    title = "{Pseudorapidity density of charged particles in $p$ + Pb collisions at $\sqrt{s_{NN}}=5.02$ TeV}",
    eprint = "1210.3615",
    archivePrefix = "arXiv",
    primaryClass = "nucl-ex",
    reportNumber = "CERN-PH-EP-2012-307",
    doi = "10.1103/PhysRevLett.110.032301",
    journal = "Phys. Rev. Lett.",
    volume = "110",
    number = "3",
    pages = "032301",
    year = "2013"
}

@article{Block:2012nj,
    author = "Block, Martin M. and Halzen, Francis",
    title = "{New experimental evidence that the proton develops asymptotically into a black disk}",
    eprint = "1208.4086",
    archivePrefix = "arXiv",
    primaryClass = "hep-ph",
    doi = "10.1103/PhysRevD.86.051504",
    journal = "Phys. Rev. D",
    volume = "86",
    pages = "051504",
    year = "2012"
}

@article{Hippert,
    author = "Hippert, Maur\'\i{}cio and Dobrigkeit Chinellato, David and Luzum, Matthew and Noronha, Jorge and Nunes da Silva, Tiago and Takahashi, Jun",
    editor = "Liu, Feng and Wang, Enke and Wang, Xin-Nian and Xu, Nu and Zhang, Ben-Wei",
    title = "{Momentum-dependent flow fluctuations as a hydrodynamic response to initial geometry}",
    eprint = "2003.01147",
    archivePrefix = "arXiv",
    primaryClass = "nucl-th",
    doi = "10.1016/j.nuclphysa.2020.121982",
    journal = "Nucl. Phys. A",
    volume = "1005",
    pages = "121982",
    year = "2021"
}

@article{ALICE:2017pcy,
    author = "Acharya, S. and others",
    collaboration = "ALICE",
    title = "{Charged-particle multiplicity distributions over a wide pseudorapidity range in proton-proton collisions at $\sqrt{s}=$ 0.9, 7, and 8 TeV}",
    eprint = "1708.01435",
    archivePrefix = "arXiv",
    primaryClass = "hep-ex",
    reportNumber = "CERN-EP-2017-192",
    doi = "10.1140/epjc/s10052-017-5412-6",
    journal = "Eur. Phys. J. C",
    volume = "77",
    number = "12",
    pages = "852",
    year = "2017"
}

@article{Denicol:2012cn,
    author = "Denicol, G. S. and Niemi, H. and Molnar, E. and Rischke, D. H.",
    title = "{Derivation of transient relativistic fluid dynamics from the Boltzmann equation}",
    eprint = "1202.4551",
    archivePrefix = "arXiv",
    primaryClass = "nucl-th",
    doi = "10.1103/PhysRevD.85.114047",
    journal = "Phys. Rev. D",
    volume = "85",
    pages = "114047",
    year = "2012",
    note = "[Erratum: Phys.Rev.D 91, 039902 (2015)]"
}

@article{Molnar:2013lta,
    author = "Moln\'ar, E. and Niemi, H. and Denicol, G. S. and Rischke, D. H.",
    title = "{Relative importance of second-order terms in relativistic dissipative fluid dynamics}",
    eprint = "1308.0785",
    archivePrefix = "arXiv",
    primaryClass = "nucl-th",
    doi = "10.1103/PhysRevD.89.074010",
    journal = "Phys. Rev. D",
    volume = "89",
    number = "7",
    pages = "074010",
    year = "2014"
}

@article{ANDERSON1974466,
title = {A relativistic relaxation-time model for the Boltzmann equation},
journal = {Physica},
volume = {74},
number = {3},
pages = {466-488},
year = {1974},
issn = {0031-8914},
doi = {https://doi.org/10.1016/0031-8914(74)90355-3},
url = {https://www.sciencedirect.com/science/article/pii/0031891474903553},
author = {J.L. Anderson and H.R. Witting},
}

@article{Chapman,
  title = {Transport coefficients for bulk viscous evolution in the relaxation-time approximation},
  author = {Jaiswal, Amaresh and Ryblewski, Radoslaw and Strickland, Michael},
  journal = {Phys. Rev. C},
  volume = {90},
  issue = {4},
  pages = {044908},
  numpages = {8},
  year = {2014},
  month = {Oct},
  publisher = {American Physical Society},
  doi = {10.1103/PhysRevC.90.044908},
  url = {https://link.aps.org/doi/10.1103/PhysRevC.90.044908}
}

@article{PrattTorrieri,
  title = {Coupling relativistic viscous hydrodynamics to Boltzmann descriptions},
  author = {Pratt, Scott and Torrieri, Giorgio},
  journal = {Phys. Rev. C},
  volume = {82},
  issue = {4},
  pages = {044901},
  numpages = {11},
  year = {2010},
  month = {Oct},
  publisher = {American Physical Society},
  doi = {10.1103/PhysRevC.82.044901},
  url = {https://link.aps.org/doi/10.1103/PhysRevC.82.044901}
}

@article{Beyer:2017gug,
    author = "Beyer, Axel and others",
    title = "{The Rydberg constant and proton size from atomic hydrogen}",
    doi = "10.1126/science.aah6677",
    journal = "Science",
    volume = "358",
    number = "6359",
    pages = "79--85",
    year = "2017"
}

@article{Fleurbaey:2018fih,
    author = "Fleurbaey, H\'el\`ene and Galtier, Sandrine and Thomas, Simon and Bonnaud, Marie and Julien, Lucile and Biraben, Fran\c{c}ois and Nez, Fran\c{c}ois and Abgrall, Michel and Gu\'ena, Jocelyne",
    title = "{New Measurement of the $1S-3S$ Transition Frequency of Hydrogen: Contribution to the Proton Charge Radius Puzzle}",
    eprint = "1801.08816",
    archivePrefix = "arXiv",
    primaryClass = "physics.atom-ph",
    doi = "10.1103/PhysRevLett.120.183001",
    journal = "Phys. Rev. Lett.",
    volume = "120",
    number = "18",
    pages = "183001",
    year = "2018"
}

@article{Bezginov:2019mdi,
    author = "Bezginov, N. and Valdez, T. and Horbatsch, M. and Marsman, A. and Vutha, A. C. and Hessels, E. A.",
    title = "{A measurement of the atomic hydrogen Lamb shift and the proton charge radius}",
    doi = "10.1126/science.aau7807",
    journal = "Science",
    volume = "365",
    number = "6457",
    pages = "1007--1012",
    year = "2019"
}

@article{Xiong:2019umf,
    author = "Xiong, W. and others",
    title = "{A small proton charge radius from an electron\textendash{}proton scattering experiment}",
    doi = "10.1038/s41586-019-1721-2",
    journal = "Nature",
    volume = "575",
    number = "7781",
    pages = "147--150",
    year = "2019"
}

@article{Noronha-Hostler,
    author = "Noronha-Hostler, Jacquelyn and Yan, Li and Gardim, Fernando G. and Ollitrault, Jean-Yves",
    title = "{Linear and cubic response to the initial eccentricity in heavy-ion collisions}",
    eprint = "1511.03896",
    archivePrefix = "arXiv",
    primaryClass = "nucl-th",
    doi = "10.1103/PhysRevC.93.014909",
    journal = "Phys. Rev. C",
    volume = "93",
    number = "1",
    pages = "014909",
    year = "2016"
}

@article{Miller,
    author = "Miller, Michael L. and Reygers, Klaus and Sanders, Stephen J. and Steinberg, Peter",
    title = "{Glauber modeling in high energy nuclear collisions}",
    eprint = "nucl-ex/0701025",
    archivePrefix = "arXiv",
    doi = "10.1146/annurev.nucl.57.090506.123020",
    journal = "Ann. Rev. Nucl. Part. Sci.",
    volume = "57",
    pages = "205--243",
    year = "2007"
}

@article{dEnterria:2020dwq,
    author = "d'Enterria, David and Loizides, Constantin",
    title = "{Progress in the Glauber Model at Collider Energies}",
    eprint = "2011.14909",
    archivePrefix = "arXiv",
    primaryClass = "hep-ph",
    doi = "10.1146/annurev-nucl-102419-060007",
    journal = "Ann. Rev. Nucl. Part. Sci.",
    volume = "71",
    pages = "315--344",
    year = "2021"
}

@article{Gardim:2011xv,
    author = "Gardim, Fernando G. and Grassi, Frederique and Luzum, Matthew and Ollitrault, Jean-Yves",
    title = "{Mapping the hydrodynamic response to the initial geometry in heavy-ion collisions}",
    eprint = "1111.6538",
    archivePrefix = "arXiv",
    primaryClass = "nucl-th",
    doi = "10.1103/PhysRevC.85.024908",
    journal = "Phys. Rev. C",
    volume = "85",
    pages = "024908",
    year = "2012"
}

@article{Loizides,
    author = "Loizides, Constantin",
    title = "{Glauber modeling of high-energy nuclear collisions at the subnucleon level}",
    eprint = "1603.07375",
    archivePrefix = "arXiv",
    primaryClass = "nucl-ex",
    doi = "10.1103/PhysRevC.94.024914",
    journal = "Phys. Rev. C",
    volume = "94",
    number = "2",
    pages = "024914",
    year = "2016"
}

@phdthesis{BernhardThesis,
    author = "Bernhard, Jonah E.",
    title = "{Bayesian parameter estimation for relativistic heavy-ion collisions}",
    eprint = "1804.06469",
    archivePrefix = "arXiv",
    primaryClass = "nucl-th",
    school = "Duke U.",
    month = "4",
    year = "2018"
}

@article{Arnold,
    author = "Arnold, Peter Brockway and Moore, Guy D. and Yaffe, Laurence G.",
    title = "{Effective kinetic theory for high temperature gauge theories}",
    eprint = "hep-ph/0209353",
    archivePrefix = "arXiv",
    doi = "10.1088/1126-6708/2003/01/030",
    journal = "JHEP",
    volume = "01",
    pages = "030",
    year = "2003"
}

@article{Heinz,
    author = "Heinz, Ulrich W.",
    editor = "Stock, R.",
    title = "{Early collective expansion: Relativistic hydrodynamics and the transport properties of QCD matter}",
    eprint = "0901.4355",
    archivePrefix = "arXiv",
    primaryClass = "nucl-th",
    doi = "10.1007/978-3-642-01539-7_9",
    journal = "Landolt-Bornstein",
    volume = "23",
    pages = "240",
    year = "2010"
}

@article{Gale,
    author = "Gale, Charles and Jeon, Sangyong and Schenke, Bjoern",
    title = "{Hydrodynamic Modeling of Heavy-Ion Collisions}",
    eprint = "1301.5893",
    archivePrefix = "arXiv",
    primaryClass = "nucl-th",
    doi = "10.1142/S0217751X13400113",
    journal = "Int. J. Mod. Phys. A",
    volume = "28",
    pages = "1340011",
    year = "2013"
}

@article{Cooper-Frye,
  title = {Single-particle distribution in the hydrodynamic and statistical thermodynamic models of multiparticle production},
  author = {Cooper, Fred and Frye, Graham},
  journal = {Phys. Rev. D},
  volume = {10},
  issue = {1},
  pages = {186--189},
  numpages = {0},
  year = {1974},
  month = {Jul},
  publisher = {American Physical Society},
  doi = {10.1103/PhysRevD.10.186},
  url = {https://link.aps.org/doi/10.1103/PhysRevD.10.186}
}

@article{iSSarticle,
title = {The iEBE-VISHNU code package for relativistic heavy-ion collisions},
journal = {Computer Physics Communications},
volume = {199},
pages = {61-85},
year = {2016},
issn = {0010-4655},
doi = {https://doi.org/10.1016/j.cpc.2015.08.039},
url = {https://www.sciencedirect.com/science/article/pii/S0010465515003392},
author = {Chun Shen and Zhi Qiu and Huichao Song and Jonah Bernhard and Steffen Bass and Ulrich Heinz},
}

@article{Huovinen,
    author = "Huovinen, Pasi and Petersen, Hannah",
    title = "{Particlization in hybrid models}",
    eprint = "1206.3371",
    archivePrefix = "arXiv",
    primaryClass = "nucl-th",
    doi = "10.1140/epja/i2012-12171-9",
    journal = "Eur. Phys. J. A",
    volume = "48",
    pages = "171",
    year = "2012"
}

@article{Bass,
    author = "Bass, S. A. and others",
    title = "{Microscopic models for ultrarelativistic heavy ion collisions}",
    eprint = "nucl-th/9803035",
    archivePrefix = "arXiv",
    doi = "10.1016/S0146-6410(98)00058-1",
    journal = "Prog. Part. Nucl. Phys.",
    volume = "41",
    pages = "255--369",
    year = "1998"
}

@article{Bleicher,
    author = "Bleicher, M. and others",
    title = "{Relativistic hadron hadron collisions in the ultrarelativistic quantum molecular dynamics model}",
    eprint = "hep-ph/9909407",
    archivePrefix = "arXiv",
    doi = "10.1088/0954-3899/25/9/308",
    journal = "J. Phys. G",
    volume = "25",
    pages = "1859--1896",
    year = "1999"
}

@misc{CODATA,
	author    = {{CODATA}},
	title     = "The NIST Reference on Constants, Units, and Uncertainty",
	url       = "https://physics.nist.gov/cgi-bin/cuu/Value?rp",
}

@article{Hofstadter1,
  title = {Elastic Scattering of 188-Mev Electrons from the Proton and the Alpha Particle},
  author = {McAllister, R. W. and Hofstadter, R.},
  journal = {Phys. Rev.},
  volume = {102},
  issue = {3},
  pages = {851--856},
  numpages = {0},
  year = {1956},
  month = {May},
  publisher = {American Physical Society},
  doi = {10.1103/PhysRev.102.851},
  url = {https://link.aps.org/doi/10.1103/PhysRev.102.851}
}

@article{GlueX,
  title = {First Measurement of Near-Threshold $J/\ensuremath{\psi}$ Exclusive Photoproduction off the Proton},
  collaboration = {GlueX Collaboration},
  journal = {Phys. Rev. Lett.},
  volume = {123},
  issue = {7},
  pages = {072001},
  numpages = {6},
  year = {2019},
  month = {Aug},
  publisher = {American Physical Society},
  doi = {10.1103/PhysRevLett.123.072001},
  url = {https://link.aps.org/doi/10.1103/PhysRevLett.123.072001}
}

@article{Borghini:2000sa,
    author = "Borghini, Nicolas and Dinh, Phuong Mai and Ollitrault, Jean-Yves",
    title = "{A New method for measuring azimuthal distributions in nucleus-nucleus collisions}",
    eprint = "nucl-th/0007063",
    archivePrefix = "arXiv",
    reportNumber = "SACLAY-SPH-T-00-110",
    doi = "10.1103/PhysRevC.63.054906",
    journal = "Phys. Rev. C",
    volume = "63",
    pages = "054906",
    year = "2001"
}

@article{Borghini:2001vi,
    author = "Borghini, Nicolas and Dinh, Phuong Mai and Ollitrault, Jean-Yves",
    title = "{Flow analysis from multiparticle azimuthal correlations}",
    eprint = "nucl-th/0105040",
    archivePrefix = "arXiv",
    doi = "10.1103/PhysRevC.64.054901",
    journal = "Phys. Rev. C",
    volume = "64",
    pages = "054901",
    year = "2001"
}

@article{Bilandzic:2010jr,
    author = "Bilandzic, Ante and Snellings, Raimond and Voloshin, Sergei",
    title = "{Flow analysis with cumulants: Direct calculations}",
    eprint = "1010.0233",
    archivePrefix = "arXiv",
    primaryClass = "nucl-ex",
    doi = "10.1103/PhysRevC.83.044913",
    journal = "Phys. Rev. C",
    volume = "83",
    pages = "044913",
    year = "2011"
}

@article{PHOBOS:2006dbo,
    author = "Alver, B. and others",
    collaboration = "PHOBOS",
    title = "{System size, energy, pseudorapidity, and centrality dependence of elliptic flow}",
    eprint = "nucl-ex/0610037",
    archivePrefix = "arXiv",
    doi = "10.1103/PhysRevLett.98.242302",
    journal = "Phys. Rev. Lett.",
    volume = "98",
    pages = "242302",
    year = "2007"
}

@article{DUKE1,
    author = "Bernhard, Jonah E. and Moreland, J. Scott and Bass, Steffen A. and Liu, Jia and Heinz, Ulrich",
    title = "{Applying Bayesian parameter estimation to relativistic heavy-ion collisions: simultaneous characterization of the initial state and quark-gluon plasma medium}",
    eprint = "1605.03954",
    archivePrefix = "arXiv",
    primaryClass = "nucl-th",
    doi = "10.1103/PhysRevC.94.024907",
    journal = "Phys. Rev. C",
    volume = "94",
    number = "2",
    pages = "024907",
    year = "2016"
}

@article{Schenke:2020mbo,
    author = "Schenke, Bjoern and Shen, Chun and Tribedy, Prithwish",
    title = "{Running the gamut of high energy nuclear collisions}",
    eprint = "2005.14682",
    archivePrefix = "arXiv",
    primaryClass = "nucl-th",
    doi = "10.1103/PhysRevC.102.044905",
    journal = "Phys. Rev. C",
    volume = "102",
    number = "4",
    pages = "044905",
    year = "2020"
}

@article{DUKE2,
    author = "Bernhard, Jonah E. and Moreland, J. Scott and Bass, Steffen A.",
    title = "{Bayesian estimation of the specific shear and bulk viscosity of quark\textendash{}gluon plasma}",
    doi = "10.1038/s41567-019-0611-8",
    journal = "Nature Phys.",
    volume = "15",
    number = "11",
    pages = "1113--1117",
    year = "2019"
}

@article{JETSCAPE:2020mzn,
    author = "Everett, D. and others",
    collaboration = "JETSCAPE",
    title = "{Multisystem Bayesian constraints on the transport coefficients of QCD matter}",
    eprint = "2011.01430",
    archivePrefix = "arXiv",
    primaryClass = "hep-ph",
    doi = "10.1103/PhysRevC.103.054904",
    journal = "Phys. Rev. C",
    volume = "103",
    number = "5",
    pages = "054904",
    year = "2021"
}

@article{Nijs:2020ors,
    author = {Nijs, Govert and van der Schee, Wilke and G\"ursoy, Umut and Snellings, Raimond},
    title = "{Transverse Momentum Differential Global Analysis of Heavy-Ion Collisions}",
    eprint = "2010.15130",
    archivePrefix = "arXiv",
    primaryClass = "nucl-th",
    reportNumber = "CERN-TH-2020-174, MIT-CTP/5250",
    doi = "10.1103/PhysRevLett.126.202301",
    journal = "Phys. Rev. Lett.",
    volume = "126",
    number = "20",
    pages = "202301",
    year = "2021"
}

@article{Nijs:2020roc,
    author = {Nijs, Govert and van der Schee, Wilke and G\"ursoy, Umut and Snellings, Raimond},
    title = "{Bayesian analysis of heavy ion collisions with the heavy ion computational framework Trajectum}",
    eprint = "2010.15134",
    archivePrefix = "arXiv",
    primaryClass = "nucl-th",
    reportNumber = "CERN-TH-2020-175, MIT-CTP/5251",
    doi = "10.1103/PhysRevC.103.054909",
    journal = "Phys. Rev. C",
    volume = "103",
    number = "5",
    pages = "054909",
    year = "2021"
}

@article{Nijs:2021clz,
    author = "Nijs, Govert and van der Schee, Wilke",
    title = "{Predictions and postdictions for relativistic lead and oxygen collisions with $Trajectum$}",
    eprint = "2110.13153",
    archivePrefix = "arXiv",
    primaryClass = "nucl-th",
    reportNumber = "CERN-TH-2021-160, MIT-CTP/5333",
    month = "10",
    year = "2021"
}

@article{Hofstadter2,
    author = "Hofstadter, R.",
    title = "{Nuclear and nucleon scattering of high-energy electrons}",
    doi = "10.1146/annurev.ns.07.120157.001311",
    journal = "Ann. Rev. Nucl. Part. Sci.",
    volume = "7",
    pages = "231--316",
    year = "1957"
}

@article{PhysRev.86.288,
  title = {Electrodynamic Displacement of Atomic Energy Levels. II. Lamb Shift},
  author = {Karplus, Robert and Klein, Abraham and Schwinger, Julian},
  journal = {Phys. Rev.},
  volume = {86},
  issue = {3},
  pages = {288--301},
  numpages = {0},
  year = {1952},
  month = {May},
  publisher = {American Physical Society},
  doi = {10.1103/PhysRev.86.288},
  url = {https://link.aps.org/doi/10.1103/PhysRev.86.288}
}

@article{Pohl,
    author = "Pohl, Randolf and Gilman, Ronald and Miller, Gerald A. and Pachucki, Krzysztof",
    title = "{Muonic hydrogen and the proton radius puzzle}",
    eprint = "1301.0905",
    archivePrefix = "arXiv",
    primaryClass = "physics.atom-ph",
    doi = "10.1146/annurev-nucl-102212-170627",
    journal = "Ann. Rev. Nucl. Part. Sci.",
    volume = "63",
    pages = "175--204",
    year = "2013"
}

@article{Hammer,
    author = "Hammer, Hans-Werner and Mei\ss{}ner, Ulf-G",
    title = "{The proton radius: From a puzzle to precision}",
    eprint = "1912.03881",
    archivePrefix = "arXiv",
    primaryClass = "hep-ph",
    doi = "10.1016/j.scib.2019.12.012",
    journal = "Sci. Bull.",
    volume = "65",
    pages = "257--258",
    year = "2020"
}

@article{JPsiScattering,
  title = {Investigating the gluonic structure of nuclei via $J/\ensuremath{\psi}$ scattering},
  author = {Caldwell, A. and Kowalski, H.},
  journal = {Phys. Rev. C},
  volume = {81},
  issue = {2},
  pages = {025203},
  numpages = {17},
  year = {2010},
  month = {Feb},
  publisher = {American Physical Society},
  doi = {10.1103/PhysRevC.81.025203},
  url = {https://link.aps.org/doi/10.1103/PhysRevC.81.025203}
}

@article{Kharzeev,
    author = "Kharzeev, Dmitri E.",
    title = "{Mass radius of the proton}",
    eprint = "2102.00110",
    archivePrefix = "arXiv",
    primaryClass = "hep-ph",
    doi = "10.1103/PhysRevD.104.054015",
    journal = "Phys. Rev. D",
    volume = "104",
    number = "5",
    pages = "054015",
    year = "2021"
}

@misc{TRENToWebsite,
	author    = {{Bernhard, J. E., Moreland, J. S., Bass, S. A.}},
	title     = "Reduced Thickness Event-by-event Nuclear Topology (TRENTo) official website",
	url       = "http://qcd.phy.duke.edu/trento/"
}

@article{TRENToArticle,
	title = {Alternative ansatz to wounded nucleon and binary collision scaling in high-energy nuclear collisions},
	author = {{Bernhard, J. E., Moreland, J. S., Bass, S. A.}},
	journal = {Phys. Rev. C},
	volume = {92},
	issue = {1},
	pages = {011901},
	numpages = {6},
	year = {2015},
	month = {Jul},
	publisher = {American Physical Society},
	doi = {10.1103/PhysRevC.92.011901}
}

@misc{MUSICWebsite,
	author    = "Gabriel Denicol et. al.",
	title     = "A (3+1)D hydrodynamic code for heavy-ion collisions (MUSIC) official website",
	note       = "\url{http://www.physics.mcgill.ca/music/}"
}

@misc{iSSWebsite,
	author    = "Chun Shen",
	title     = "iSS, Monte Carlo sampler for particle distribution from Cooper-Frye freeze-out procedure",
	note       = "\url{https://github.com/chunshen1987/iSS}"
}

@misc{UrQMDWebsite,
	author    = {{Frankfurt Institute for Advanced Studies}},
	title     = "\textit{Ultrarelativistic Quantum Molecular Dynamics (UrQMD) official website}",
	url       = {https://urqmd.org/}
}

@article{KompostArticle,
	author         = "Kurkela, Aleksi and Mazeliauskas, Aleksas and Paquet,
	Jean-François and Schlichting, Sören and Teaney, Derek",
	title          = "{Effective kinetic description of event-by-event
	pre-equilibrium dynamics in high-energy heavy-ion
	collisions}",
	journal        = "Phys. Rev.",
	volume         = "C99",
	year           = "2019",
	number         = "3",
	pages          = "034910",
	doi            = "10.1103/PhysRevC.99.034910",
	eprint         = "1805.00961",
	archivePrefix  = "arXiv",
	primaryClass   = "hep-ph",
	SLACcitation   = "%%CITATION = ARXIV:1805.00961;%%"
}

@article{Snellings:2011sz,
    author = "Snellings, Raimond",
    title = "{Elliptic Flow: A Brief Review}",
    eprint = "1102.3010",
    archivePrefix = "arXiv",
    primaryClass = "nucl-ex",
    doi = "10.1088/1367-2630/13/5/055008",
    journal = "New J. Phys.",
    volume = "13",
    pages = "055008",
    year = "2011"
}

@article{Giacalone:2021clp,
    author = {Giacalone, Giuliano and Schenke, Bj\"orn and Shen, Chun},
    title = "{Constraining the Nucleon Size with Relativistic Nuclear Collisions}",
    eprint = "2111.02908",
    archivePrefix = "arXiv",
    primaryClass = "nucl-th",
    doi = "10.1103/PhysRevLett.128.042301",
    journal = "Phys. Rev. Lett.",
    volume = "128",
    number = "4",
    pages = "042301",
    year = "2022"
}

@article{Nijs:2022rme,
    author = "Nijs, Govert and van der Schee, Wilke",
    title = "{The hadronic nucleus-nucleus cross section and the nucleon size}",
    eprint = "2206.13522",
    archivePrefix = "arXiv",
    primaryClass = "nucl-th",
    reportNumber = "CERN-TH-2022-103;MIT-CTP/5445",
    month = "6",
    year = "2022"
}

@article{NunesdaSilva:2020bfs,
    author = "Nunes da Silva, Tiago and Chinellato, David and Hippert, Mauricio and Serenone, Willian and Takahashi, Jun and Denicol, Gabriel S. and Luzum, Matthew and Noronha, Jorge",
    title = "{Pre-hydrodynamic evolution and its signatures in final-state heavy-ion observables}",
    eprint = "2006.02324",
    archivePrefix = "arXiv",
    primaryClass = "nucl-th",
    doi = "10.1103/PhysRevC.103.054906",
    journal = "Phys. Rev. C",
    volume = "103",
    pages = "054906",
    year = "2021"
}

@article{Gale:2021emg,
    author = {Gale, Charles and Paquet, Jean-Fran\c{c}ois and Schenke, Bj\"orn and Shen, Chun},
    title = "{Multimessenger heavy-ion collision physics}",
    eprint = "2106.11216",
    archivePrefix = "arXiv",
    primaryClass = "nucl-th",
    doi = "10.1103/PhysRevC.105.014909",
    journal = "Phys. Rev. C",
    volume = "105",
    number = "1",
    pages = "014909",
    year = "2022"
}

@article{Kurkela:2018qwx,
    author = {Kurkela, Aleksi and Mazeliauskas, Aleksas and Paquet, Jean-Fran\c{c}ois and Schlichting, S\"oren and Teaney, Derek},
    editor = "d'Enterria, David and Morsch, Andreas and Crochet, Philippe",
    title = "{2+1D simulations of pre-equilibrium stage with QCD kinetic theory}",
    doi = "10.22323/1.345.0111",
    journal = "PoS",
    volume = "HardProbes2018",
    pages = "111",
    year = "2018"
}

@article{MCMC,
    author = {Hastings, W. K.},
    title = "{Monte Carlo sampling methods using Markov chains and their applications}",
    journal = {Biometrika},
    volume = {57},
    number = {1},
    pages = {97-109},
    year = {1970},
    month = {04},
    issn = {0006-3444},
    doi = {10.1093/biomet/57.1.97},
    url = {https://doi.org/10.1093/biomet/57.1.97},
    eprint = {https://academic.oup.com/biomet/article-pdf/57/1/97/23940249/57-1-97.pdf},
}

\end{document}